\documentclass[12pt]{iopart}
\usepackage{iopams}
\usepackage{setstack}
\usepackage{graphicx}
\begin{document}
\title[Dark energy interacting with dark matter] {Reconstruction
of the interaction term between dark matter and dark energy using
SNe Ia}
\author{Freddy Cueva Solano and Ulises Nucamendi}

\address{Instituto de F\'{\i}sica y Matem\'aticas\\
Universidad Michoacana de San Nicol\'as de Hidalgo \\
Edificio C-3, Ciudad Universitaria, CP. 58040\\
Morelia, Michoac\'an, M\'exico\\ }


\eads{\mailto{freddy@ifm.umich.mx}, \mailto{ulises@ifm.umich.mx}}

%

\begin{abstract}
We apply a parametric reconstruction method to a homogeneous,
isotropic and spatially flat Friedmann-Robertson-Walker (FRW)
cosmological model filled of a fluid of dark energy (DE) with
constant equation of state (EOS) parameter interacting with dark
matter (DM). The reconstruction method is based on expansions of the
general interaction term and the relevant cosmological variables in
terms of Chebyshev polynomials which form a complete set orthonormal
functions. This interaction term describes an exchange of energy
flow between the DE and DM within dark sector. To show how the
method works we do the reconstruction of the interaction function
expanding it in terms of only the first six Chebyshev polynomials
and obtain the best estimation for the coefficients of the expansion
assuming three models: (a) a DE equation of the state parameter $w
=-1$ (an interacting cosmological $\Lambda$), (b) a DE equation of
the state parameter $w =$ constant with a dark matter density parameter
fixed, (c) a DE equation of the state parameter $w =$ constant with
a free constant dark matter density parameter to be estimated, and using the
Union2 SNe Ia data set from ``The Supernova Cosmology Project'' (SCP)
composed by 557 type Ia supernovae. In both cases, the preliminary reconstruction
shows that in the best scenario there exist the possibility of a
crossing of the noninteracting line $Q=0$ in the recent past within
the $1\sigma$ and $2\sigma$ errors from positive values at early
times to negative values at late times. This means that, in this
reconstruction, there is an energy transfer from DE to DM at early
times and an energy transfer from DM to DE at late times. We
conclude that this fact is an indication of the possible existence
of a crossing behavior in a general interaction coupling between
dark components.
\end{abstract}

\pacs{95.36.+x, 98.80.-k, 98.80.Es} \submitto{Journal of Cosmology
and Astroparticle physics} \maketitle

\section{Introduction}
In the last years the accelerated expansion of the universe has now
been confirmed by several independent observations including those
of high redshift $(z\leq 1)$ type Ia Supernovae (SNeIa) data at
cosmological distances \cite{Riess1998}-\cite{AmanullahUnion22010}.
This has been verified by precise measurements of the power spectrum
of the cosmic microwave background (CMB) anisotropies
\cite{WMAP}-\cite{Komatsu2011}, the galaxy power spectrum detection
and the baryon acoustic peak in the large-scale correlation function
of luminous red galaxies in the experiment Sloan Digital Sky Survey
(SDSS) \cite{SDSS}-\cite{ReidSDSS}. To explain these observations,
it has been postulated the existence of a new and enigmatic
component of the universe so-called dark energy (DE)
\cite{Peebles1988}-\cite{Copeland-Sami-2006} from which the
cosmological constant is the simplest model \cite{Peebles2003},
\cite{Weinberg1989}-\cite{Padmanabhan2003}. Recent observations
\cite{AmanullahUnion22010}, \cite{Komatsu2011}, \cite{ReidSDSS},
\cite{Vikhlinin2009}-\cite{Rozo2010} show that if it is assumed a
dark energy (DE) equation of state (EOS) with constant parameter $w
= P_{DE}/ \rho_{DE}$, then there remains little room for departure
of DE from the cosmological constant. In addition these observations
indicate that our universe is flat and it consists of approximately
$70\%$ of Dark Energy (DE) in the form of a cosmological constant,
$25\%$ of Dark Matter and $5\%$ of baryonic matter.

However the cosmological constant model has two serious problems:
the first of them is the \textit{cosmological constant problem}
\cite{Sahni-2002}, \cite{Peebles2003},
\cite{Weinberg1989}-\cite{Padmanabhan2003} which consists in why the
observed value of the Cosmological Constant $\rho_{\Lambda}^{obs}
\sim (10^{-12} \,\, \rm{Gev})^{4}$ is so-small compared with the
theoretical value $\rho_{\Lambda}^{Pl} \sim (10^{18} \,\,
\rm{Gev})^{4}$ predicted from local quantum field theory if we are
confident in its application to the Planck scale?. The second
problem is the so named \textit{The Cosmic Coincidence problem}
\cite{Steinhardt1997}-\cite{Copeland-Sami-2006} consisting in why,
in the present, the energy density of DE is comparable with the
density of dark matter (DM) while the first one is subdominant
during almost all the past evolution of the universe?.

In the last decade, in order to solve the \textit{The Cosmic
Coincidence problem}, several researchers have considered a possible
phenomenological interaction between the DE and DM components
\cite{Amendola2000}-\cite{Lip2011}. As far as we know, the first
models of dark energy coupled with dark matter were proposed by
Wetterich \cite{Wetterich1988}, \cite{Wetterich1995} in the
framework of a scalar field with an exponential potential (named The
Cosmon) coupled with the matter. Some years later, with the
discovery of the recent accelerated expansion of the universe
\cite{Riess1998}-\cite{AmanullahUnion22010} and in order to solve
the coincidence problem, several authors put forward the idea of a
coupled scalar field with dark matter named Coupled Quintessence
\cite{Amendola2000}-\cite{amendola2004},
\cite{campo-olivares2006}-\cite{Huey-Wandelt-2006},
\cite{campo-herrera-pavon2008},
\cite{jesus-santos-alcaniz-2008}-\cite{pettorino2008},
\cite{Lopez-Mena-Grigoris2010}. On the other hand, the theory of
dynamical systems have been applied to different models of coupled
dark energy in order to clarify the cosmological evolution of the
solutions of every model with emphasis in the study of the critical
points \cite{Curbelo-Tame-Quiros2006}-\cite{cabral-maartens2010}.

Some recent studies have claimed that, for reasonable and suitably
chosen interaction terms, the coincidence problem can be
significantly ameliorated in the sense that the rate of densities $r
\equiv \rho_{DM}/ \rho_{DE}$ either tends to a constant or varies
more slowly than the scale factor, $a(t)$, in late times
\cite{campo-herrera-pavon2008}, \cite{campo-herrera-pavon2009}.
However, the existence or not of some class of interaction between
dark components is to be discerned observationally. To this respect,
constraints on the strength of such interaction have been put using
different observations \cite{pasqui}-\cite{He-Zhang}.

Recently, it has been suggested that an interacting term $Q(z)$
dependent of the redshift crosses the noninteracting line $Q(z)=0$
\cite{Cai-Su}-\cite{He-Zhang}. In \cite{Cai-Su}, this conclusion
have been obtained using observational data samples in the range $z
\in [0, 1.8]$ in order to fit a scenario in which the whole redshift
range is divided into a determined numbers of bins and the
interaction function is set to be a constant in each bin. They found
an oscillatory behavior of the interaction function $Q(z)$ changing
its sign several times during the evolution of the universe. On the
other hand, in \cite{He-Zhang} is reported a crossing of the
noninteracting line $Q(z) = 0$ under the assumption that the
interacting term $Q(z)$ is a linearly dependent interacting function
of the scale factor with two free parameters to be estimated. They
found a crossing from negative values at the past (energy transfers
from dark matter to dark energy) to positive values at the present
(energy transfers from dark energy to dark matter) at $z \simeq
0.2-0.3$.

While it is not totally clear if an interaction term can solved the
\textit{The Cosmic Coincidence problem} or if such crossing really
exists, we can yet put constraints on the size of such assumed
general interaction and on the probability of existence of such
crossing using recent cosmological data. We will do this postulating
the existence of an general nongravitational interaction between the
two dark components. We introduce phenomenologically this general
interaction term $Q$ into the equations of motion of DE and DM,
which describes an energy exchange between these components
\cite{Amendola2000}-\cite{Lip2011}. In order to reconstruct the
interaction term $Q$ as a function of the redshift we expand it in
terms of Chebyshev Polynomials which constitute a complete
orthonormal basis on the finite interval [-1,1] and have the nice
property to be the minimax approximating polynomial (this technique
has been applied to the reconstruction of the DE potential in
\cite{Simon2005}-\cite{Martinez2008}). At the end, we do the
reconstruction using the observations of ``The Supernova Cosmology
Project'' (SCP) composed by 557 type Ia supernovae
\cite{AmanullahUnion22010}.

Due to that in this paper our principal goals are: (i) the
development of the formalism of reconstruction of the interaction
and (ii) the recent reconstruction of the evolution of that
interaction, we do not include another data sets like CMB
anisotropies, the galaxy power spectrum or the baryon acoustic peak
(BAO) measured in the experiment SDSS. Clearly the use of these data
sets implies special considerations such as the application of the
cosmological perturbation theory in the reconstruction method which
is beyond the scope of this paper. We will do the total
reconstruction in the evolution of the interaction in our future
work.

In our reconstruction process we assume two interacting models: (a)
a DE equation of the state parameter $w =-1$ (an interacting
cosmological $\Lambda$) and (b) a DE equation of the state parameter
$w =$ constant (as far as know the only reference proposing a
reconstruction process of coupled dark energy using
parameterizations of the coupling function is \cite{Rosenfeld2007}).

The organization of this paper is a follow. In the second section we
introduce the general equations of motion of the DE model
interacting with DM. In the third section, we write the cosmological
equations for both interacting dark components. In the forth section
we develop the reconstruction scheme of the interaction term in
terms of a expansion of Chebyshev polynomials. In the fifth section,
we briefly describe the application of the type Ia Supernova data
cosmological test and the priors used on the free parameters of the
reconstruction together with a brief discussion of the results of
our reconstruction and the best estimated values of the parameters
fitting the observations. Finally, in the last section we discuss
our main results and present our conclusions.

\section{General equations of motion for dark energy interacting with dark matter.}
\label{General equations of motion for dark energy interacting with dark
matter}

We assume an universe formed by four components: the baryonic matter
fluid $(b)$, the radiation fluid $(r)$, the dark matter fluid $(DM)$
and the dark energy fluid $(DE)$. Moreover all these constituents
are interacting gravitationally and additionally only the dark
components interact nongravitationally through an energy exchange
between them mediated by the interaction term defined below.

The gravitational equations of motion are the Einstein field
equations
\\
\begin{equation}
\label{EinsteinEquations} G_{\mu\nu} = 8\pi G \left[ T^{b}_{\mu\nu}
+ T^{r}_{\mu\nu} + T^{DM}_{\mu\nu} + T^{DE}_{\mu\nu} \right] ,
\end{equation}
whereas that the equations of motion for each fluid are
\begin{equation}
\label{FluidsEquationsb} \nabla^{\nu} T^{b}_{\mu\nu} = 0 ,
\end{equation}
\begin{equation}
\label{FluidsEquationsr} \nabla^{\nu} T^{r}_{\mu\nu} = 0 ,
\end{equation}
\begin{equation}
\label{FluidsEquationsDM} \nabla^{\nu} T^{DM}_{\mu\nu} = - F_{\mu} ,
\end{equation}
\begin{equation}
\label{FluidsEquationsDE} \nabla^{\nu} T^{DE}_{\mu\nu} = F_{\mu} ,
\end{equation}
where the respective energy-momentum tensor for the fluid $i$ is
defined as $(i = b, r, DM, DE)$,
\begin{equation}\label{Energy_momentum_tensor}
T^{i}_{\mu\nu}=\rho_{i} \,u_\mu u_\nu + (g_{\mu\nu}+u_\mu
u_\nu)P_{i}
\end{equation}
here $u_\mu$ is the velocity of the fluids (assumed to be the same
for each one) where as $\rho_{i}$ and $P_{i}$ are respectively the
density and pressure of the fluid $i$ measured by an observer with
velocity $u^\mu$. $F_{\mu}$ is the cuadrivector of interaction
between dark components and its form is not known a priori because
in general we do not have fundamental theory, in case of existing,
to predict its structure. \\
\\
We project the equations
(\ref{FluidsEquationsb})-(\ref{FluidsEquationsDE}) in a part
parallel to the velocity $u^\mu$,
\begin{equation}
\label{FluidsEquationsbP} u^\mu \nabla^{\nu} T^{b}_{\mu\nu} = 0 ,
\end{equation}
\begin{equation}
\label{FluidsEquationsrP} u^\mu \nabla^{\nu} T^{r}_{\mu\nu} = 0 ,
\end{equation}
\begin{equation}
\label{FluidsEquationsDMP} u^\mu \nabla^{\nu} T^{DM}_{\mu\nu} = -
u^\mu F_{\mu} ,
\end{equation}
\begin{equation}
\label{FluidsEquationsDEP} u^\mu \nabla^{\nu} T^{DE}_{\mu\nu} =
u^\mu F_{\mu} ,
\end{equation}
and in other part orthogonal to the velocity using the projector
$h_{\beta\mu} = g_{\beta\mu} + u_{\beta}u_{\mu}$ acting on the
hypersurface orthogonal to the velocity $u^\mu$,
\begin{equation}
\label{FluidsEquationsbO} h^{\mu\beta} \nabla^{\nu} T^{b}_{\mu\nu} =
0 ,
\end{equation}
\begin{equation}
\label{FluidsEquationsrO} h^{\mu\beta} \nabla^{\nu} T^{r}_{\mu\nu} =
0 ,
\end{equation}
\begin{equation}
\label{FluidsEquationsDMO} h^{\mu\beta} \nabla^{\nu} T^{DM}_{\mu\nu}
= - h^{\mu\beta} F_{\mu} ,
\end{equation}
\begin{equation}
\label{FluidsEquationsDEO} h^{\mu\beta} \nabla^{\nu} T^{DE}_{\mu\nu}
= h^{\mu\beta} F_{\mu} ,
\end{equation}
using (\ref{Energy_momentum_tensor}) in
(\ref{FluidsEquationsbP})-(\ref{FluidsEquationsDEP}) we obtain the
mass energy conservation equations for each fluid,
\begin{equation}
\label{MECB} u^{\mu} \nabla_{\mu} \rho_{b} + \left( \rho_{b} + P_{b}
\right) \nabla_{\mu} u^{\mu} = 0 ,
\end{equation}
\begin{equation}
\label{MECr} u^{\mu} \nabla_{\mu} \rho_{r} + \left( \rho_{r} + P_{r}
\right) \nabla_{\mu} u^{\mu} = 0  ,
\end{equation}
\begin{equation}
\label{MECDM} u^{\mu} \nabla_{\mu} \rho_{DM} + \left( \rho_{DM} +
P_{DM} \right) \nabla_{\mu} u^{\mu} = u^{\mu}F_{\mu}  ,
\end{equation}
\begin{equation}
\label{MECDE} u^{\mu} \nabla_{\mu} \rho_{DE} + \left( \rho_{DE} +
P_{DE} \right) \nabla_{\mu} u^{\mu} = - u^{\mu}F_{\mu} ,
\end{equation}
at the other hand it introducing (\ref{Energy_momentum_tensor}) in
(\ref{FluidsEquationsbO})-(\ref{FluidsEquationsDEO}) it permits to
have the Euler equations for every fluid,
\begin{equation}
\label{EEB} h^{\mu\beta} \nabla_{\mu} P_{b} + \left( \rho_{b} +
P_{b} \right) u^{\mu} \nabla_{\mu} u^{\beta} = 0,
\end{equation}
\begin{equation}
\label{EEr} h^{\mu\beta} \nabla_{\mu} P_{r} + \left( \rho_{r} +
P_{r} \right) u^{\mu} \nabla_{\mu} u^{\beta} = 0,
\end{equation}

\begin{equation}
\label{EEDM} h^{\mu\beta} \nabla_{\mu} P_{DM} + \left( \rho_{DM} +
P_{DM} \right) u^{\mu} \nabla_{\mu} u^{\beta} = -
h^{\mu\beta}F_{\mu}  ,
\end{equation}
\begin{equation}
\label{EEDE} h^{\mu\beta} \nabla_{\mu} P_{DE} + \left( \rho_{DE} +
P_{DE} \right) u^{\mu} \nabla_{\mu} u^{\beta} = h^{\mu\beta} F_{\mu}
,
\end{equation}
Finally we closed the system of equations assuming the following
state equations for the respectively baryonic, dark matter,
radiation components,
\\
\begin{eqnarray}\label{State EquationB}
P_{b} &=& 0 \,
\\
\label{State EquationDM} P_{DM} &=& 0  \,
\\
\label{State EquationR} P_{r} &=& \frac{1}{3} \, \rho_{r} \,
\end{eqnarray}
while for the dark energy we assume a state equation with constant
parameter $w$,
\begin{eqnarray}\label{State EquationDE}
P_{DE} &=& w \rho_{DE} \,
\end{eqnarray}

\section{Cosmological Equations of motion for dark energy interacting with dark matter.}
\label{Equations of motion for dark energy interacting with dark
matter}

We assumed that the background metric is described by the flat
Friedmann-Robertson-Walker (FRW) metric written in comoving
coordinates as supported by the anisotropies of the cosmic microwave
background (CMB) radiation measured by the WMAP experiment
\cite{WMAP}
\begin{equation}
\label{metricFRW} ds^{2}=-dt^{2} + a^{2}(t) \left( dr^{2} +
r^{2}d\Omega^2 \right ),
\end{equation}
where $a(t)$ is the scale factor and $t$ is the cosmic time.

In these coordinates we choose for the normalized velocity,
\begin{equation}
\label{velocity}
u^{\mu} = (1,0,0,0)
\end{equation}
and therefore we have,
\begin{eqnarray}
\label{expansion}
\nabla_{\mu} u^{\mu} &=& 3\, \frac{\dot{a}}{a} \equiv 3H
\\
\label{derivative velocity}
u^{\mu} \nabla_{\mu} u^{\beta} &=& 0
\end{eqnarray}
where $H$ is the Hubble parameter and the point means derivative respect to the cosmic time.
In congruence with the symmetries of spatial isotropy and homogeneity of the FRW spacetime,
the densities and pressures of the fluids are depending only of the cosmic time, $\rho_{i}(t)$, $P_{i}(t)$,
and at the same time the parallel and orthogonal components of the cuadrivector of interaction
with respect to the velocity are respectively,
\begin{eqnarray}
\label{nonullparallel}
u^{\mu} F_{\mu} &=& Q(a)
\\
\label{nullorthogonal}
 h^{\mu\beta} F_{\mu} &=& 0
\end{eqnarray}
where $Q(a)$ is known as the interaction function depending on the scale factor.
The introduction of the state equations (\ref{State EquationB})-(\ref{State EquationDE}),
the metric (\ref{metricFRW}) and the expressions (\ref{velocity})-(\ref{nullorthogonal})
in the equations of mass energy conservation for the fluids (\ref{MECB})-(\ref{MECDE}) produces,
\begin{equation}
\label{EoFB}
{\dot \rho}_{b} + 3 H \rho_{b} = 0  ,
\end{equation}
\begin{equation}
\label{EoFr}
{\dot \rho}_{r} + 4 H \rho_{r} = 0  ,
\end{equation}
\begin{equation}
\label{EoFDM}
{\dot \rho}_{DM} + 3 H \rho_{DM} = Q  ,
\end{equation}
\begin{equation}
\label{EoFDE}
{\dot \rho}_{DE} + 3 \left(1 + w \right) H \rho_{DE} = - Q  ,
\end{equation}
At the other hand, the Euler equations (\ref{EEB})-(\ref{EEDE}) are
satisfied identically and do not produce any new equation. From the
Einstein equation (\ref{EinsteinEquations}) we complete the
equations of motion with the first Friedmann equation,
\begin{equation}
\label{eq:hubble1} H^{2}\left(a \right) = \frac{8\pi G}{3}
\left(\rho_{b} + \rho_{r} + \rho_{DM} + \rho_{DE} \right).
\end{equation}
Its convenient to define the following dimensionless density
parameters $\Omega^{\star}_{i}$, for $i=b, r, DM, DE$, as the energy
densities normalized by the critical density at the actual epoch,
\begin{equation}
\label{densityparameters} \Omega^{\star}_{i} \equiv
\frac{\rho_{i}}{\rho_{crit}^{0}} ,\\
\end{equation}
and the corresponding dimensionless density parameters at the
present,
\begin{equation}
\label{densityparameterstoday} \Omega^{0}_{i} \equiv
\frac{\rho^{0}_{i}}{\rho_{crit}^{0}} ,\\
\end{equation}
where $\rho_{crit}^{0} \equiv 3H^2_0/8\pi G$ is the critical density
today and $H_0$ is the Hubble constant. Solving (\ref{EoFB}) and
(\ref{EoFr}) in terms of the redshift $z$, defined as $a = 1/(1+z)$,
we obtain the known solutions for the baryonic matter and radiation
density parameters respectively:
\begin{equation}
\label{Omegabs} \Omega^{\star}_{b}(z) =
\Omega_{b}^{0}{(1+z)}^{3},\\
\end{equation}
\begin{equation}
\label{Omegars} \Omega^{\star}_{r}(z) =
\Omega_{r}^{0}{(1+z)}^{4},\\
\end{equation}
The energy conservation equations (\ref{EoFDM}) and (\ref{EoFDE})
for both dark components are rewritten in terms of the redshift as:
\begin{eqnarray}
\label{EDFDM} \frac{\mathrm{d}{\rho}_{DM}}{\mathrm{d}z} -
\frac{3}{1+z} \,\rho_{DM}= -
\frac{Q(z)}{(1 + z) \cdot H(z)},\\
\label{EDFDE} \frac{\mathrm{d}{\rho}_{DE}}{\mathrm{d}z} -
\frac{3(1+w)}{1 + z}\, \rho_{DE} = \frac{Q(z)}{(1+z) \cdot H(z)},
\end{eqnarray}
Phenomenologically, we choose to describe the interaction between
the two dark fluids as an exchange of energy at a rate proportional
to the Hubble parameter: \\
\begin{eqnarray}
\label{Interaction} Q(z) &\equiv& \rho_{crit}^{0}\cdot(1+z)^{3}
\cdot H(z)\cdot {\rm I}_{\rm Q}(z),
\end{eqnarray}
\\
The term $\rho_{crit}^{0}\cdot(1+z)^{3}$ has been introduced by
convenience in order to mimic a rate proportional to the behavior of
a matter density without interaction. Let be note that the
dimensionless interaction function ${\rm I}_{\rm Q}(z)$ depends
of the redshift and it will be the function to be reconstructed.
With the help of (\ref{Interaction}) we rewrite the equations for
the dark fluids (\ref{EDFDM})-(\ref{EDFDE}) as,
\begin{eqnarray}
\label{EDFDMOmega} \frac{\mathrm{d}
\Omega^{\star}_{DM}}{\mathrm{d}z} -
\frac{3}{1+z} \,\Omega^{\star}_{DM}= - (1 + z)^2 \cdot {\rm I}_{\rm Q}(z),\\
\label{EDFDEOmega} \frac{\mathrm{d}
\Omega^{\star}_{DE}}{\mathrm{d}z} - \frac{3(1+w)}{1 + z}\,
\Omega^{\star}_{DE}= (1 + z)^2 \cdot {\rm I}_{\rm Q}(z),
\end{eqnarray}

\section{General Reconstruction of the interaction using Chebyshev
polynomials.} \label{General Reconstruction}

We do the parametrization of the dimensionless coupling ${\rm
I}_{\rm Q}(z)$ in terms of the Chebyshev polynomials, which form a
complete set of orthonormal functions on the interval $[-1, 1]$.
They also have the property to be the minimax approximating
polynomial, which means that has the smallest maximum deviation from
the true function at any given order
\cite{Simon2005}-\cite{Martinez2008}). Without loss of generality,
we can then expand the coupling ${\rm I}_{\rm Q}(z)$ in the redshift
representation as:
\begin{equation}
\label{eq:Coupling} {\rm I}_{\rm Q}(z) \equiv
\sum_{n=0}^{N}\lambda_{n} \cdot T_{n}(z),
\end{equation}
where $T_{n}(z)$ denotes the Chebyshev polynomials of order $n$ with
$n \in [0,N]$ and $N$ a positive integer. The coefficients of the
polynomial expansion $\lambda_{n}$ are real free dimensionless
parameters. Then the interaction function can be rewritten as
\begin{equation}
\label{Coupling} Q(z) = \rho_{crit}^{0}\cdot(1+z)^{3} \cdot
H(z)\cdot \sum_{n=0}^{N}\lambda_{n}\cdot T_{n}(z),
\end{equation}
We introduce (\ref{eq:Coupling}) in
(\ref{EDFDMOmega})-(\ref{EDFDEOmega}) and integrate both equations
obtaining the solutions,
\begin{eqnarray}
\label{eq:Omega12} \Omega^{\star}_{DM}(z) &=& (1+z)^{3}\left[
{\Omega_{DM}^0} - \frac{z_{max}}{2} \sum_{n=0}^{N}
\lambda_{n} \, \cdot K_{n}(x, 0) \right],\\
\label{eq:Omega13} \Omega^{\star}_{DE}(z) &=& (1+z)^{3(1+w)}\left[
{\Omega_{DE}^0} + \frac{z_{max}}{2} \sum_{n=0}^{N} \lambda_{n} \,
\cdot K_{n}(x, w) \right]  \,,
\end{eqnarray}
where we have defined the integrals
\begin{eqnarray}
\label{IntegralK} K_{n}(x, w) \equiv \int_{-1}^{x}
\frac{T_{n}(\tilde{x})}{(a+b\tilde{x})^{(1+3w)}} \, d\tilde{x} \,\,,
\end{eqnarray}
and the quantities,
\begin{eqnarray}
x \equiv \frac{2 \,z}{z_{max}} - 1 , \\
a \equiv 1\,+\, \frac{z_{max}}{2} , \\
b \equiv \frac{z_{max}}{2},
\end{eqnarray}
here $z_{max}$ is the maximum redshift at which observations are
available so that $x \in [-1, 1]$ and $ \vert T_{n}(x)
\vert\leq 1$, \,for all $n \in [0,N]$.\\
Finally, using the solutions (\ref{Omegabs})-(\ref{Omegars}) and
(\ref{eq:Omega12})-(\ref{eq:Omega13}) we rewrite the Friedmann
equation (\ref{eq:hubble1}) as \vspace{0.5cm}
\begin{equation}
\label{eq:hubble2} H^{2}\left(z \right) = H^2_{0}
\left[\Omega_{b}^{0}{(1+z)}^{3} + \Omega_{r}^{0}{(1+z)}^{4} +
\Omega^{\star}_{DM}(z) + \Omega^{\star}_{DE}(z) \right],
\end{equation}
\vspace{0.5cm}
The Hubble parameter depends of the
parameters ($H_0$, $\Omega_{b}^0$, $\Omega_{r}^0$, $\Omega_{DM}^0$,
$\Omega_{DE}^0$, $w$) and the dimensionless coefficients
$\lambda_{n}$. However one of the parameters depends of the
others due to the Friedmann equation evaluated at the present,
\begin{equation}
\label{HubblePresent} \Omega^{0}_{DE} = 1 - \Omega_{b}^{0} -
\Omega_{r}^{0} - \Omega^{0}_{DM}
\end{equation}
At the end, for the reconstruction, we have the five parameters
($H_0$, $\Omega_{b}^0$, $\Omega_{r}^0$, $\Omega_{DM}^0$, $w$) and
the dimensionless coefficients $\lambda_{n}$.

To do a general reconstruction in
(\ref{eq:Omega12})-(\ref{eq:Omega13}) we must take $N\rightarrow
\infty$ and to obtain the solutions in a closed form. The details of
the calculation of the integrals $K_{n}(x, w)$ in the right hand
side of (\ref{eq:Omega12})-(\ref{eq:Omega13}) are shown in detail in
the Appendix A which shows the closed forms
(\ref{closedsolutionKodd})-(\ref{closedsolutionKeven}) for the
integrals with odd and even integer $n$ subindex, and valid for
$w\neq n/3$ where $n\geq 0$.

Finally, we point out the formula we use for the reconstruction of
other important cosmological property of the universe:

\begin{itemize}
    \item The deceleration parameter
\begin{eqnarray}
q(z) =-1 + \frac{(1 + z)}{H(z)} \cdot \frac{d H(z)}{dz}
\end{eqnarray}
\end{itemize}


\section{Reconstruction of the interaction up to order $N=5$ using the type Ia Supernovae test.}
\label{SectionSNeTest}

To simplify our analysis and to show how the method works, in this
section we reconstruct the coupling function ${\rm I}_{\rm Q}(z)$ to
different orders ($N=1,2,3,4,5$), up to order $N=5$, using the type
Ia Supernovae test. The details of this reconstruction are described
in the Appendix B. We test and constrain the coupling function ${\rm
I}_{\rm Q}(z)$ using the ``Union2'' SNe Ia data set from ``The
Supernova Cosmology Project'' (SCP) composed by 557 type Ia
supernovae \cite{AmanullahUnion22010}. \noindent As it is usual, we
use the definition of luminosity distance $d_L$ (see
\cite{Riess1998}) in a flat cosmology,
\begin{eqnarray}\label{luminosity_distance1}
d_L(z, \mathbf{X}) &=& c(1+z) \int_0^z \frac{dz'}{H(z', \mathbf{X})}
\end{eqnarray}

\noindent where $H(z, \mathbf{X})$ is the Hubble parameter, i.e.,
the expression (\ref{eq:hubble2}), ''$c$'' is the speed of light given
in units of km/sec and $\mathbf{X}$ represents the parameters of the
model,
\begin{eqnarray}\label{parametersm}
\mathbf{X} & \equiv & (H_0, \Omega_{b}^0, \Omega_{r}^0,
\Omega_{DM}^0, w, \lambda_{1},..., \lambda_{N})
\end{eqnarray}
\noindent The \emph{theoretical distance moduli} for the $k$-th
supernova with redshift $z_k$ is defined as
\begin{eqnarray}\label{distanceModuli}
\mu^{{\rm th}}(z_k, \mathbf{X}) & \equiv & m(z)-M = 5\log_{10}
\left[\frac{d_L(z_k, \mathbf{X})}{{\rm Mpc}} \right] +25
\end{eqnarray}
\noindent where $m$ and $M$ are the apparent and absolute magnitudes
of the SNe Ia respectively, and the superscript ``th'' stands for
\textit{``theoretical''}. We construct the statistical $\chi^2$
function as
\begin{eqnarray}\label{ChiSquareDefinition}
\chi^2 (\mathbf{X}) & \equiv & \sum_{k = 1}^n \frac{\left[\mu^{{\rm
t}} (z_k, \mathbf{X}) - \mu_k \right]^2}{\sigma_k^2}
\end{eqnarray}

\noindent where $\mu_k$ is the \emph{observational} distance moduli
for the $k$-th supernova, $\sigma_k^2$ is the variance of the
measurement and $n$ is the amount of supernova in the data set. In
this case $n=557$, using the ``Union2'' SNe Ia data set
\cite{AmanullahUnion22010}.

With this $ \chi^2$ function we construct the probability density
function (\textbf{pdf}) as
\begin{equation}\label{expChi}
{{\rm {\bf pdf}}}(\mathbf{X}) = \rm{A} \cdot{{\rm e}}^{-\chi^2 \,/2}
\end{equation}

\noindent where $\rm A$ is a integration constant.



\subsection{Priors on the the probability density
function (\bf{pdf}).} \label{PriorsSS}

In the models I, II and III shown in the Table \ref{tablemodels} we
marginalize the parameters $\mathbf{Y} = (H_0, \Omega_{DM}^0,
\Omega_{b}^0, \Omega_{r}^0$) in the \textbf{pdf} (\ref{expChi})
choosing priors on them. In order to it, we must compute the
following integration,

\begin{eqnarray}\label{MarginalizationIntegral2}
 {{\rm {\bf pdf}}}(\mathbf{V}) = \int^{\infty}_{0}
 \int^{\infty}_{0} \int^{\infty}_{0} \int^{\infty}_{0}
 {{\rm {\bf pdf}}}(\mathbf{X}) \,{{\rm {\bf pdf}}}
 (\mathbf{Y}) \, dH_0 \,
 d\Omega_{DM}^0 \, d\Omega_{b}^0 \, d\Omega_{r}^0
\end{eqnarray}

\noindent where $\mathbf{V} = (w, \lambda_{1},..., \lambda_{N})$
represents the nonmarginalized parameters, ${{\rm {\bf
pdf}}}(\mathbf{X}) $ is given by (\ref{expChi}) and ${{\rm {\bf
pdf}}} (\mathbf{Y})$ is the \textit{prior} probability distribution
function for the parameters ($H_0, \Omega_{DM}^0, \Omega_{b}^0,
\Omega_{r}^0$) which are chosen as Dirac delta priors around the
specific values $\mathbf{\tilde{Y}} = (\tilde{H}_0,
\tilde{\Omega}_{DM}^0, \tilde{\Omega}_{b}^0, \tilde{\Omega}_{r}^0$)
measured by some other independent observations,

\begin{equation}\label{Priors}
{{\rm {\bf pdf}}} (\mathbf{Y}) = \delta(H_0 - \tilde{H}_0)
 \cdot \delta(\Omega_{DM}^0 -  \tilde{\Omega}_{DM}^0)
 \cdot \delta(\Omega_{b}^0 -  \tilde{\Omega}_{b}^0)
 \cdot \delta(\Omega_{r}^0 -  \tilde{\Omega}_{r}^0)
\end{equation}

\noindent Introducing (\ref{Priors}) in
((\ref{MarginalizationIntegral2}) it produces,
\begin{equation}\label{expChinew}
{{\rm {\bf pdf}}}(\mathbf{V}) = \rm{A} \cdot{{\rm
e}}^{-\tilde{\chi}^2 \,/2}
\end{equation}

\noindent where we have defined a new function $\tilde{\chi}^2$
depending only on the parameters $\mathbf{V} = (w, \lambda_{1},...,
\lambda_{N})$ as,
\begin{eqnarray}\label{ChiSquareew}
\tilde{\chi}^2 (\mathbf{V}) & \equiv & \sum_{k = 1}^n
\frac{\left[\mu^{{\rm th}} (z_k, \mathbf{V}, \mathbf{\tilde{Y}}) -
\mu_k \right]^2}{\sigma_k^2}
\end{eqnarray}
The specific values chosen for the Dirac delta priors are,

\begin{itemize}
\item
$\tilde{H}_0 = 72 \; ({{\rm km}}/{{\rm s}}){{\rm Mpc}}^{-1}$ as
suggested by the observations of the Hubble Space Telescope (HST)
\cite{Freedman2001}.
\item $\tilde{\Omega}_{DM}^0 = 0.233$
\item $\tilde{\Omega}_{b}^0 = 0.0462$
\item $\tilde{\Omega}_{r}^0 = 4.62 \times 10^{-5}$
\end{itemize}

Once constructed the function $\tilde{\chi}^2$ (\ref{ChiSquareew}),
we numerically minimize it to compute the ``\textit{best
estimates}'' for the free parameters of the model: $\mathbf{V} = (w,
\lambda_{1},..., \lambda_{N})$. The minimum value of the
$\tilde{\chi}^2$ function gives the best estimated values of
$\mathbf{V}$ and measures the goodness-of-fit of the model to data.
\\
\\
For the Model IV, we leave too the parameter $\Omega_{DM}^0$ free to
vary and estimated it from the minimization of the $\tilde{\chi}^2$
function. In this case, the parameters to be marginalized are
$\mathbf{Y} = (H_0, \Omega_{b}^0, \Omega_{r}^0$). Then, the
marginalization will be as,

\begin{eqnarray}\label{MarginalizationIntegral3}
 {{\rm {\bf pdf}}}(\mathbf{V}) =
 \int^{\infty}_{0} \int^{\infty}_{0} \int^{\infty}_{0}
 {{\rm {\bf pdf}}}(\mathbf{X}) \,{{\rm {\bf pdf}}}
 (\mathbf{Y}) \, dH_0 \, d\Omega_{b}^0 \, d\Omega_{r}^0
\end{eqnarray}

\noindent where now $\mathbf{V} = (w, \Omega_{DM}^0,
\lambda_{1},..., \lambda_{N})$ represents the nonmarginalized
parameters to be estimated, ${{\rm {\bf pdf}}}(\mathbf{X}) $ is
given by (\ref{expChi}) and ${{\rm {\bf pdf}}} (\mathbf{Y})$ is the
\textit{prior} probability distribution function for the parameters
($H_0, \Omega_{b}^0, \Omega_{r}^0$) which are chosen as Dirac delta
priors around the specific values $\mathbf{\tilde{Y}} =
(\tilde{H}_0, \tilde{\Omega}_{b}^0, \tilde{\Omega}_{r}^0$) given
above. \\
\\
In the models II, III and IV the interaction function ${\rm I}_{\rm
Q}(z)$ will be reconstructed up to order $N=5$ in the expansion in
terms of Chebyshev polynomials.


\begin{table}
  \centering
\begin{tabular}{| c | c | c | c | }
 \multicolumn{4}{c}{\textbf{Models}} \\

\hline

Models & $\Omega_{DM}^0$ & EOS parameter $w$ & Interaction function \\
\hline \hline

Model I & 0.233 (fixed) &  constant &  ${\rm I}_{\rm Q}(z) \equiv 0$ \\
Model II & 0.233 (fixed) & -1 &  ${\rm I}_{\rm Q}(z) \neq 0$ \\
Model III & 0.233 (fixed) & constant & ${\rm I}_{\rm Q}(z) \neq 0$ \\
Model IV & free parameter & constant & ${\rm I}_{\rm Q}(z) \neq 0$ \\

\hline \hline

\end{tabular}
\caption{Summary of the models studied in this work. In the models
II,  III and IV the interaction function ${\rm I}_{\rm Q}(z)$ will
be reconstructed. Additionally, in the model IV the parameter
$\Omega_{DM}^0$ is estimated.} \label{tablemodels}
\end{table}

\subsection{Results of the reconstruction of the interaction function.}
\label{Results}

\begin{center}
\begin{figure}
  \includegraphics[width=8cm, height=60mm, scale=0.90]{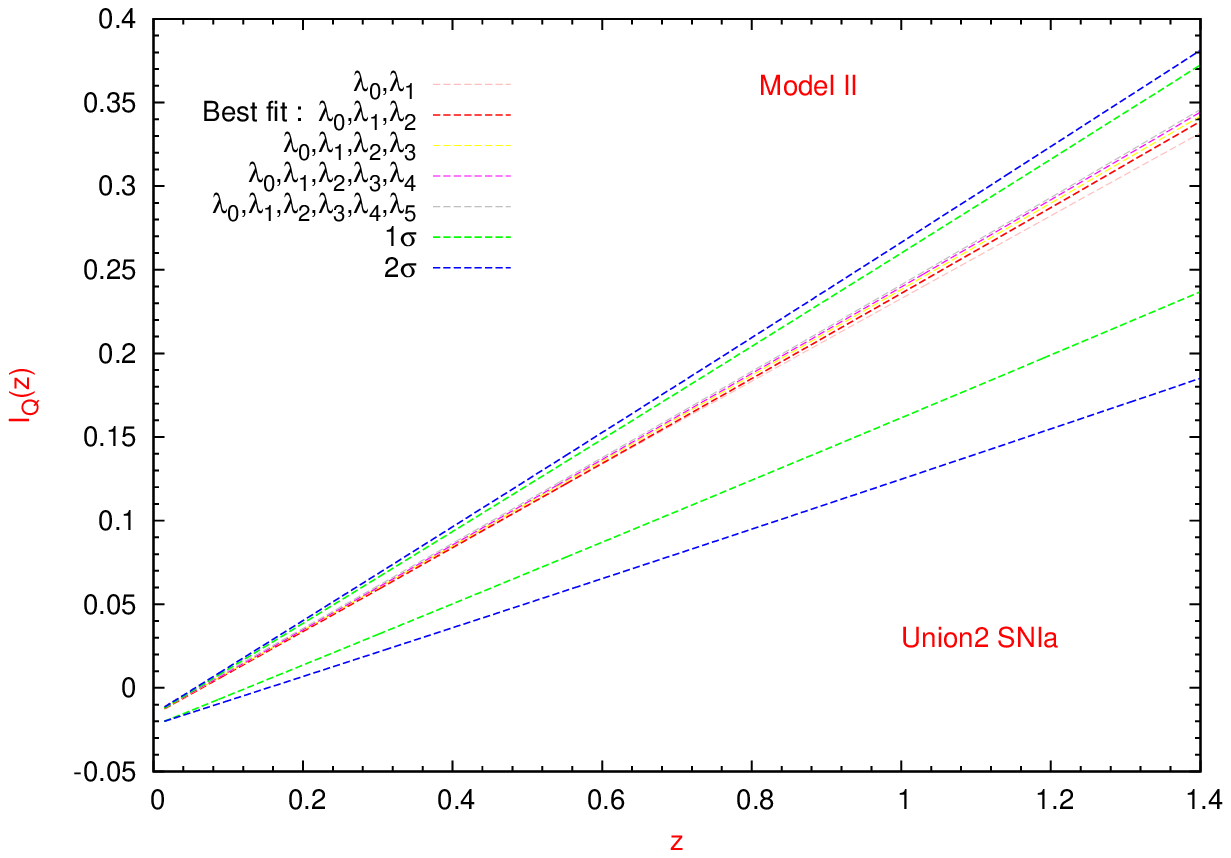}
  \includegraphics[width=8cm, height=60mm, scale=0.90]{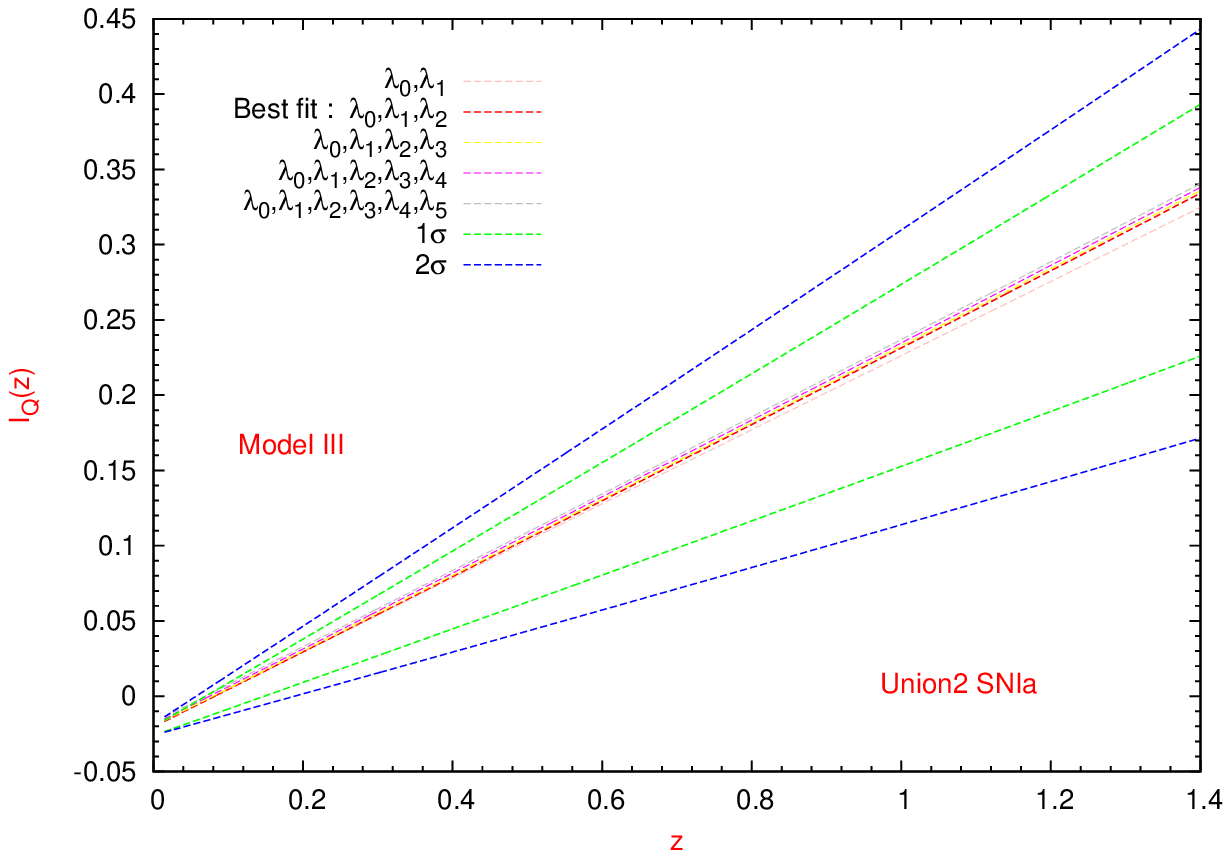}
  \includegraphics[width=8cm, height=60mm, scale=0.90]{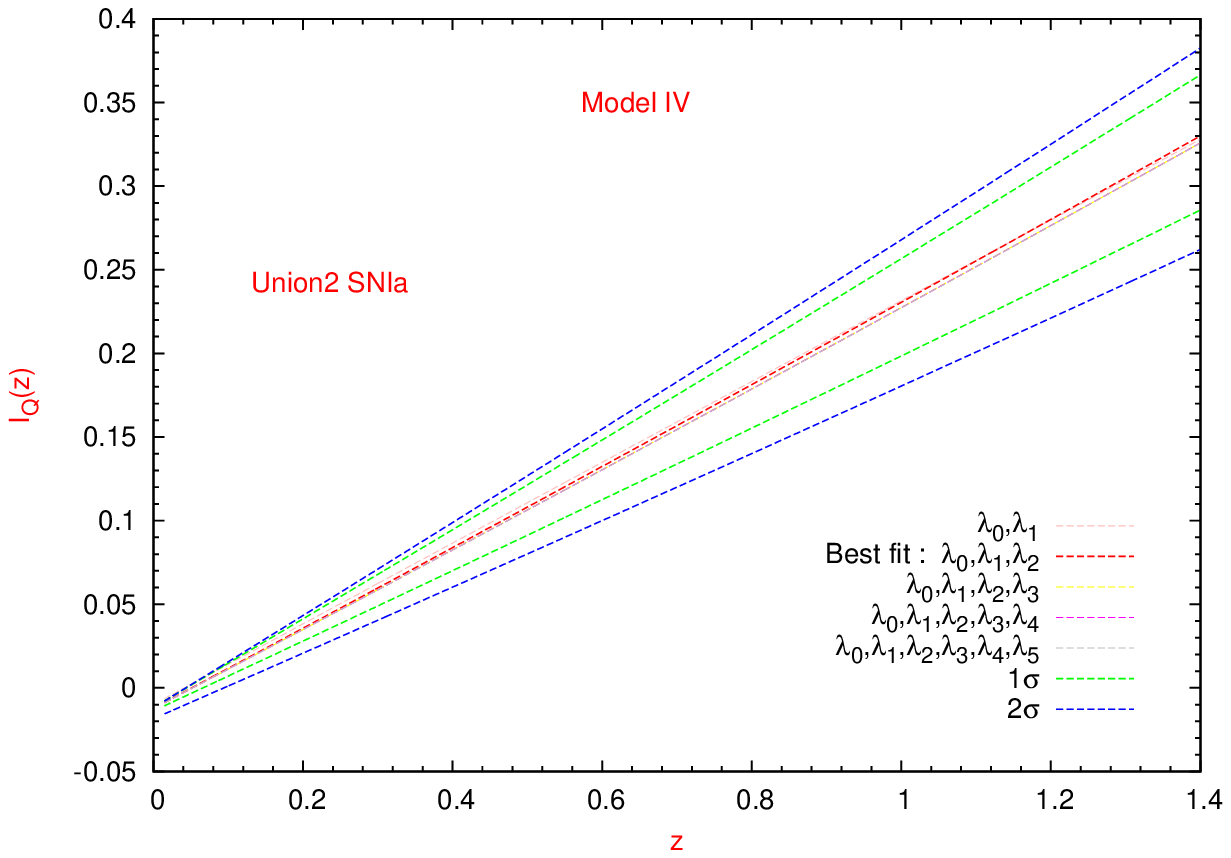}
  \caption{Reconstruction for the dimensionless interaction function
  ${\rm I}_{\rm Q}(z)$ as a function of the redshift for the models
  II (left above panel), III (right above panel) and IV (left below panel) corresponding to
  a dark energy equation of state parameter (II) $w=-1$, (III) $w=$ constant
  (both with $\Omega_{DM}^0=0.233$),
  and (IV) $w=$ constant, $\Omega_{DM}^0=$ constant, respectively.
  The curves with different colours show the best estimates using the expansion
  of ${\rm I}_{\rm Q}(z)$ in terms of the parameters $\lambda_n$
  corresponding to the Chebyshev polynomial expansion ranging from $N=1$ to $5$.
  Note the fast convergence of the curves when the number of polynomials
  $N$ involved in the expansion increases.
  We show the best estimates (red lines), the $1\sigma$ (green lines)
  and $2\sigma$ errors (blue lines) for the dimensionless interaction function
  ${\rm I}_{\rm Q}(z)$ as a function of the redshift.
  The reconstruction is derived from the best estimation obtained using the type
  Ia Supernova SCP Union2 data set sample. Note that the best estimates of
  the strength of the interaction cross marginally the noninteracting line
  ${\rm I}_{\rm Q}(z)=0$ only at the present changing sign from positive
  values at the past (energy transfers from dark energy to dark matter) to negative
  values almost at the present (energy transfers from dark matter to
  dark energy). Moreover, note that within the $1\sigma$ and $2\sigma$ errors it could be the possibility that
  the crossing the noninteracting line ${\rm I}_{\rm Q}(z)=0$ happens before
  the present and the dark energy density is positive in all the redshift range.}
  \label{ExchangeModel}
\end{figure}
\end{center}

\begin{center}
\begin{figure}
   \includegraphics[width=8cm, height=60mm, scale=0.90]{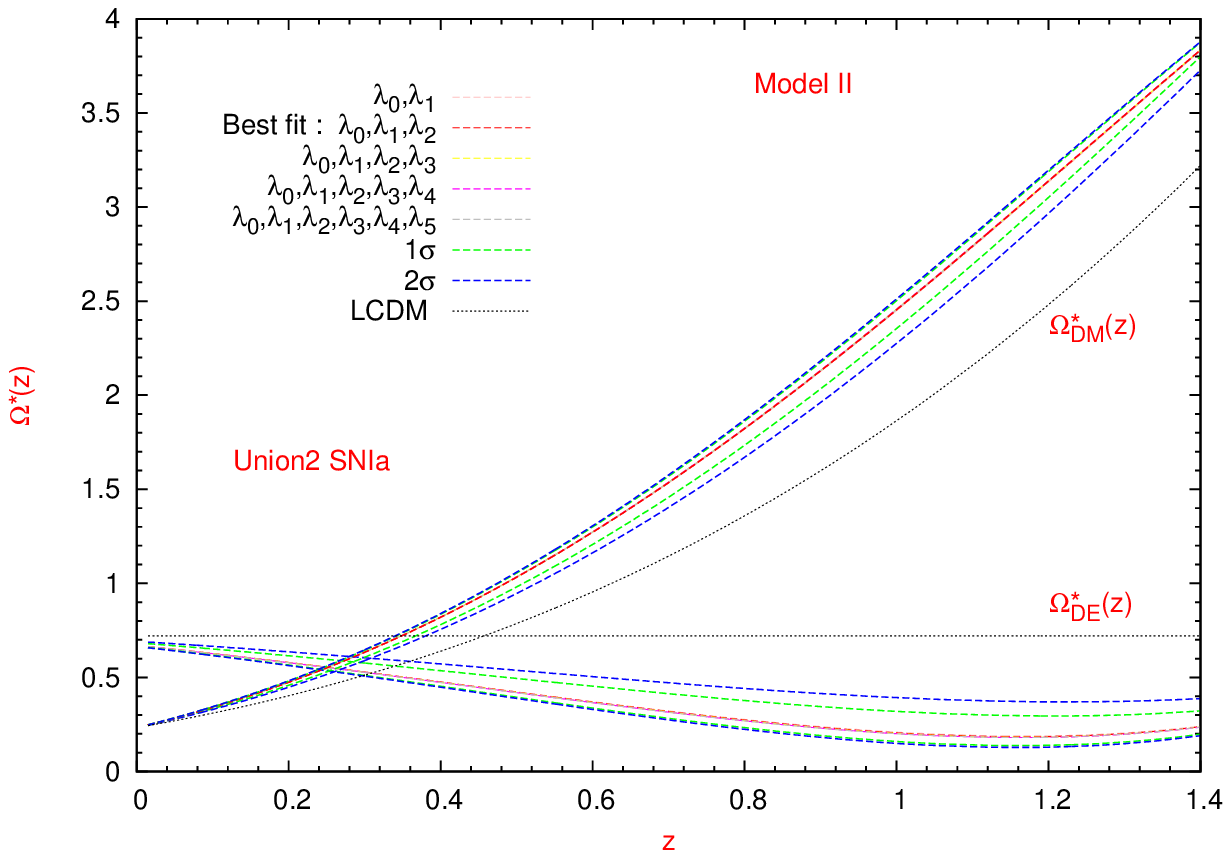}
   \includegraphics[width=8cm, height=60mm, scale=0.90]{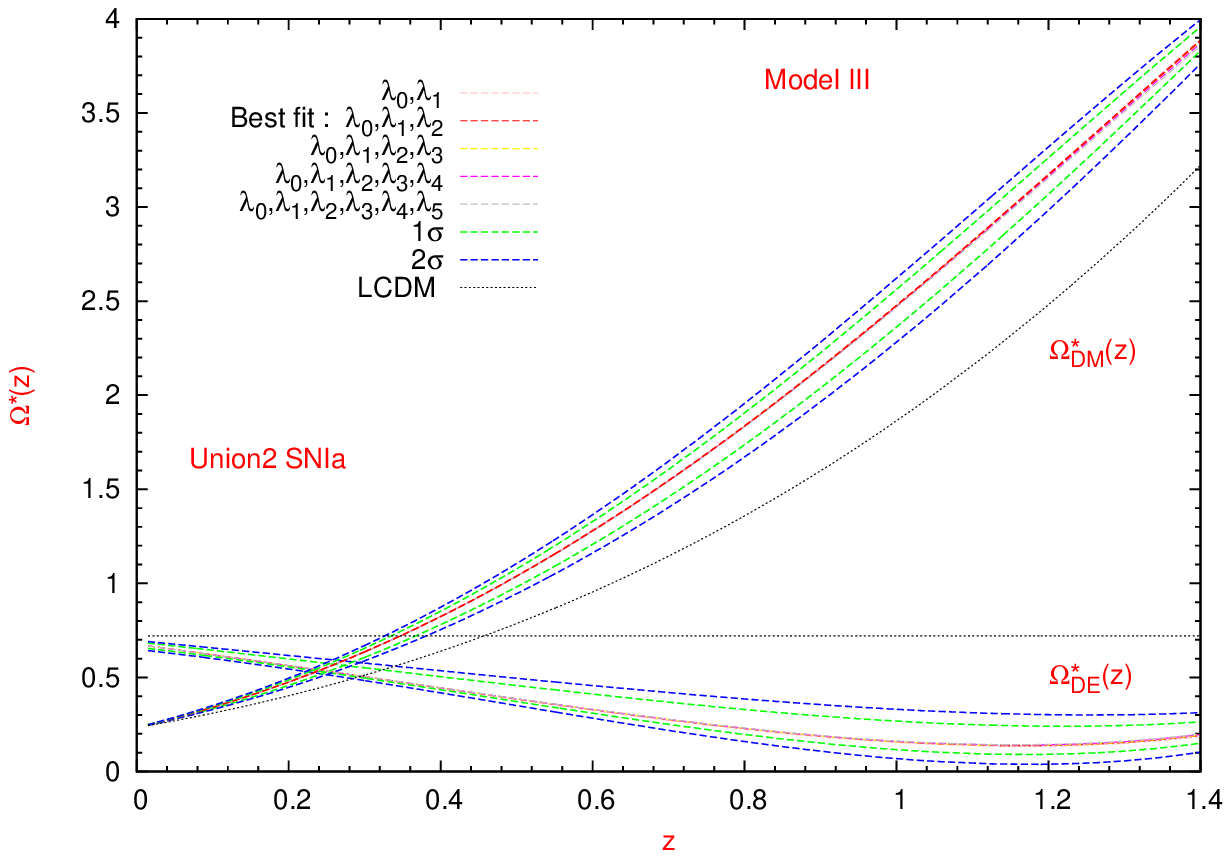}
   \includegraphics[width=8cm, height=60mm, scale=0.90]{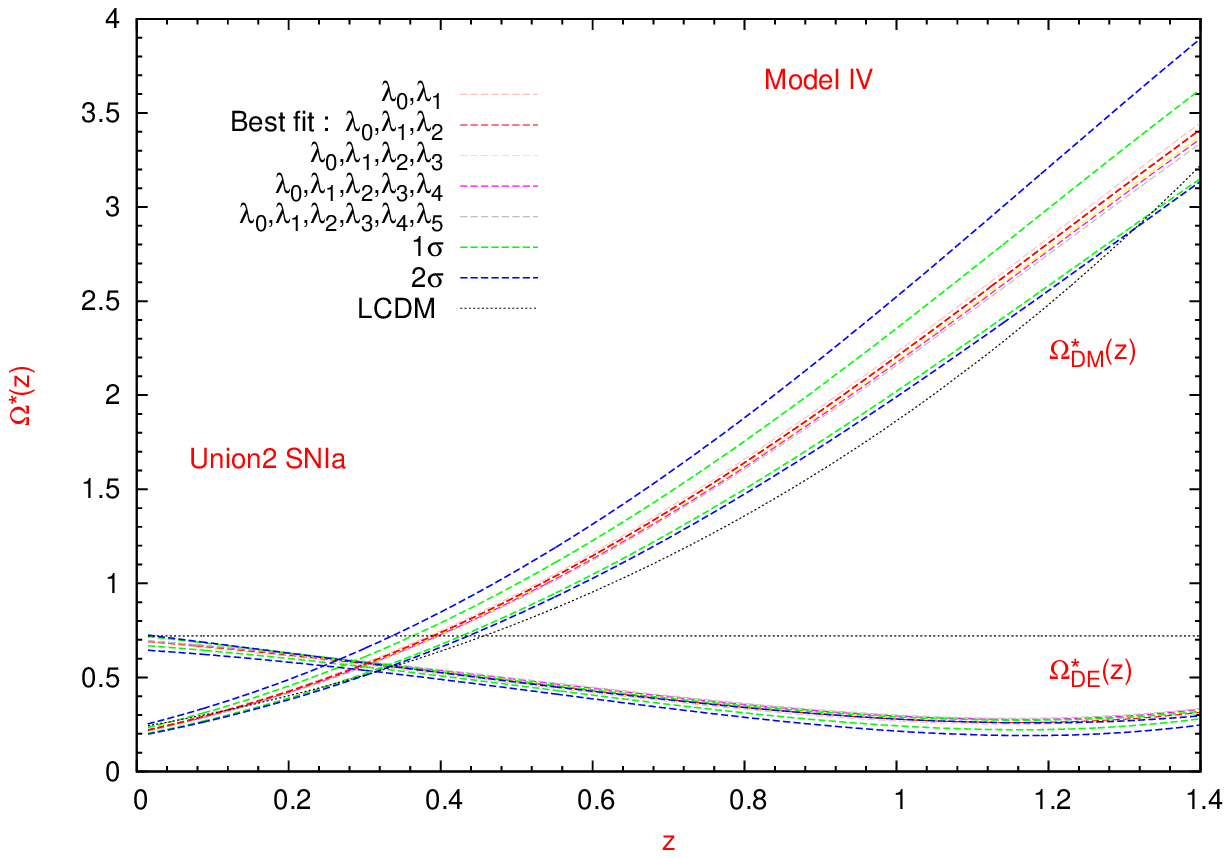}
   \caption{The same as the Figure \ref{ExchangeModel} but now for the reconstruction
   of the dark matter and dark energy density parameters,
   $\Omega^{\star}_{DM}(z)$, $\Omega^{\star}_{DE}(z)$,
   as a function of the redshift for the model II (left above panel), III (right above panel)
   and IV (left below panel) respectively.
   These curves also show the best estimates (red lines), the $1\sigma$ (green lines)
   and $2\sigma$ errors (blue lines) for the density parameters
   $\Omega^{\star}_{DM}(z)$, and $\Omega^{\star}_{DE}(z)$.
   Note that the density parameter of dark energy is definite positive for all the range of redshift
   considered in the reconstruction.}
  \label{CoincidenceModel}
\end{figure}
\end{center}

\begin{center}
\begin{figure}
   \includegraphics[width=8cm, height=60mm, scale=0.90]{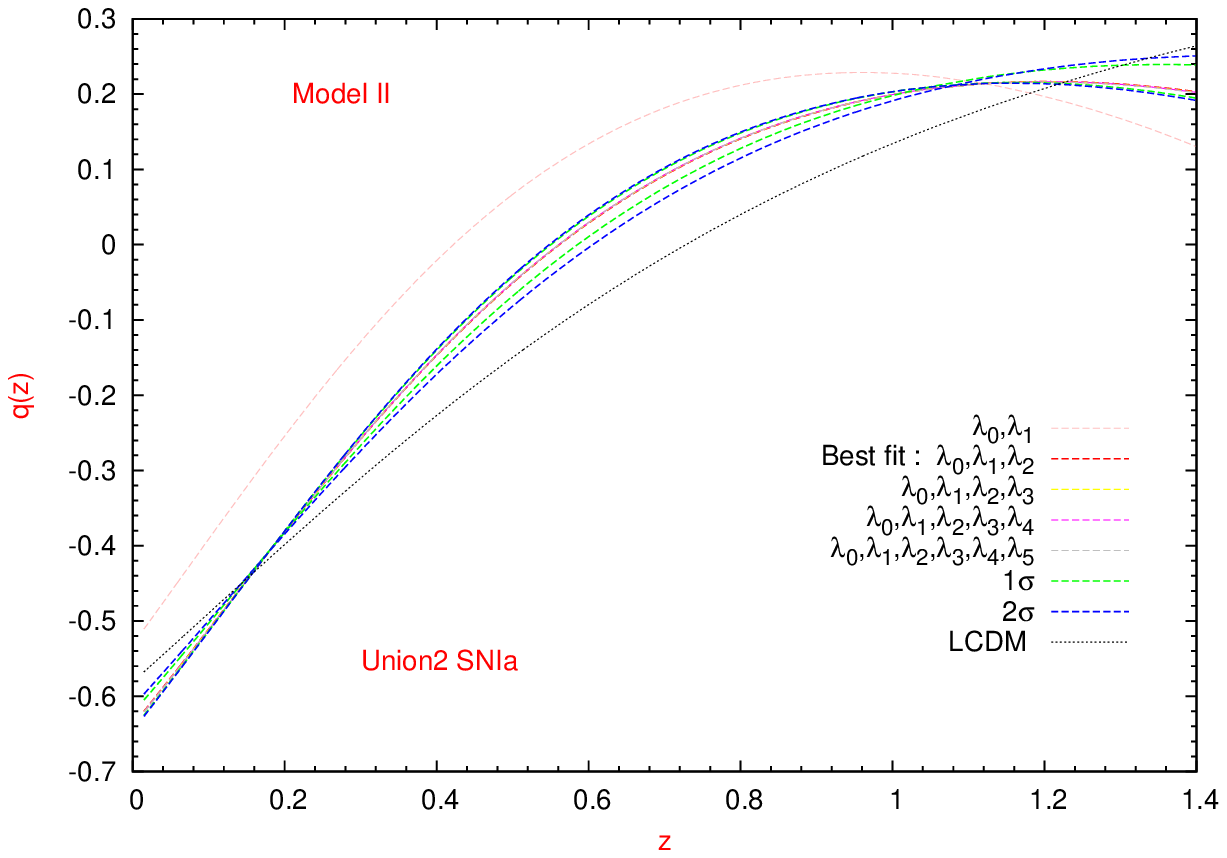}
   \includegraphics[width=8cm, height=60mm, scale=0.90]{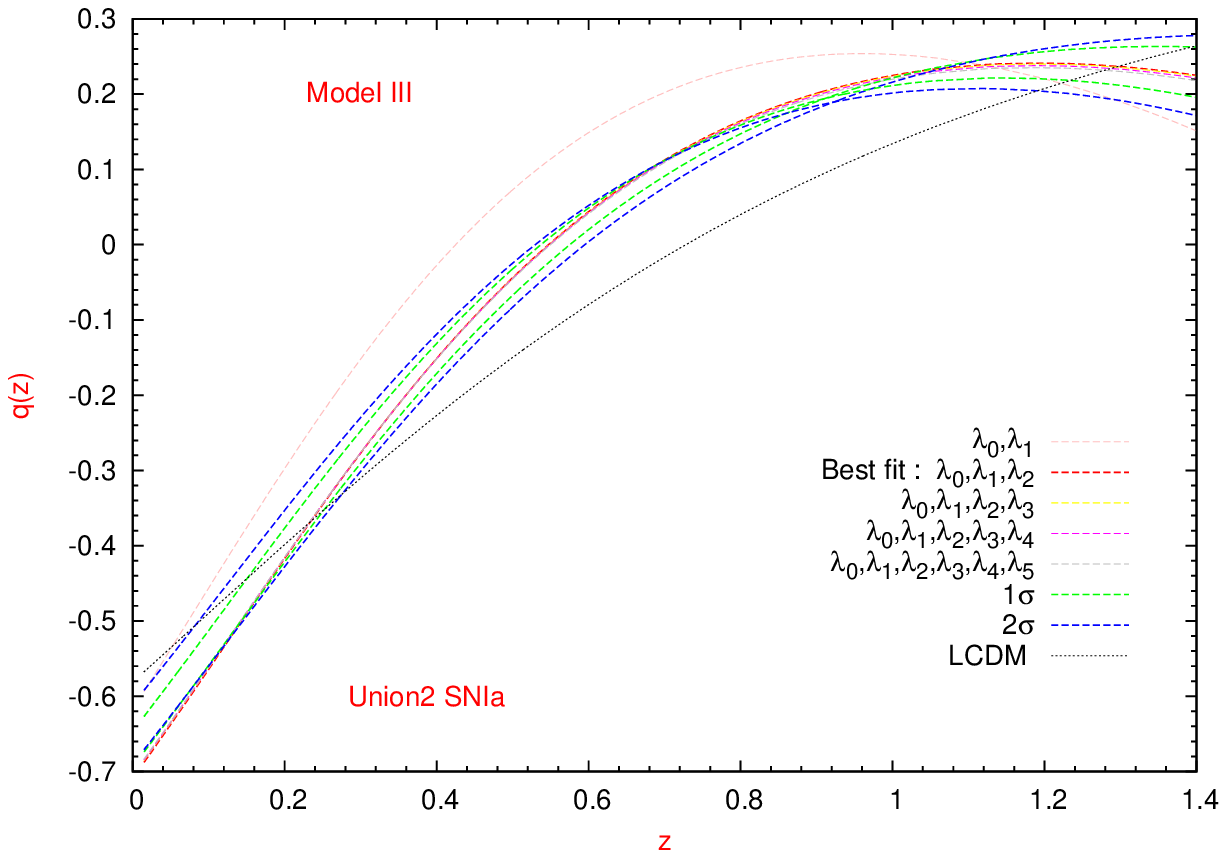}
   \includegraphics[width=8cm, height=60mm, scale=0.90]{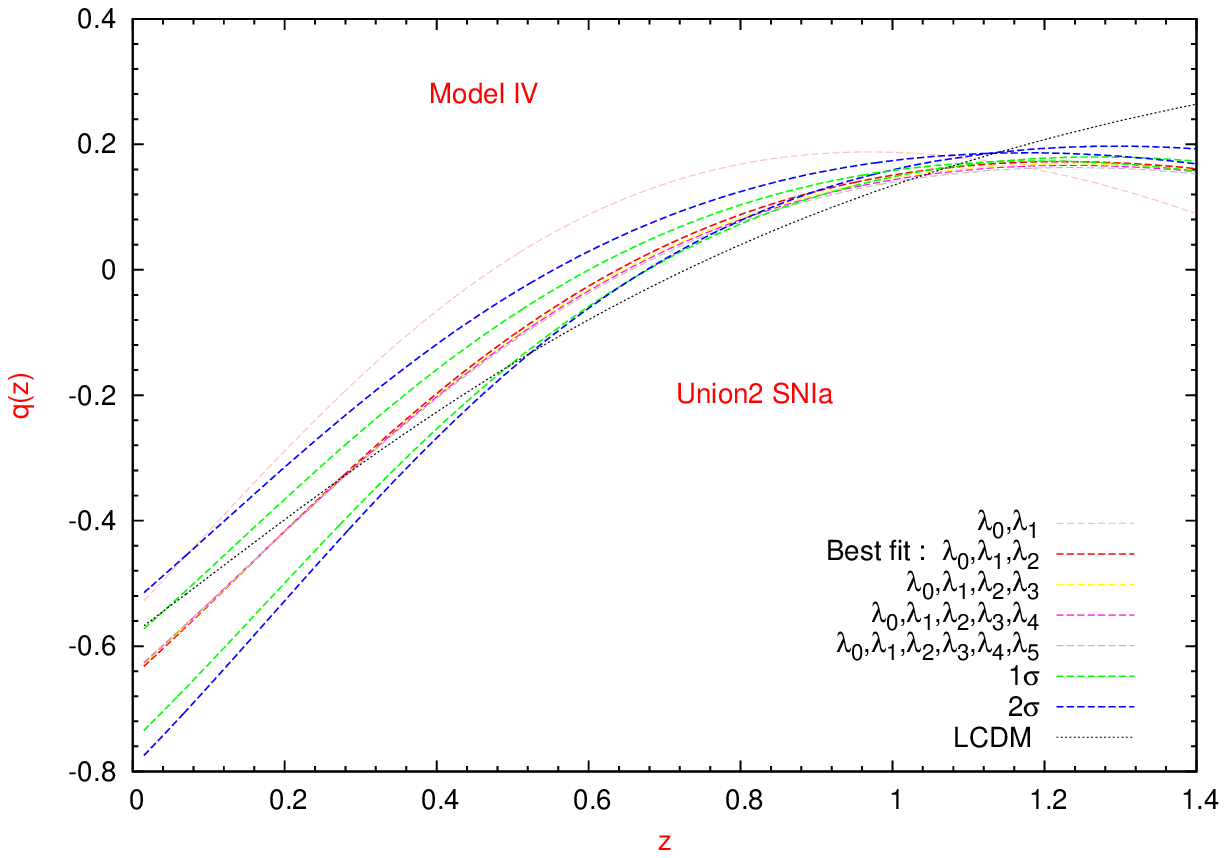}
  \caption{The same as the Figure \ref{ExchangeModel}. for the reconstruction
  of the deceleration parameter $q(z)$ as a function of the redshift for the
  model II (left above panel), III (right above panel) and IV (left below panel).
  These curves also show the best estimates (red lines), the $1\sigma$ (green lines)
  and $2\sigma$ errors (blue lines) for the deceleration parameter $q(z)$ as a function
  of the redshift.
  By comparison them with the corresponding curve for the LCDM (Lambda Cold Dark Matter)
  model.}
  \label{DecelerateModel}
\end{figure}
\end{center}

Now, we present the results of the fit of the models listed in the
Table \ref{tablemodels} with the the ``Union2'' SNe Ia data set
\cite{AmanullahUnion22010} and the priors described in the Section
\ref{PriorsSS}. For the noninteracting model I, the only free
parameter to be estimated is $\theta = \{w\}$, whilst for the
interacting models II, III and IV the free parameters are $\theta =
\{\lambda_{0},..., \lambda_{N}\}$, $\theta = \{w, \lambda_{0},...,
\lambda_{N}\}$ and $\theta = \{w, \Omega_{DM}^0, \lambda_{0},...,
\lambda_{N}\}$ respectively, where $N$ is taking the values $N = 1,
2, 3, 4, 5$. In every case, we obtain the best fit parameters and
the corresponding $\tilde{\chi}^2_{min}$.
\\
\\
The Figure \ref{ExchangeModel} show the reconstruction of the
dimensionless interaction function ${\rm I}_{\rm Q}(z)$ as a
function of the redshift for the models II (corresponding to a dark
energy EOS parameter $w=-1$), III (corresponding to a dark energy
EOS parameter $w=$ constant and $\Omega_{DM}^0$ fixed) and IV
(corresponding to a dark energy EOS parameter $w=$ constant and
$\Omega_{DM}^0$ as a free parameter to be estimated) respectively.
\\
\\
The Figure \ref{CoincidenceModel} show the reconstruction of the the
dark matter and dark energy density parameters
$\Omega^{\star}_{DM}(z)$, $\Omega^{\star}_{DE}(z)$ as a function of
the redshift for the models II, III and IV described above. The same
applies to the Figure \ref{DecelerateModel} which shows the
reconstruction of the deceleration parameter $q(z)$.
\\
\\
All the Figures show the superposition of the best estimates for
every cosmological variable using the expansion of the interaction
function ${\rm I}_{\rm Q}(z)$ in terms of the parameters $\lambda_n$
corresponding to the Chebyshev polynomial expansion
(\ref{eq:Coupling}) ranging from $N=1$ to $5$. The Tables
\ref{BestestimationII}, \ref{BestestimationIII} and
\ref{BestestimationIV}, show the best fit parameters and the minimum
of the function $\tilde{\chi}_{min}^2$ for the models II, III and IV
respectively. From these tables, we can note the fast convergence of
the best estimates when the numbers of parameters $N$
is increased in the expansion (\ref{eq:Coupling}).
\\
\\
From the Figure \ref{ExchangeModel}, we first notice that, for all
interacting models, the best estimates for the interaction function
${\rm I}_{\rm Q}(z)$ cross marginally the noninteracting line ${\rm
I}_{\rm Q}(z) = 0$ during the present cosmological evolution (at
around $z \approx 0.08$) changing sign from positive values at the
past (energy transfers from dark energy to dark matter) to negative
values at the present (energy transfers from dark matter to dark
energy). However, taking in account the errors corresponding to the
fit using three parameters ($N=2$), we see that within the $1\sigma$
and $2\sigma$ errors, it exists the possibility of crossing of the
noninteraction line in the recent past at around the range $z \in
(0.04, 0.16)$.
\\
\\
Crossings of the noninteracting line $Q(z) = 0$ have been recently
reported at the references \cite{Cai-Su} (with an interacting term
$Q(z)$ proportional to the Hubble parameter) and \cite{He-Zhang}
(where the assumption that the interacting term $Q(z)$ is
proportional to the Hubble parameter is abandoned), although the
direction of the change found in our work is contrary to the results
published by Li and Zhang in the reference \cite{He-Zhang} where a
crossing (from negative values at the past to positive values at the
present) was found at $z \simeq 0.2-0.3$ with a linearly dependent
interacting function of the scale factor with two free parameters.
On the other hand, we did not find the oscillatory behavior of the
interaction function found by Cai and Su \cite{Cai-Su} who, using
observational data samples in the range $z \in [0, 1.8]$, fitted a
scheme in which the whole redshift range is divided into a
determined numbers of bins and the interaction function set to be a
constant in each bin. \\
\\
Lets mention that, recently, it has been mentioned that a crossing
of the noninteracting line $Q(z) = 0$ implying a transfer of energy
from DM to DE may well conflict with the second law of
thermodynamics \cite{pavon1999GRGThermo}. Too, in the context of
bouncing coupled dark energy scenarios, a possible inversion of the
flux of energy from DM to DE has been found in terms of a change of
sign of the gradient of a scalar field \cite{Mbaldi} which, of
course, it is not the case considered here.
\\
\\
The Figure \ref{CoincidenceModel} shows that, for all the
interacting models studied in this work, the best estimates for the
dark energy density parameter $\Omega^{\star}_{DE}(z)$ become to be
definite positive at all the range of redshifts considered in the
data sample. However, this statement is conclusive which shows that within the $1\sigma$
and $2\sigma$ errors for the fit with three parameters ($N=2$), the
$\Omega^{\star}_{DE}(z)$ becomes positive in all the range of
redshifts considered (remember that we are using only the SNe Ia
data set which has been attributed to have systematic errors
\cite{nesseris1}-\cite{wei}. However, let us mention that $1\sigma$
error becomes greater if the number of free parameter to be fitted
increases, a fact already described in \cite{kunz-corasaniti} as the cost of the compression. \\
\\
The Figure \ref{DecelerateModel} shows the reconstruction
for the deceleration parameter $q(z)$ for the fit with $N=2$
parameters and their respective $1\sigma$ and $2\sigma$ constraints.
It also shows that, for all models, a transition from a deceleration
era at early times dominated by the dark and baryonic matter density
to an acceleration era at late times corresponding to the present
domain of the dark energy density. At the present, the dark energy density parameter
becomes $\Omega^{0}_{DE}(z) \approx 0.7$ as we can see from the Figure \ref{CoincidenceModel},
which is sufficiently large to generate a non negligible dimensionless
interacting term of the order of ${\rm I}^{0}_{\rm Q} \approx
-10^{-2}$, as is shown in the Figure \ref{ExchangeModel}. In fact, in this same figure we can appreciate that in the interval of
redshifts $z \in [0, 1.4]$, the dimensionless interaction is in the
range ${\rm I}_{\rm Q} \in [-0.02, 0.4]$ corresponding to a
$2\sigma$ error. The order of magnitude of this interaction is in
agreement with the local constraints put on the strength of a
constant dimensionless interaction derived from the fit to a data
sample of virial masses of relaxed galaxies clusters obtained using
weak lensing, x-ray and optical data \cite{abramo}(for previous
constraints to see the references \cite{pasqui}-\cite{He-Zhang}).
\begin{table}
  \centering
\begin{tabular}{| c | c | c |}
\multicolumn{3}{c}{\textbf{Model I: $w=$ constant.}} \\
\multicolumn{3}{c}{Best estimate for the EOS parameter $w$.} \\
\hline
    Errors & $\pm 1\sigma$ & $\pm 2\sigma$\\
\hline
$\omega$&${-1.24}^{+0.03}_{-0.03}$&${-1.24}^{+0.06}_{-0.07}$\, \\
\hline
\end{tabular}
  \caption{The best estimate of the dark energy EOS parameter $w$
  for the Model I. The best estimate was computed through a
  Bayesian statistical analysis using the ``Union2'' SNe Ia data set giving
  $\tilde{\chi}^{2}_{min} = 562.51$.}
  \label{BestestimationI}
\end{table}
\begin{table}
  \centering
\begin{tabular}{| c | c | c | c | c | c |}
\multicolumn{6}{c}{\textbf{Model II: $w=-1$.}} \\
\multicolumn{6}{c}{Best estimates for the parameters $\lambda_n$.} \\
\hline  & $N=1$ & $N=2$ & $N=3$ & $N=4$  & $N=5$ \\
\hline \hline
$\lambda_{0}$ & $-1.46\times 10^{-2}$ & $-1.43\times 10^{-2}$ &
$-1.40\times 10^{-2}$ &
$-1.37\times 10^{-2}$ & $-1.32\times 10^{-2}$ \\
$\lambda_{1}$ & $2.47\times 10^{-1}$ & $2.48\times 10^{-1}$ & $2.50\times 10^{-1}$ & $2.51\times 10^{-1}$ & $2.52\times 10^{-1}$ \\
$\lambda_{2}$ & $0.0$ & $1.810\times 10^{-3}$ & $1.800\times 10^{-3}$ & $2.0\times 10^{-3}$ & $1.801\times 10^{-3}$ \\
$\lambda_{3}$ & $0.0$ & $0.0$ & $1.07 \times 10^{-5}$ & $2.60 \times 10^{-5}$ & $4.98 \times 10^{-5}$ \\
$\lambda_{4}$ & $0.0$ & $0.0$ & $0.0$ & $1.07 \times 10^{-6}$ & $3.29 \times 10^{-6}$ \\
$\lambda_{5}$ & $0.0$ & $0.0$ & $0.0$ & $0.0$ & $1.0 \times 10^{-7}$ \\
\hline \hline $\tilde{\chi}^{2}_{min}$ & $544.80$ & $544.72$ &
$544.58$ &$544.48$ & $544.36$ \\
\hline
\end{tabular}
  \caption{Summary of the best estimates of the dimensionless coefficients
  $\lambda_n$ of the expansion of the interaction function
  for the Model II corresponding to a interacting dark energy EOS parameter $w=-1$.
  The best estimates were computed through a Bayesian statistical
  analysis using the ``Union2'' SNe Ia data set. The number $N$ in the
  top of every column indicates the maximum number of Chebyshev
  polynomials used in the expansion (\ref{eq:Coupling}) of the
  coupling function ${\rm I}_{\rm Q}(z)$ starting from $N=1$. From the
  Figure \ref{ExchangeModel} to Figure \ref{DecelerateModel}, show the reconstruction of several cosmological variables using the best
  estimates for $N=5$. Moreover, they also show the reconstruction of
  several cosmological variables using the best
  estimates for $N=2$ and their confidence intervals at $1\sigma$ and $2\sigma$.}
  \label{BestestimationII}
\end{table}

\begin{table}
  \centering
\begin{tabular}{| c | c | c |}
\multicolumn{3}{c}{\textbf{Model II: $w=-1$.}} \\
\hline Errors & $\pm 1\sigma$ & $\pm 2\sigma$\\
\hline \hline
$\lambda_{0}$&${-1.43\times 10^{-2}}^{+0.008\times 10^{-2}}_{-0.689\times 10^{-2}}$&${-1.43\times 10^{-2}}^{+0.06\times 10^{-2}}_{-0.63\times 10^{-2}}$\, \\
$\lambda_{1}$&${+2.48\times 10^{-1}}^{+0.23\times 10^{-1}}_{-0.67\times 10^{-1}}$&${+2.48\times 10^{-1}}^{+0.29\times 10^{-1}}_{-1.04\times 10^{-1}}$\, \\
$\lambda_{2}$&${+1.81\times 10^{-3}}^{+0.03\times 10^{-3}}_{-0.31\times 10^{-3}}$&${+1.81\times 10^{-3}}^{+0.08\times 10^{-3}}_{-0.35\times 10^{-3}}$\, \\
\hline
\end{tabular}
  \caption{Summary of the $1\sigma$ and $2\sigma$ errors of the best estimate for $N=2$.}
  \label{BestestimationIIerrors}
\end{table}
\begin{table}
  \centering
\begin{tabular}{| c | c | c | c | c | c | }
\multicolumn{6}{c}{\textbf{Model III: $w=$} constant.} \\
\multicolumn{6}{c}{Best estimates for the parameters $\lambda_n$ and $w$.} \\
\hline  & $N=1$ & $N=2$ & $N=3$ & $N=4$  & $N=5$ \\
\hline \hline
$\lambda_{0}$ & $-2.32\times 10^{-2}$ & $-1.85\times 10^{-2}$ &
$-1.80\times 10^{-2}$ &
$-1.70\times 10^{-2}$ & $-1.58\times 10^{-2}$ \\
$\lambda_{1}$ & $2.46\times 10^{-1}$ & $2.48\times 10^{-1}$ & $2.49\times 10^{-1}$ & $2.50\times 10^{-1}$
& $2.51\times 10^{-1}$ \\
$\lambda_{2}$ & $0.0$ & $1.80\times 10^{-3}$ & $1.78\times 10^{-3}$ & $1.70\times 10^{-3}$ & $1.62\times 10^{-3}$ \\
$\lambda_{3}$ & $0.0$ & $0.0$ & $1.06 \times 10^{-5}$ & $1.07 \times 10^{-5}$ & $1.05 \times 10^{-5}$ \\
$\lambda_{4}$ & $0.0$ & $0.0$ & $0.0$ & $1.0 \times 10^{-6}$ & $1.19 \times 10^{-6}$ \\
$\lambda_{5}$ & $0.0$ & $0.0$ & $0.0$ & $0.0$ & $1.0 \times 10^{-7}$ \\
$w$ & $-1.080$ & $-1.068$ & $-1.065$ & $-1.064$ & $-1.063$ \\
\hline \hline $\tilde{\chi}^{2}_{min}$ & $543.52$ & $543.42$
& $543.40$ & $543.38$ & $543.36$ \\
\hline
\end{tabular}
  \caption{The same as the Table \ref{BestestimationII}.
  for the Model III corresponding to a interacting dark energy EOS
  parameter $w=$ constant and $\Omega_{DM}^0 = 0.233$. From the
  Figure \ref{ExchangeModel} to Figure \ref{DecelerateModel}, show the reconstruction of several cosmological
  variables using the best estimates for $N=5$. Moreover, they also show the reconstruction of
  several cosmological variables using the best estimates for $N=2$ and their confidence intervals at $1\sigma$ and $2\sigma$.}
  \label{BestestimationIII}
\end{table}
\begin{table}
  \centering
\begin{tabular}{| c | c | c | }
\multicolumn{3}{c}{\textbf{Model III: $w=$constant.}} \\
\hline Errors & $\pm 1\sigma$ & $\pm 2\sigma$\\
\hline \hline
$\lambda_{0}$&${-1.85\times 10^{-2}}^{+0.008\times 10^{-2}}_{-0.608\times 10^{-2}}$&${-1.85\times 10^{-2}}^{+0.19\times 10^{-2}}_{-0.59\times 10^{-2}}$\, \\
$\lambda_{1}$&${+2.48\times 10^{-1}}^{+0.42\times 10^{-1}}_{-0.72\times 10^{-1}}$&${+2.48\times 10^{-1}}^{+0.76\times 10^{-1}}_{-1.11\times 10^{-1}}$\, \\
$\lambda_{2}$&${+1.80\times 10^{-3}}^{+0.01\times 10^{-3}}_{-0.30\times 10^{-3}}$&${+1.80\times 10^{-3}}^{+0.12\times 10^{-3}}_{-0.34\times 10^{-3}}$\, \\
$\omega$ & ${-1.068}^{+0.070}_{-0.001}$&${-1.068}^{+0.114}_{-0.006}$\, \\
\hline
\end{tabular}
  \caption{Summary of the $1\sigma$ and $2\sigma$ errors of the best estimate for $N=2$.}
  \label{BestestimationIIIerrors}
\end{table}
\begin{table}
  \centering
\begin{tabular}{| c | c | c | c | c | c |}
\multicolumn{6}{c}{\textbf{Model IV: $w=$constant and $\Omega_{DM}^0=$constant }} \\
\multicolumn{6}{c}{Best estimates for the parameters $\lambda_n$, $w$ and $\Omega_{DM}^0$.} \\
\hline  & $N=1$ & $N=2$ & $N=3$ & $N=4$ & $N=5$ \\
\hline
$\lambda_{0}$&$-1.0041\times 10^{-2}$&$-1.0045\times 10^{-2}$&$-1.0050\times 10^{-2}$&$-1.0046\times 10^{-2}$&$-1.0042\times 10^{-2}$ \\
$\lambda_{1}$&$2.41\times 10^{-1}$&$2.38\times 10^{-1}$&$2.351\times 10^{-1}$&$2.3509\times 10^{-1}$&$2.3500\times 10^{-1}$ \\
$\lambda_{2}$&$0.0$&$2.17\times 10^{-3}$&$2.25\times 10^{-3}$&$2.292\times 10^{-3}$&$2.294\times 10^{-3} $ \\
$\lambda_{3}$&$0.0$&$0.0$&$1.06\times 10^{-5}$&$1.06\times 10^{-5}$&$1.90\times 10^{-5}$ \\
$\lambda_{4}$&$0.0$&$0.0$&$0.0$&$1.0\times 10^{-6}$&$1.80\times 10^{-6}$ \\
$\lambda_{5}$&$0.0$&$0.0$&$0.0$&$0.0$&$1.0\times 10^{-7}$ \\
$w$&$-0.979$&$-0.973$&$-0.968$&$-0.964$&$-0.96$ \\
$\Omega_{DM}^0$&$0.21$&$0.207$&$0.205$&$0.204$&$0.202$ \\
\hline $\tilde{\chi}^{2}_{min}$&$542.77$&$542.73$&$542.71$&$542.70$&$542.69$ \\
\hline
\end{tabular}
 \caption{The same as the Table \ref{BestestimationII}.
  for the Model IV corresponding to a interacting dark energy EOS
  parameter $w=$ constant and $\Omega_{DM}^0$ as a free parameter to be estimated.
  From the Figure \ref{ExchangeModel} to Figure \ref{DecelerateModel}, show the reconstruction of several cosmological variables using the best
  estimates for $N=5$. Moreover, They also show the reconstruction of
  several cosmological variables using the best estimates for $N=2$ and their confidence intervals at $1\sigma$ and $2\sigma$.}
  \label{BestestimationIV}
 \end{table}

\begin{table}
  \centering
\begin{tabular}{| c | c | c |}
\multicolumn{3}{c}{\textbf{Model IV: $w=$constant and $\Omega_{DM}^0=$constant.}} \\
\hline Errors & $\pm 1\sigma$ & $\pm 2\sigma$\\
\hline \hline
$\lambda_{0}$&${-1.00\times 10^{-2}}^{+0.07\times 10^{-2}}_{-0.16\times 10^{-2}}$&${-1.00\times 10^{-2}}^{+0.04\times 10^{-2}}_{-0.65\times 10^{-2}}$\, \\
$\lambda_{1}$&${+2.38\times 10^{-1}}^{+0.25\times 10^{-1}}_{-0.30\times 10^{-1}}$&${+2.38\times 10^{-1}}^{+0.36\times 10^{-1}}_{-0.43\times 10^{-1}}$\, \\
$\lambda_{2}$&${+2.17\times 10^{-3}}^{+0.13\times 10^{-3}}_{-0.03\times 10^{-3}}$&${+2.17\times 10^{-3}}^{+0.22\times 10^{-3}}_{-0.28\times 10^{-3}}$\, \\
$\omega$&${-0.973}^{+0.045}_{-0.078}$&${-0.973}^{+0.079}_{-0.112}$\, \\
$\Omega^{0}_{DM}$&${+0.207}^{+0.013}_{-0.016}$&${+0.207}^{+0.031}_{-0.020}$\, \\
\hline
\end{tabular}
  \caption{Summary of the $1\sigma$ and $2\sigma$ errors of the best estimate for $N=2$.}
  \label{BestestimationIVerrors}
\end{table}

\begin{center}
\begin{figure}
   \includegraphics[width=8cm, height=60mm, scale=0.90]{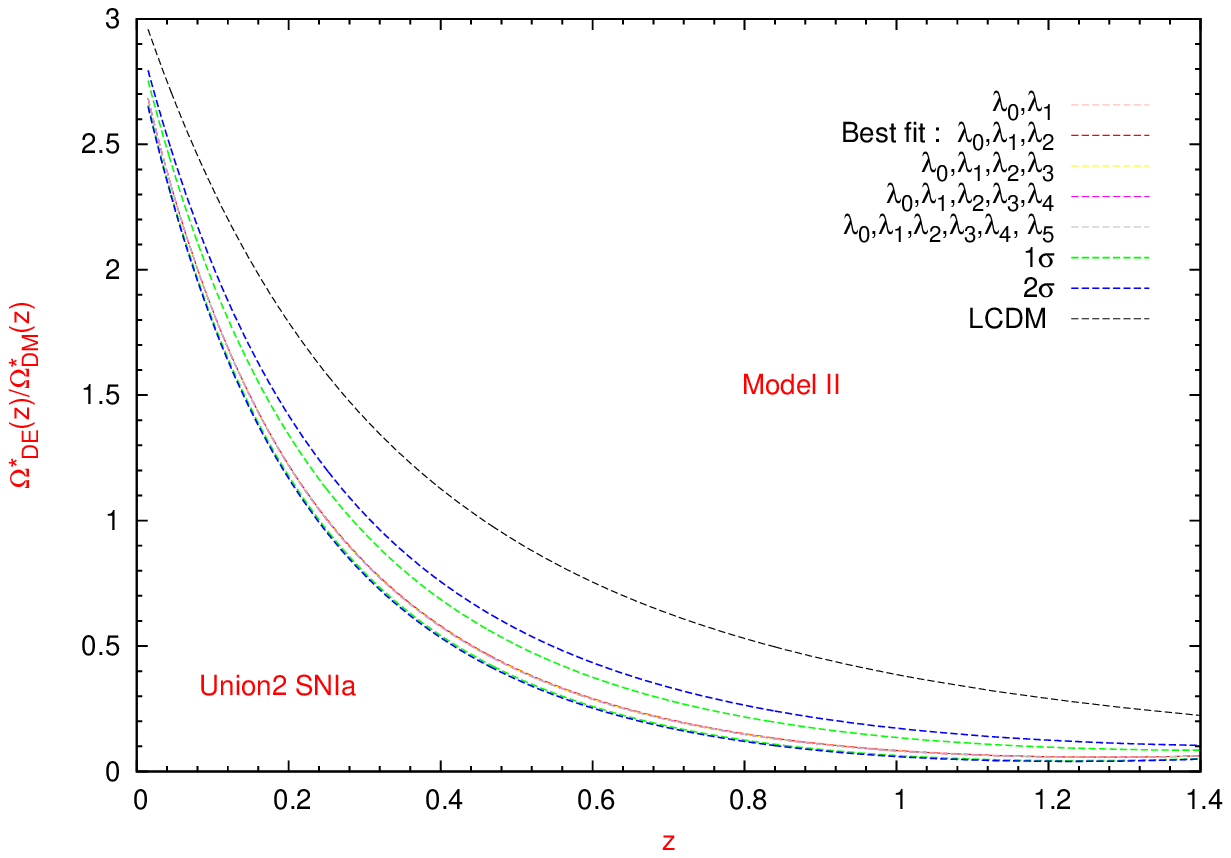}
   \includegraphics[width=8cm, height=60mm, scale=0.90]{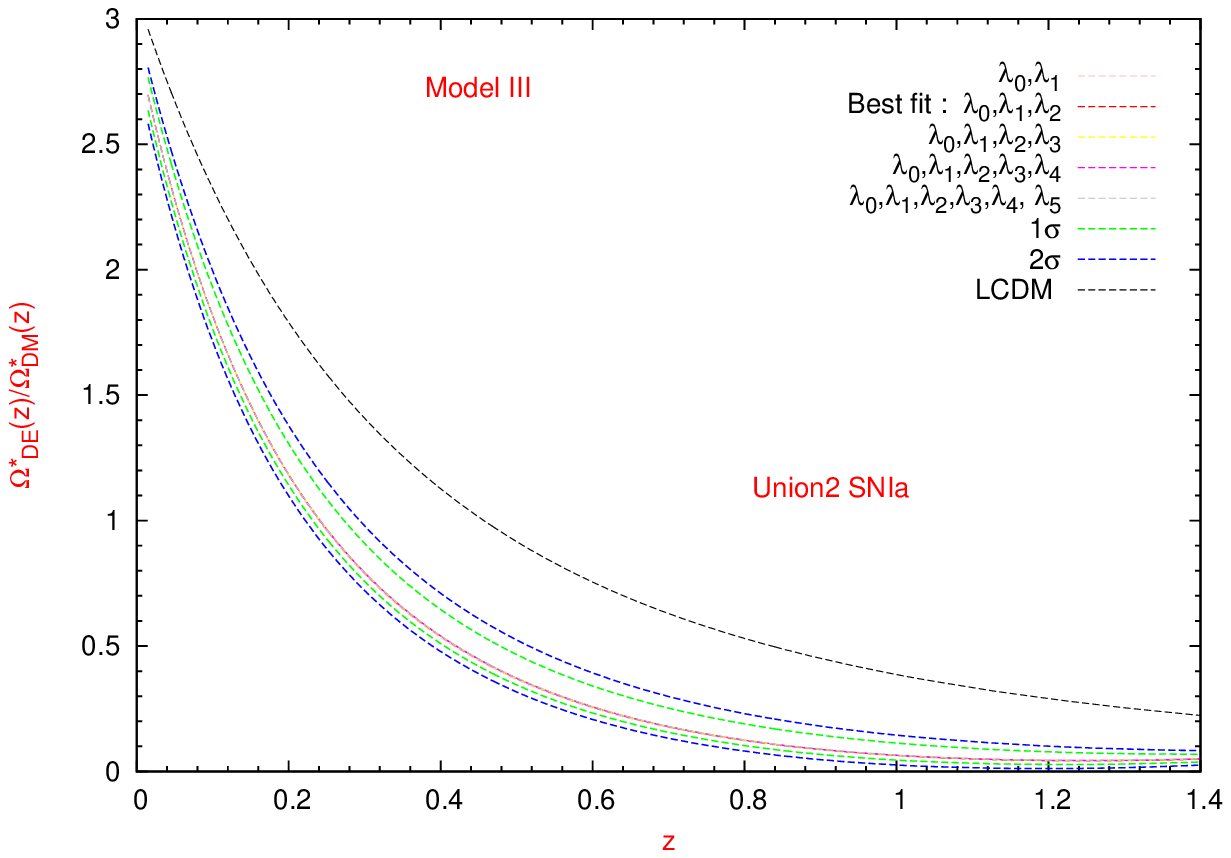}
   \includegraphics[width=8cm, height=60mm, scale=0.90]{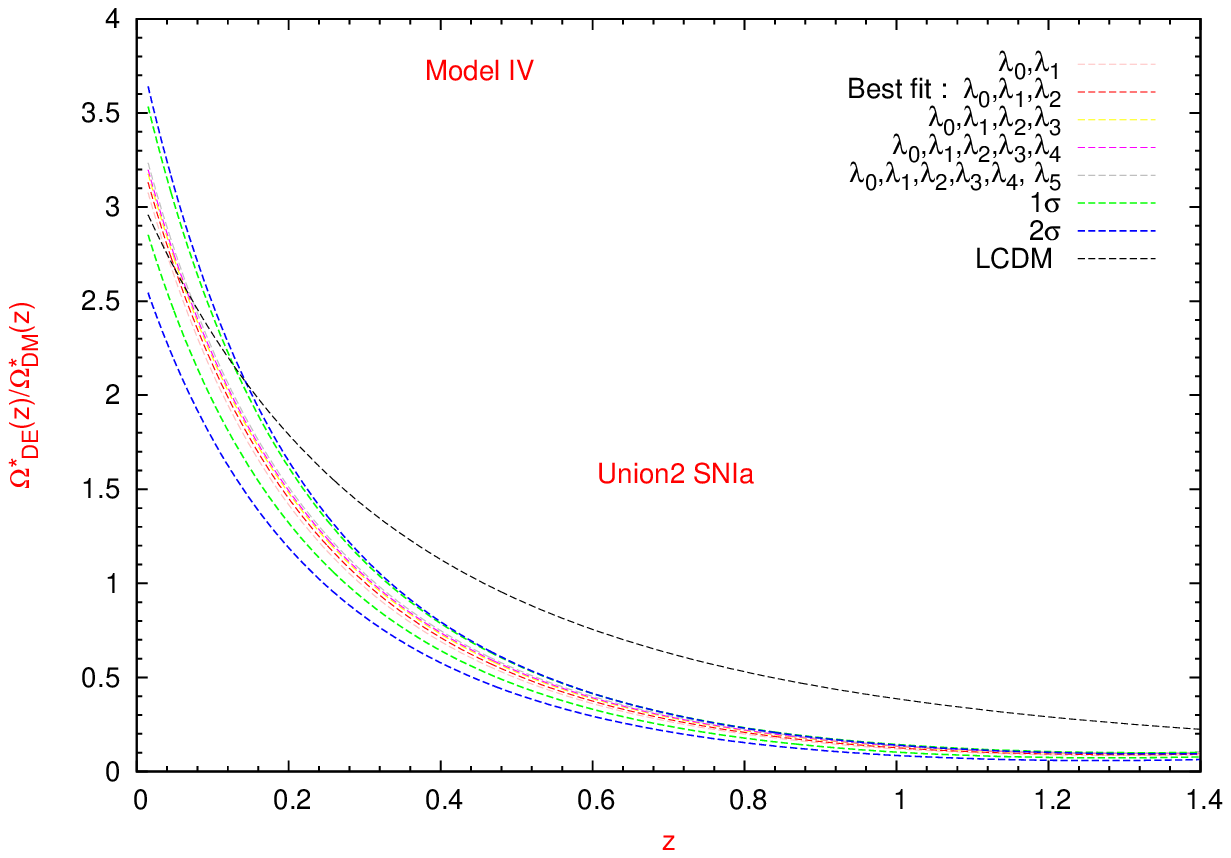}
   \includegraphics[width=8cm, height=60mm, scale=0.90]{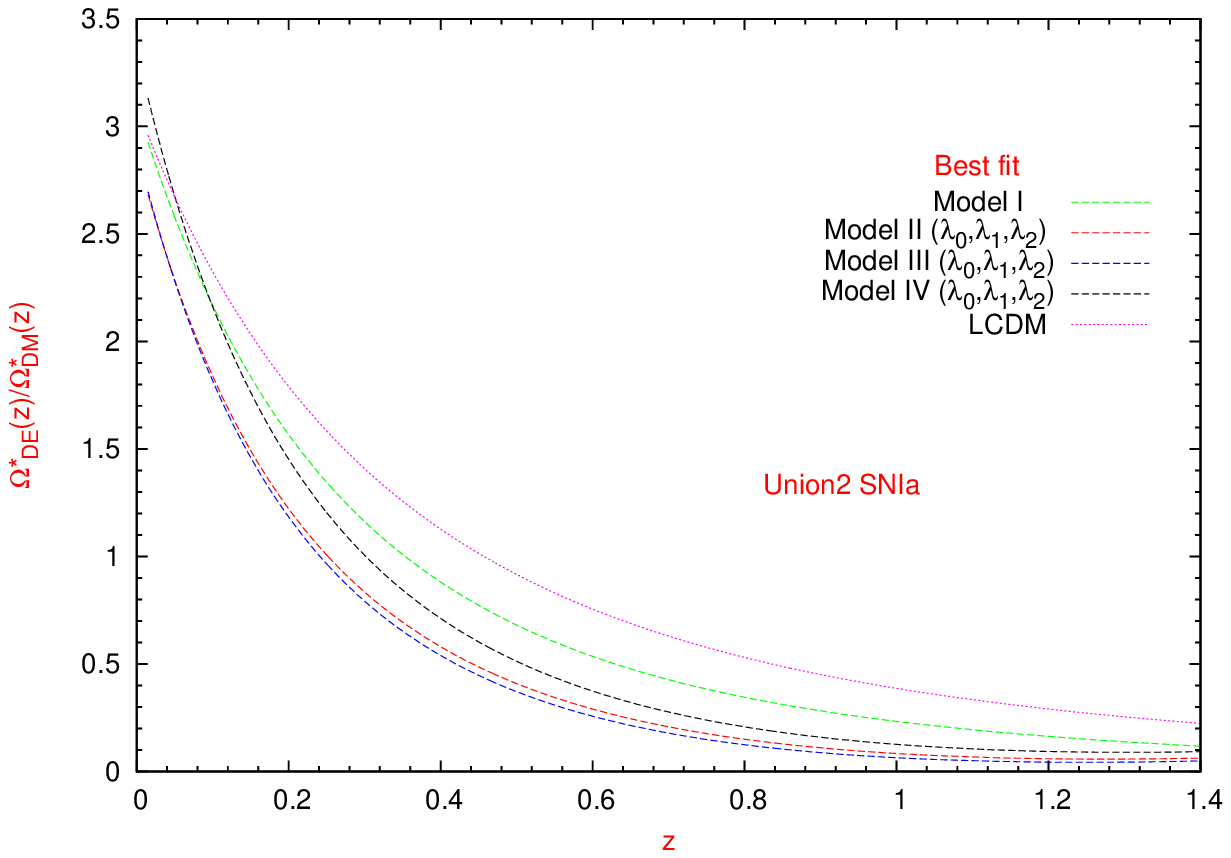}
  \caption{Superposition of the best estimates for the rate between dark
  density parameters $\Omega^{\star}_{DE}(z)/\Omega^{\star}_{DM}(z)$ for
  the model II (left above panel), III (right above panel), IV (left below panel).
  In the above figures, the different colored curves
  show the best estimates using the expansion in terms of the first
  $N=1,2,3,4,5$ Chebyshev polynomials respectively. For comparison purposes, the
  figure of the right below panel shows the best estimates for the same variable
  for models I, II, III and IV using the expansion in terms of the first
  $N=2$ Chebyshev polynomials and the corresponding curve for the LCDM model.
  Here  we also show the best estimates (red lines), the $1\sigma$ (green lines)
  and $2\sigma$ errors (blue lines) for the rate between dark
  density parameters $\Omega^{\star}_{DE}(z)/\Omega^{\star}_{DM}(z)$ as a function of the redshift.}
  \label{RelationModels}
\end{figure}
\end{center}

\begin{center}
\begin{figure}
   \includegraphics[width=8cm, height=60mm, scale=0.90]{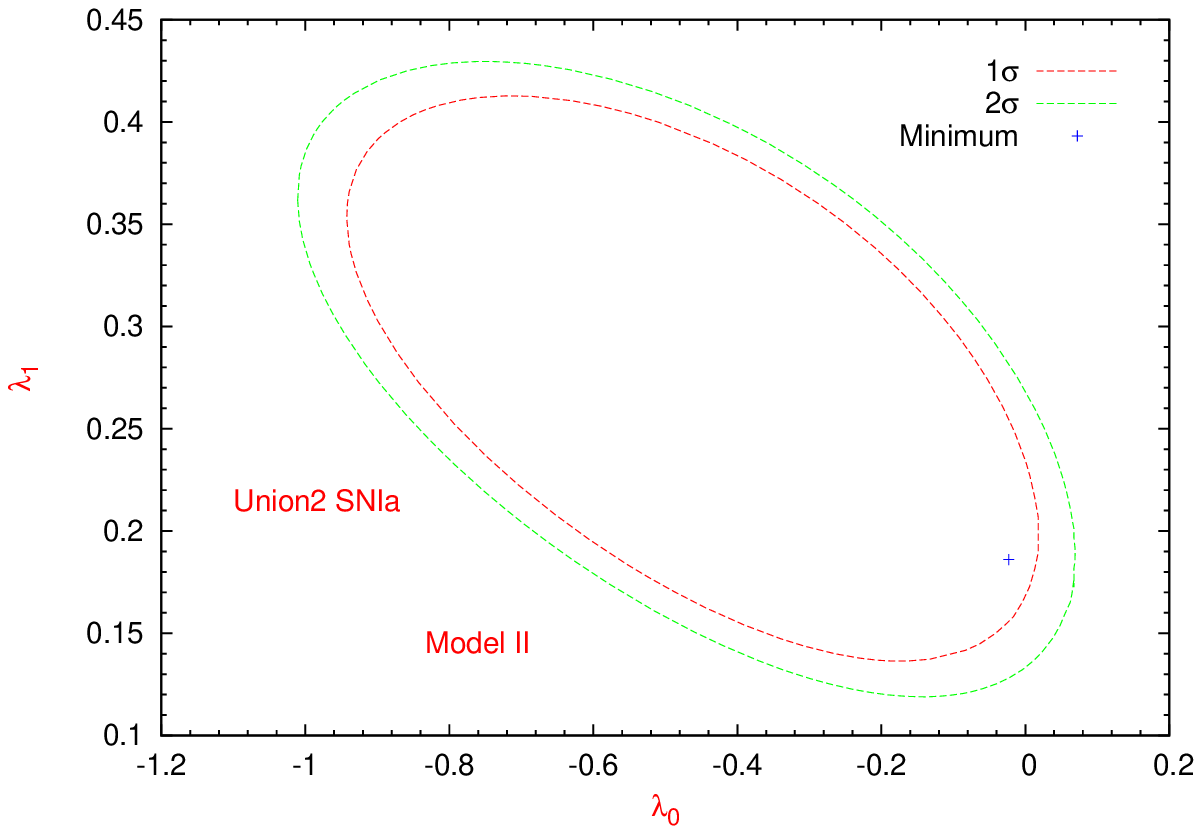}
   \includegraphics[width=8cm, height=60mm, scale=0.90]{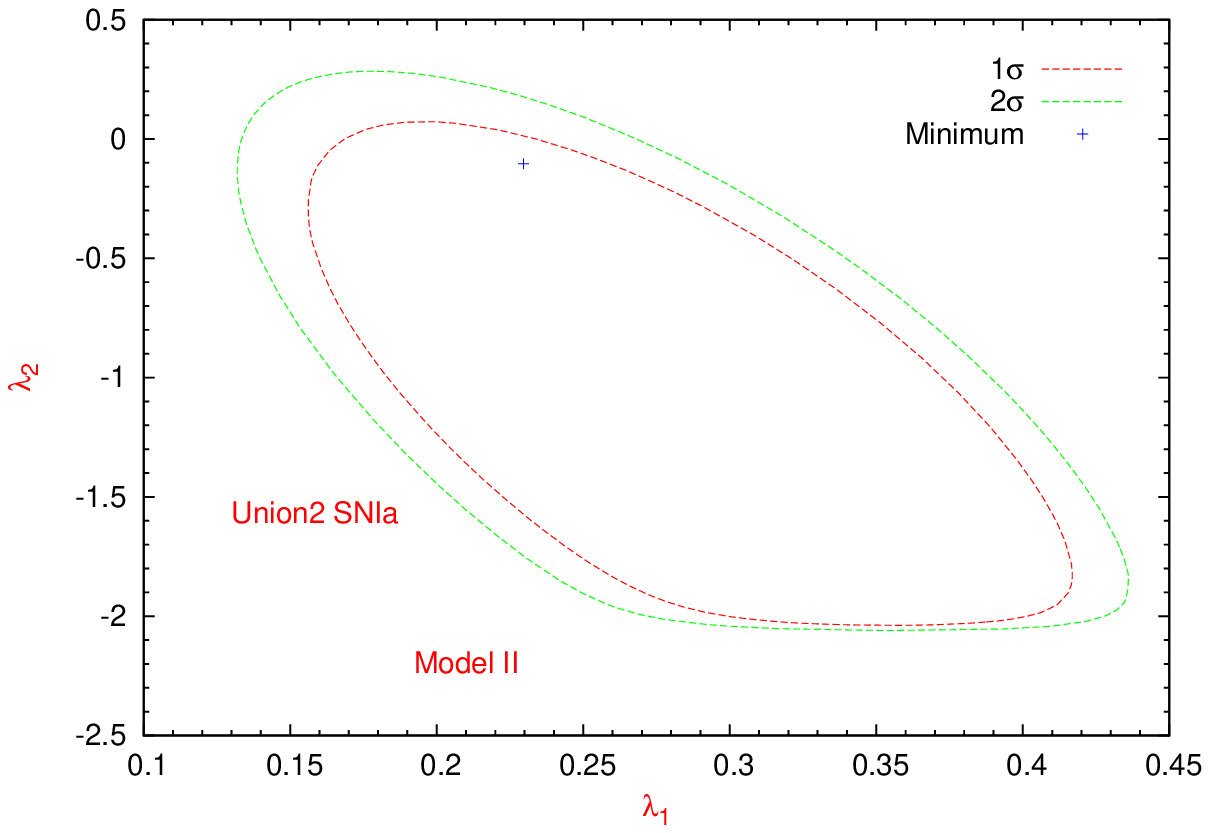}
  \caption{Best estimate and the $1\sigma$, $2\sigma$ confidence
  intervals for the marginalized probability densities for the model II
  using an expansion in terms of the first $N=2$ Chebyshev polynomials.
  Before marginalization, we have three free parameters
  ($\lambda_0$,$\lambda_1$,$\lambda_2$).
  In every figure, we marginalized on one of the parameters.}
  \label{NewLCDM2}
\end{figure}
\end{center}

\begin{center}
\begin{figure}
   \includegraphics[width=8cm, height=60mm, scale=0.90]{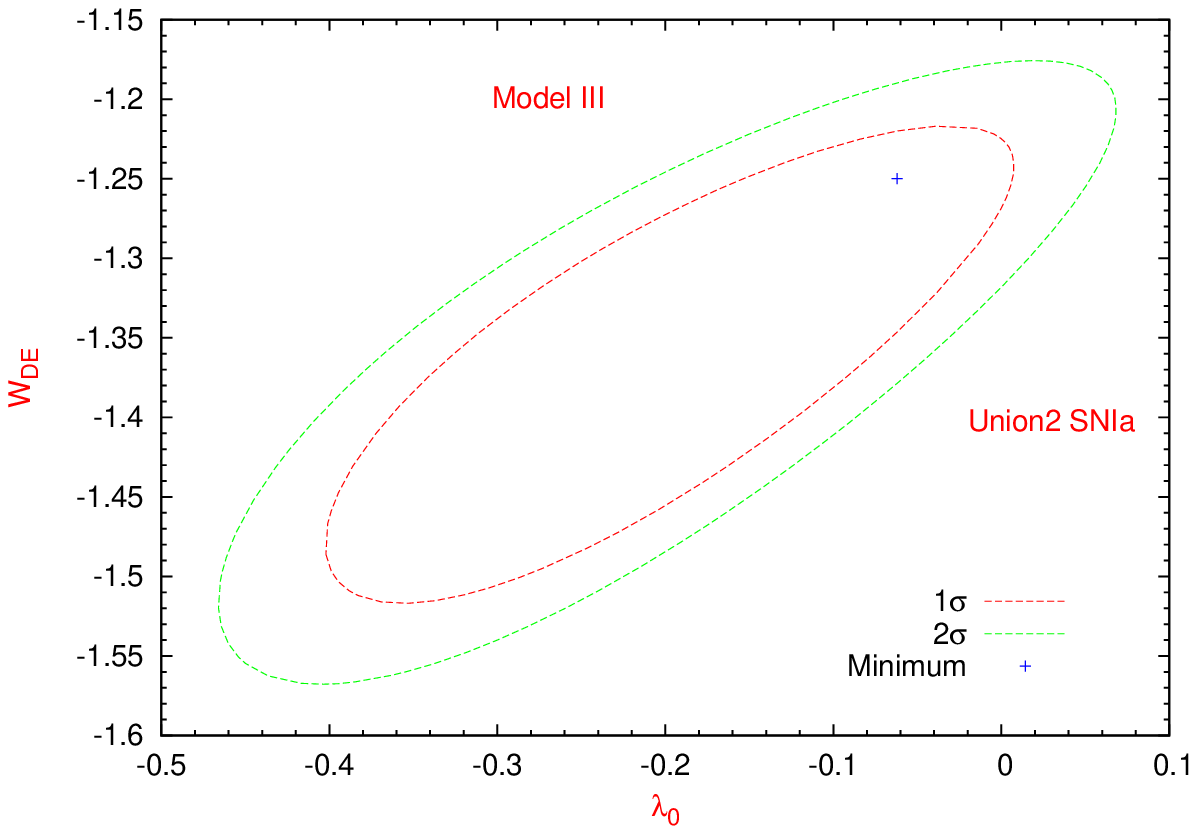}
   \includegraphics[width=8cm, height=60mm, scale=0.90]{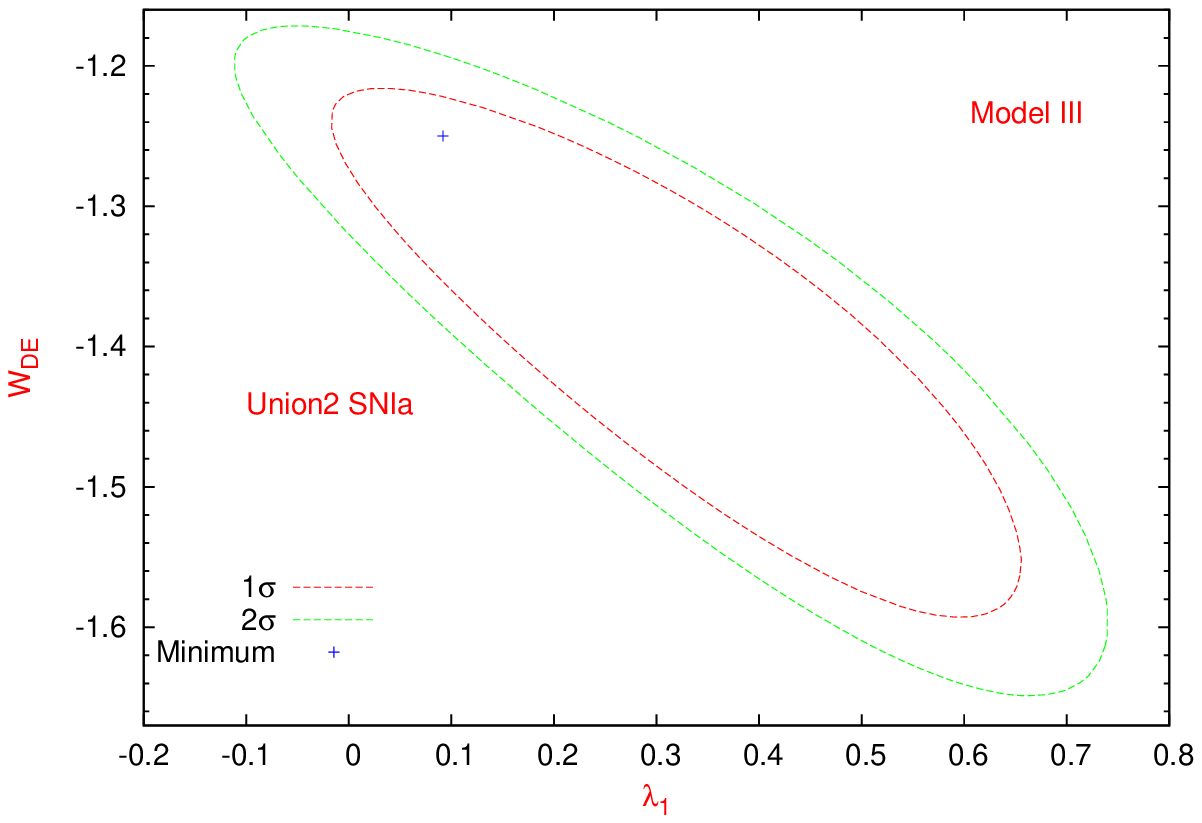}
   \includegraphics[width=8cm, height=60mm, scale=0.90]{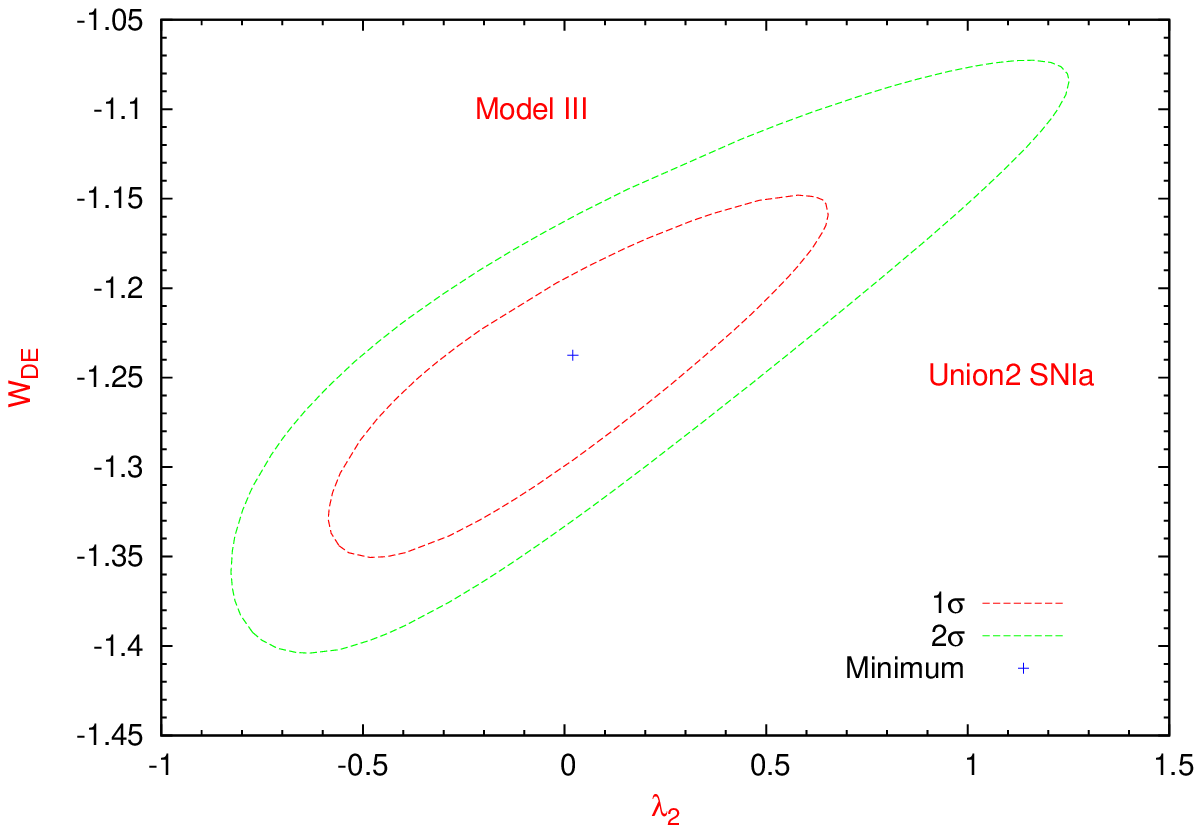}
   \includegraphics[width=8cm, height=60mm, scale=0.90]{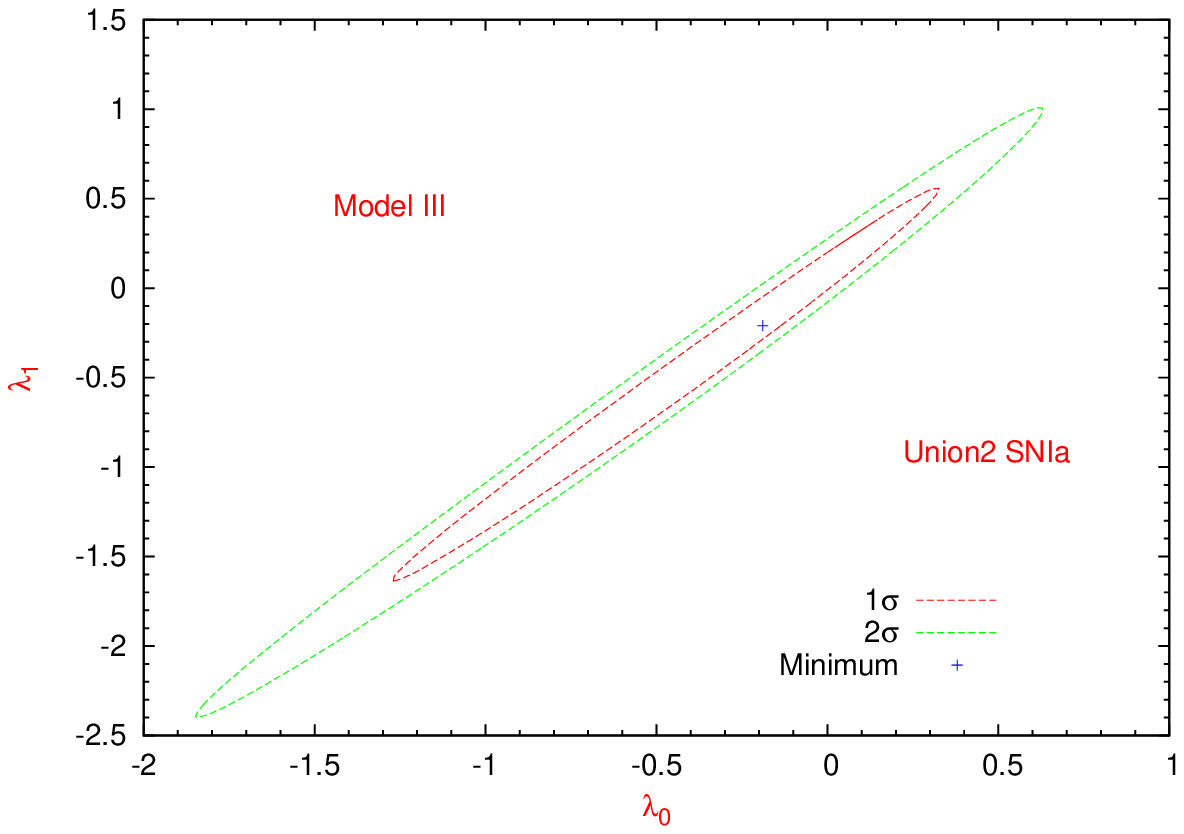}
   \includegraphics[width=8cm, height=60mm, scale=0.90]{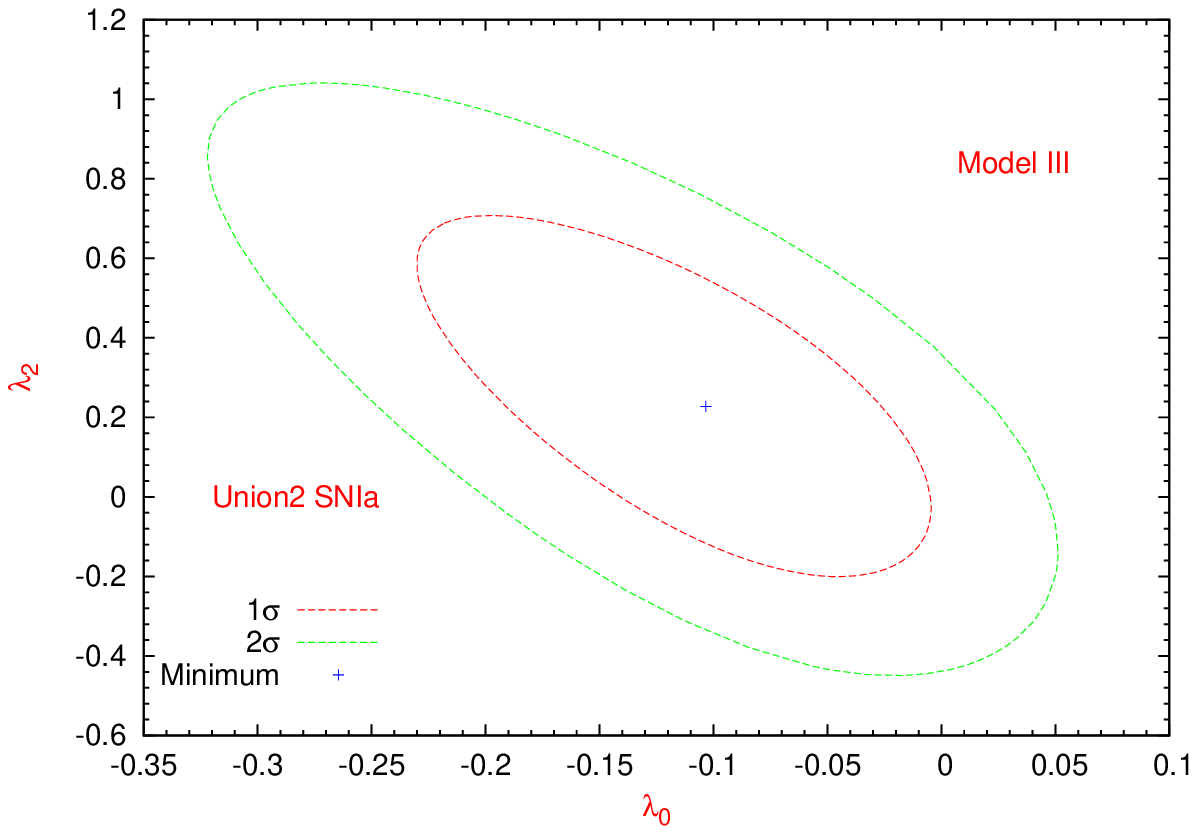}
   \includegraphics[width=8cm, height=60mm, scale=0.90]{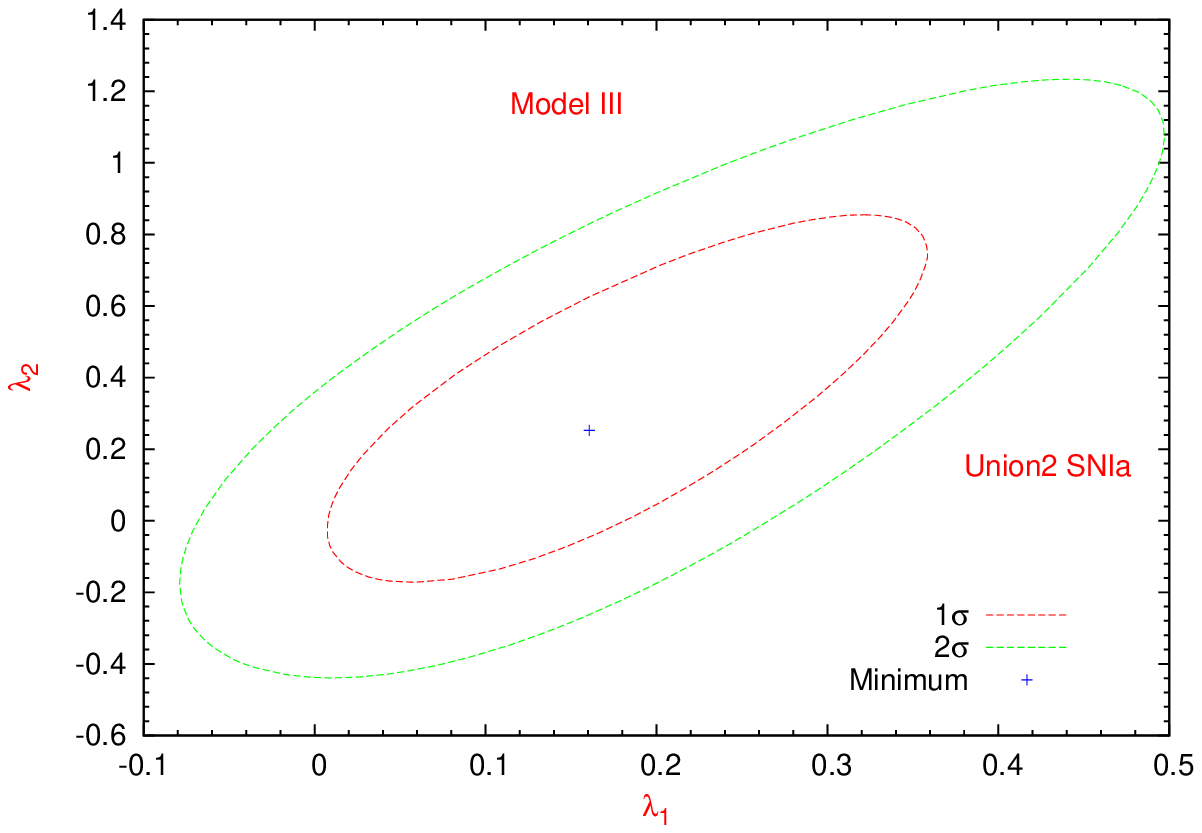}
  \caption{Best estimate and the $1\sigma$, $2\sigma$ confidence intervals
  for the marginalized probability densities for the model III using an
  expansion in terms of the first $N=2$ Chebyshev polynomials.
  Before marginalization, we have
  four free parameters ($\lambda_0$,$\lambda_1$,$\lambda_2$,$w$).
  In every figure, we marginalized on the last two remaining
  parameters. Note that the preferred region for the EOS parameter $w$
  is the phantom region.}
  \label{NewPhantom3}
\end{figure}
\end{center}

\begin{center}
\begin{figure}
   \includegraphics[width=8cm, height=60mm, scale=0.90]{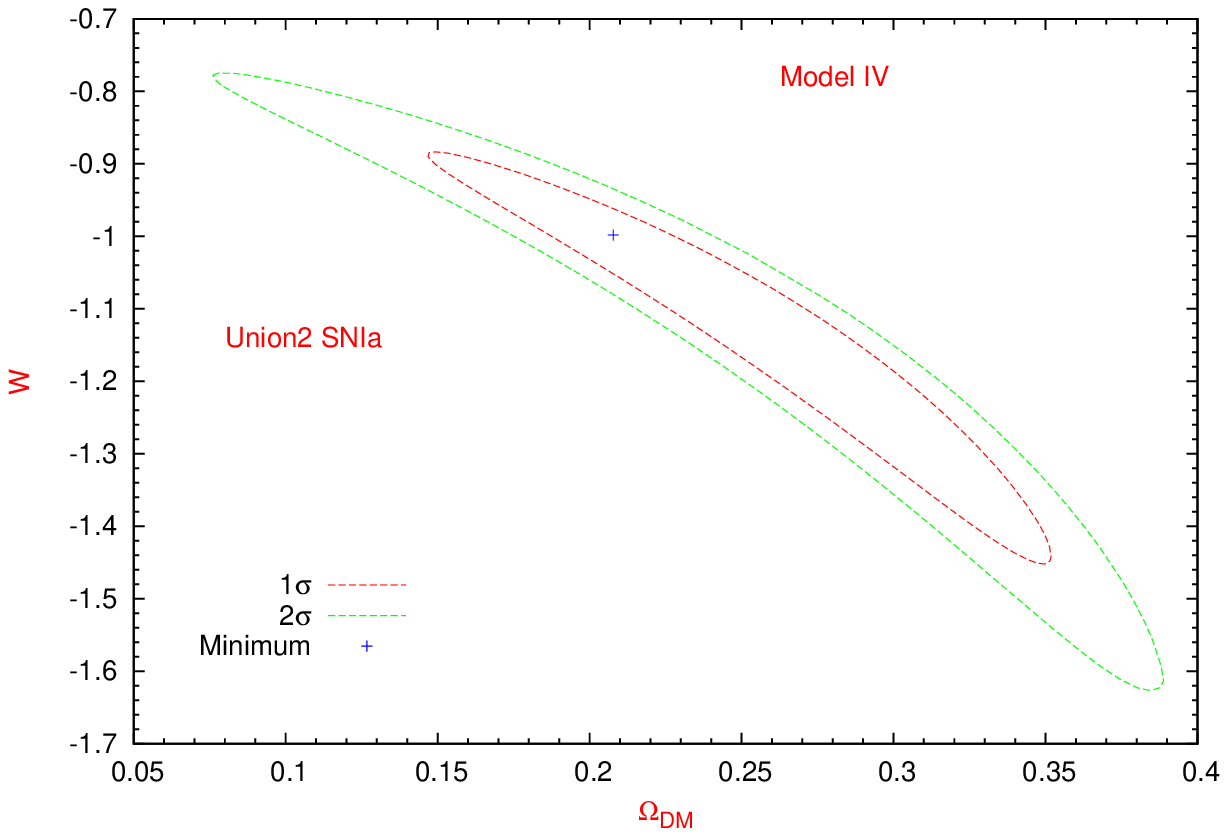}
   \includegraphics[width=8cm, height=60mm, scale=0.90]{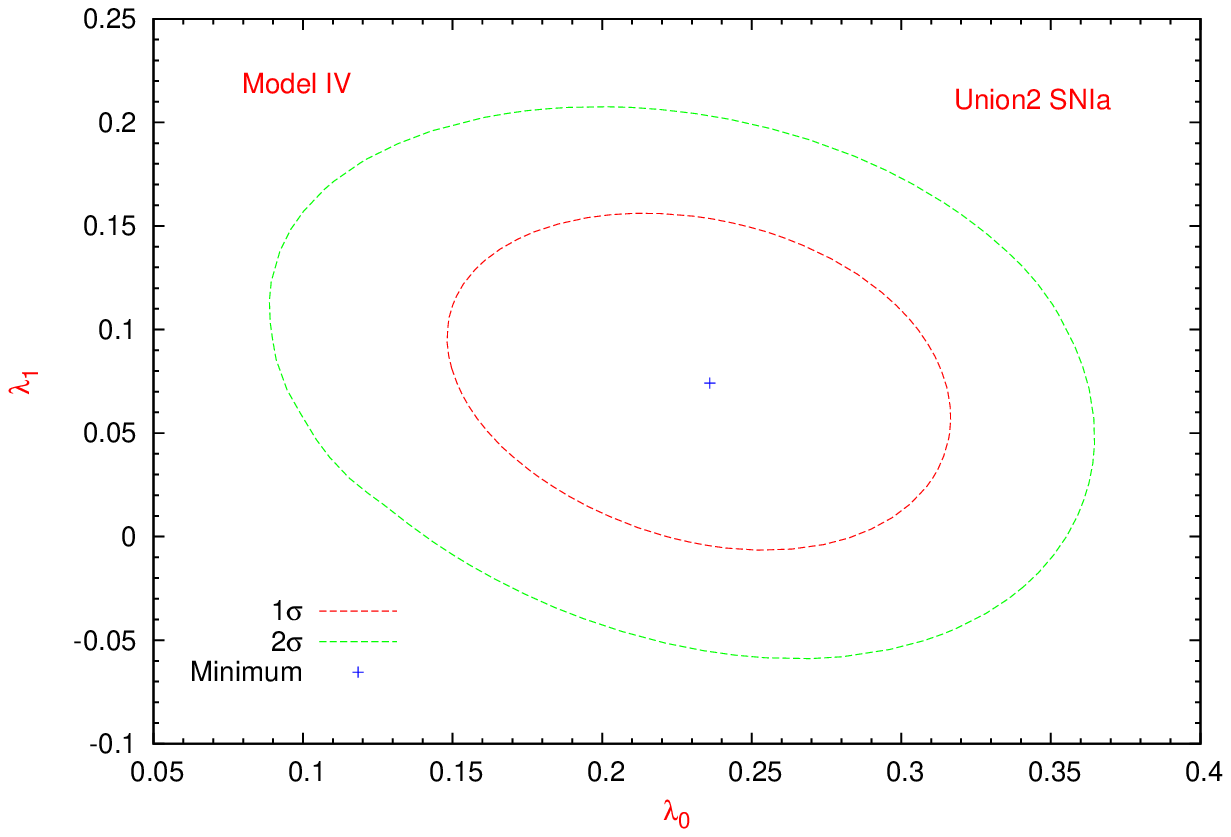}
   \includegraphics[width=8cm, height=60mm, scale=0.90]{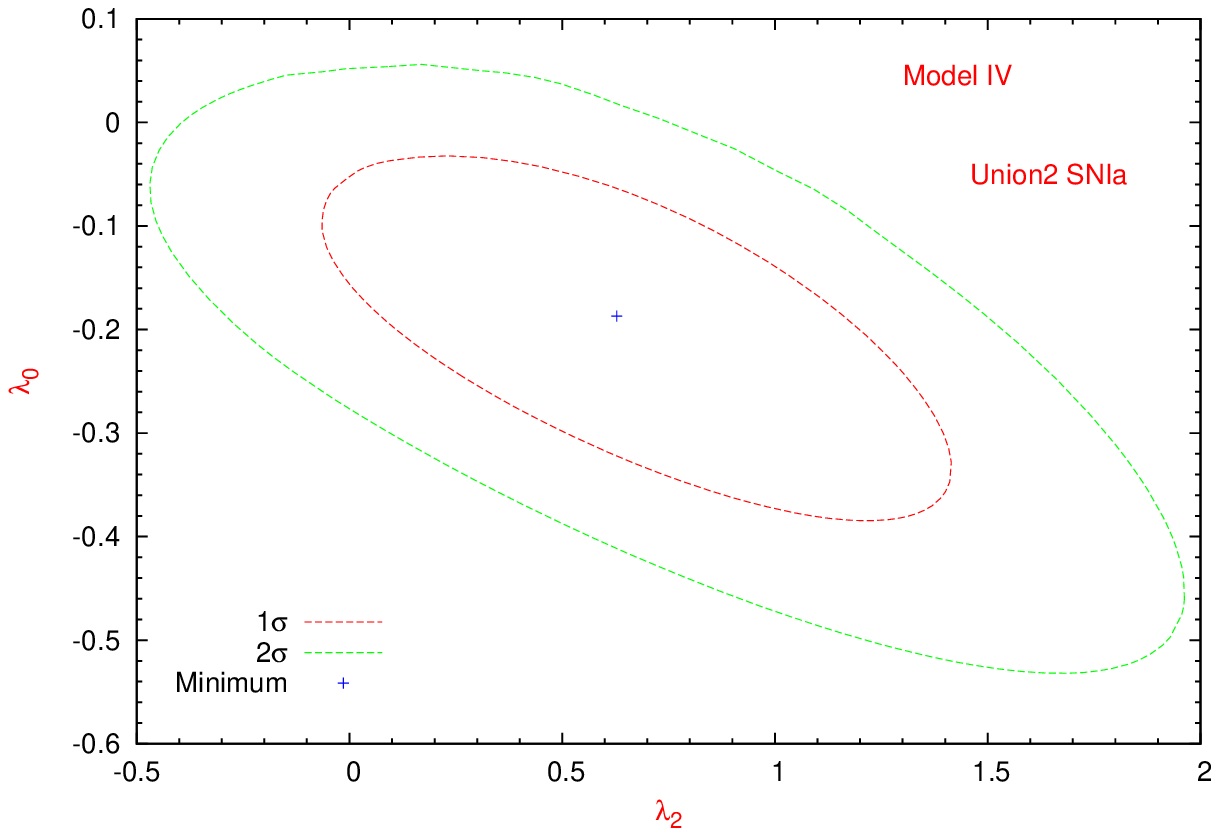}
  \caption{Best estimate and the $1\sigma$, $2\sigma$ confidence intervals
  for the marginalized probability densities for the model IV using an
  expansion in terms of the first $N=2$ Chebyshev polynomials.
  Before marginalization, we have five free parameters
  ($\lambda_0$,$\lambda_1$,$\lambda_2$, $w$, $\Omega_{DM}^0$).
  In every figure, we marginalized on the last three remaining
  parameters.}
  \label{Newmodel4}
\end{figure}
\end{center}

However, a recent study fitting CMB anisotropy data from the
seven-year Wilkinson Microwave Anisotropy Probe (WMAP)
\cite{Komatsu2011}, the BAO distance measurements \cite{ReidSDSS},
the Constitution sample of SnIa \cite{Riess1998} and constraints on
the present-day Hubble constant, put stronger constraints on the
magnitude of such dimensionless strength of the order of $\xi
\approx 10^{-2}-10^{-4}$ \cite{abramo2} (for more recent constraints
to see the references
\cite{valiviita-majerotto-maartens-2010-2}-\cite{cao-liang-zhu-2011}).
\\
In order to study the coincidence problem, in the Figure
\ref{RelationModels} we plot the best estimates for the rate between
dark density parameters
$\Omega^{\star}_{DE}(z)/\Omega^{\star}_{DM}(z)$ for the model II
(left above panel), III (right above panel) and IV (left below panel).
\\
Finally, in the Figure \ref{NewLCDM2} we plot the $1\sigma$ and
$2\sigma$ constraints on all the pair of parameters taken from the
set $\{\lambda_{0}, \lambda_{1}, \lambda_{2}\}$ for the Model II
marginalizing on one of them. The same is plotted in the Figure
\ref{NewPhantom3} for the model III but with the parameters taken
from the set $\{w, \lambda_{0}, \lambda_{1}, \lambda_{2}\}$ and
marginalizing on two of them. Finally, in the Figure
\ref{Newmodel4}, a similar procedure is applied to the model IV with
the parameters taken from the set $\{w, \Omega_{DM}^0, \lambda_{0},
\lambda_{1}, \lambda_{2}\}$ and marginalizing on three of them.
\\
We noted that for the model III, from the Figure \ref{ExchangeModel}
to Figure \ref{CoincidenceModel} and the left
medium panel of Figure \ref{NewPhantom3} show that the $1\sigma$ and
$2\sigma$ constraints imply that the preferred region for the EOS
parameter $w$ is totally contained in the phantom region. This is
due to the fact that we have considered only the fit to the SNIa
observations which are affected by systematic errors
\cite{nesseris1}-\cite{nesseris2}, \cite{wei}. In fact, some studies
show that there exists the possibility of a crossing of the phantom
divide line \cite{nesseris3}. By the contrary, for the model IV, the
above panels of Figure \ref{Newmodel4} show that the $1\sigma$ and
$2\sigma$ constraints on the EOS parameter $w$ contain high
probability of being in the quintessence region. Finally we note the
broad dispersion on the dark matter density parameter measured at
the present $\Omega_{DM}^0$.

\begin{center}
\begin{figure}
   \includegraphics[width=8cm, height=60mm, scale=0.90]{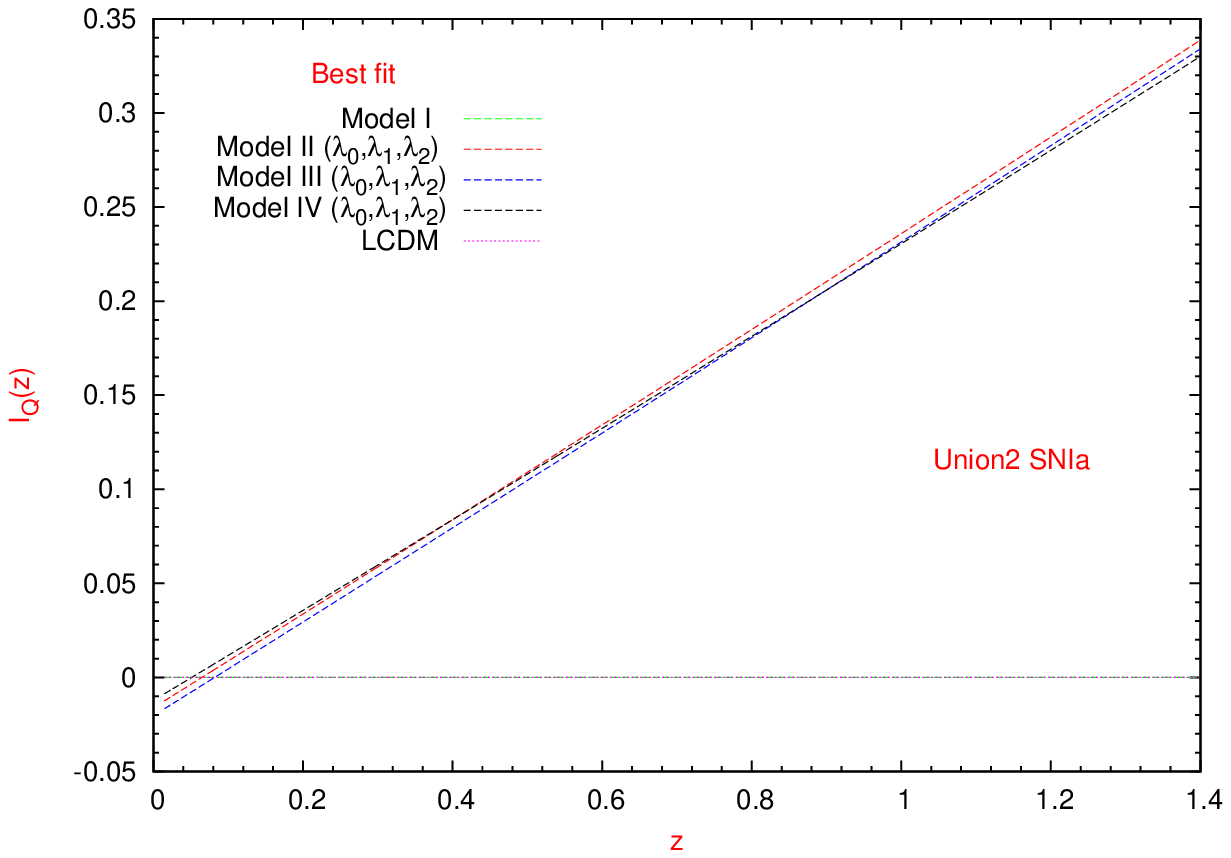}
   \includegraphics[width=8cm, height=60mm, scale=0.90]{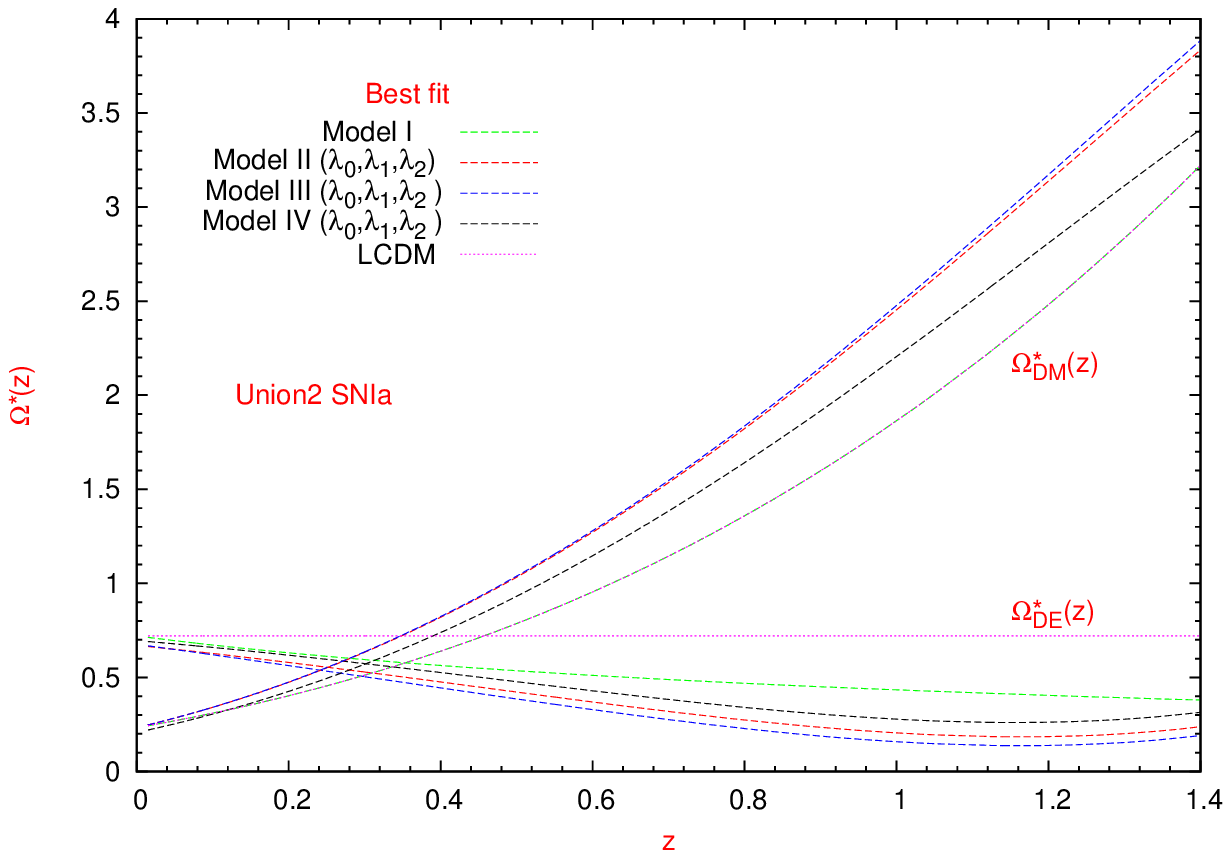}
   \includegraphics[width=8cm, height=60mm, scale=0.90]{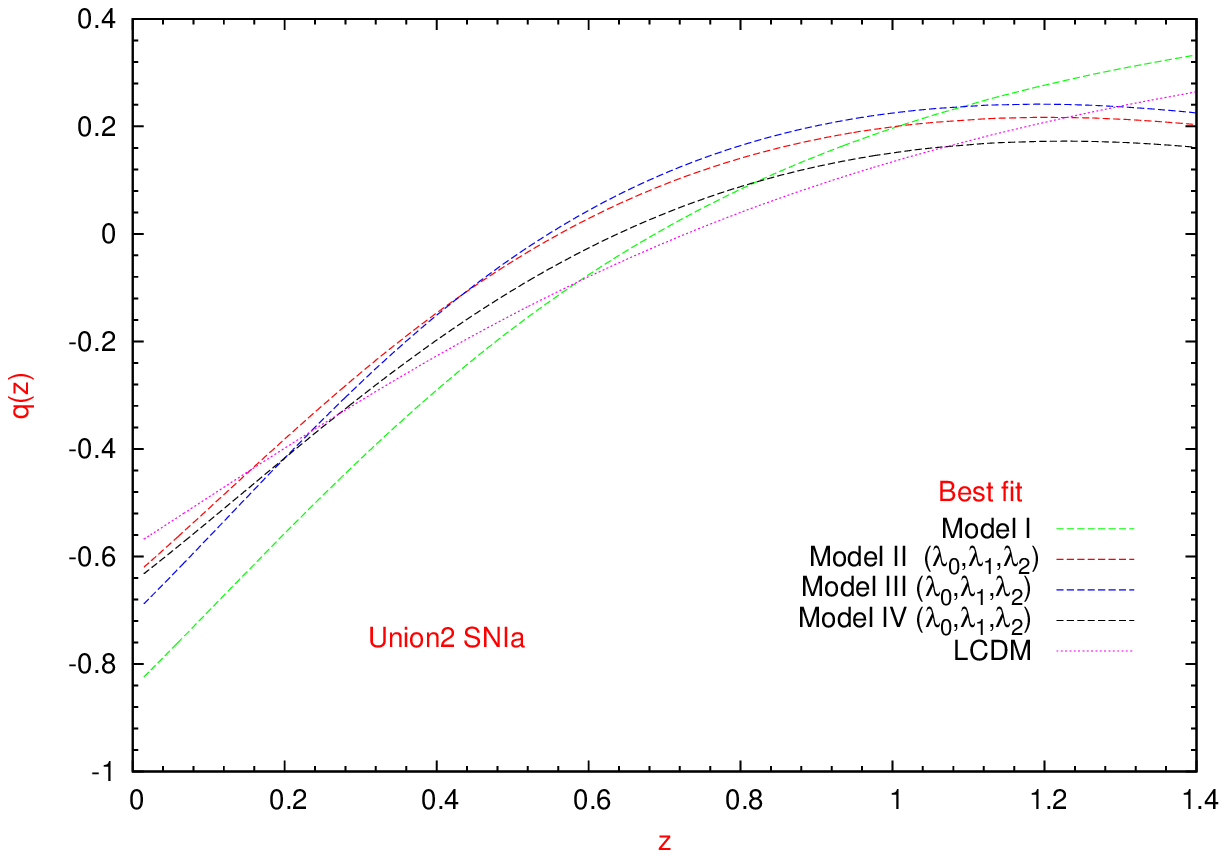}
  \caption{Superposition of the best estimates for the dimensionless interaction function
  ${\rm I}_{\rm Q}(z)$ (left above panel), the density parameters
  $\Omega^{\star}_{DM}(z), \, \Omega^{\star}_{DE}(z)$ (right above panel) and
  the deceleration parameter $q(z)$ (left below panel) as a function of the redshift
  for the models I (green line), II (red line), III (blue line) and IV (black line).
  By comparison, the LCDM model (pink line) is shown.
  The curves show the best estimates using the expansion of all the functions
  in terms of the first $N=2$ Chebyshev polynomials.
  Note that the reconstruction of the best estimate of the dimensionless
  interaction function ${\rm I}_{\rm Q}(z)$ for the models II, III and
  IV produces roughly the same curve and that the density parameter of dark
  energy is definite positive for all the range of redshift
  considered in the reconstruction.}
  \label{FourModels}
\end{figure}
\end{center}

The Figure \ref{FourModels} shows the superposition of the best
estimates for the dimensionless interaction function
${\rm I}_{\rm Q}(z)$ (left above panel), the density parameters
$\Omega^{\star}_{DM}(z), \, \Omega^{\star}_{DE}(z)$ (right above panel) and
the deceleration parameter $q(z)$ (left below panel) as a function of the redshift
for the models I, II, III and IV respectively.

\newpage

\newpage

        \section{Conclusions.}
        \label{SectionConclusions}

In this paper, we developed theoretically a novel method for the
reconstruction of the interaction function between dark matter and
dark energy assuming an expansion of the general interaction term
proportional to the Hubble parameter in terms of Chebyshev
polynomials which form a complete set of orthonormal functions. To
show how the method works, we applied it to the reconstruction of
the interaction function expanding it in terms of only the first $N$
Chebyshev polynomials (with $N=1,2,3,4,5$) and fitted for the
coefficients of the expansion  assuming two models: (a) a DE
equation of the state parameter $w =-1$ (an interacting cosmological
$\Lambda$) and (b) a DE equation of the state parameter $w =$
constant. The fit of the free parameters of every model is done
using the Union2 SNe Ia data set from ``The Supernova Cosmology
Project'' (SCP) composed by 557 type Ia supernovae
\cite{AmanullahUnion22010}.

Our principal results can be summarized as follows:
\begin{enumerate}
    \item Our fitting results show the fast convergence of the
    best estimates for the several cosmological variables considered
    in this paper when the numbers of parameters $N$ is increased
    in the expansion (\ref{eq:Coupling}).
    \item The best estimates for the interaction function
    ${\rm I}_{\rm Q}(z)$ prefer to cross the noninteracting line
    ${\rm I}_{\rm Q}(z)=0$ during the present cosmological evolution.
    This conclusion is independent of the numbers of coefficients
    (up to $N=5$ in this work) used in the expansion of
    ${\rm I}_{\rm Q}(z)$. The crossing implies a change of sign of
    ${\rm I}_{\rm Q}(z)$ from positive values at the past
    (energy transfers from dark energy to dark matter)
    to negative values at the present (energy transfers
    from dark matter to dark energy).
    The decay direction is contrary to the results found in the
    recent literature \cite{He-Zhang} and in disagreement with the
    oscillatory behavior reported in \cite{Cai-Su}.
    \item The statement above is conclusive because the
    existence of crossing of the noninteraction line
    ${\rm I}_{\rm Q}(z) = 0$ in some moment of the recent past is
    totally contained inside the $1\sigma$ and $2\sigma$ constraints given
    by SNe Ia observations.
    \item For all the interacting models studied in this work, the best estimates for the
    dark energy density parameter $\Omega^{\star}_{DE}(z)$ becomes
    positive definite in the range of redshifts considered in this work.
    This statement is conclusive because, within
    the $1\sigma$ and $2\sigma$ errors for the fit with three parameters
    ($N=2$), $\Omega^{\star}_{DE}(z)$
    becomes positive in all the range of redshifts considered in the sample of SnIa.
    \item The $1\sigma$ and $2\sigma$ confidence intervals, for the
    EOS parameter $w$ considered in the marginalized model III, are
    totally contained in the phantom region ($w<-1$).
    This is not totally true for the model IV which presents
    high regions of probability for the EOS parameter $w$ of being
    in the phantom or in the quintessence region. This is
    because we have performed the fit with only the SNe Ia data set
    which, for some data samples, is known to fit preferably in the phantom region
    \cite{nesseris1}-\cite{nesseris2}, \cite{wei}.
    This last problem can be corrected if we use
    more cosmological observations that can provide information of
    the late and early universe as such observations of the Cosmic
    Microwave Background (CMB) anisotropies from Microwave
    Anisotropy Probe (WMAP) experiment and the large scale structure
    (LSS) from Sloan Digital Sky Survey (SDSS) experiment.
\end{enumerate}

In order to clarify the last points mentioned above (the crossing or
not of the noninteracting line and the appearance or not of phantom
regions for the EOS parameter $w$) in our reconstruction, we must
repeat our analysis fitting with more data samples covering a
broader range of the history of the universe: the Cosmic Microwave
Background (CMB) shift parameters from the test 7-year Wilkinson
Microwave Anisotropy Probe (WMAP) \cite{Komatsu2011}, the Baryon
Acoustic Oscillation (BAO) from experiment SDSS DR7 \cite{ReidSDSS},
a spectroscopic catalog of red galaxies in galaxy clusters
\cite{Stern2010-2}, the Hubble expansion rate (15 data)
\cite{Riess-Hubble}- \cite{Stern2010-1} and the X-ray gas mass
fraction (42 data) \cite{Sasaki1996}-\cite{Rapetti2008}. We will go
ahead with this analysis in our next future work
\cite{Cueva-Nucamendi}.

       \section{Acknowledgements.}
This work was in part supported by grants SNI-20733, CIC-UMSNH
No.4.8, UMSNH-CA-22. F. C. thanks support by CONACYT-SEP.


                \appendix

        \section{Calculation of the integrals $K_n(x,w)$ and $J_n(x,w)$.}
        \label{integration}

        \subsection{Calculation of the integrals $J_n(x,w)$.}
        \label{integrationJ}

In order to calculate the integrals $K_n(x,w)$ defined in
(\ref{IntegralK}) we need to obtain, for $m \geq 0$, closed
expressions for the integrals,
\begin{eqnarray}
\label{Integral-J} J_{m}(x, w) \equiv \int_{-1}^{x}
\frac{\tilde{x}^{m}}{(a+b\tilde{x})^{(1+3w)}} \, d\tilde{x}
\end{eqnarray}
To this goal, we use the recurrence relation valid for integers $m
\geq 1$, $m \neq 3w$,
\begin{eqnarray}
\label{Recurrence-Relation-J} J_{m}(x, w) = \frac{1}{(m-3w)b} \cdot
\Biggl\{ \Biggl[ \frac{x^m}{(a+bx)^{3w}} - (-1)^{m} \Biggr] - a
m J_{m-1}(x, w) \Biggr\}   \nonumber \\
\end{eqnarray}
where we have for the initial integrals $J_{0}(x, w)$,
\begin{equation}
\label{Integral-initialK} J_{0}(x, w) = \left\{
\begin{array}{rl}
\frac{1}{b} \ln (a+bx) & \mbox{if} \quad w = 0,\\
\frac{1}{3wb} \Biggl[ 1 - \frac{1}{(a+bx)^{3w}} \Biggr] & \mbox{if}
\quad w \neq 0.
\end{array} \right.
\end{equation}
From (\ref{Recurrence-Relation-J}) we can guess the series for
$J_{m}(x, w)$ in terms of $J_{0}(x, w)$ and $m \geq 1$,
\begin{eqnarray}
\label{Series-wd0} J_{m}(x, w) & = \frac{1}{b} \cdot \sum_{l=1}^{m}
(-1)^{(m-l)} \cdot \frac{m!}{l!} \cdot
\Biggl(\frac{a}{b}\Biggr)^{(m-l)} \cdot \Biggl[ \prod_{k=l}^{m}
\frac{1}{(k-3w)} \Biggr] \cdot \nonumber
\\
& \quad \qquad \quad \cdot \Biggl[ \frac{x^l}{(a+bx)^{3w}} -(-1)^{l}
\Biggr] \nonumber
\\
& \quad + (-1)^{m} \cdot \Biggl(\frac{a}{b}\Biggr)^{m} \cdot
\Biggl[\prod_{k=1}^{m} \frac{k}{(k-3w)} \Biggr] \cdot J_{0}(x,w)
\end{eqnarray}
For the case $w=0$, the above formula reduces to
\begin{eqnarray}
\label{Series-w=0} J_{m}(x, 0) = \sum_{l=1}^{m}
\frac{(-1)^{(m-l)}}{l b} \Biggl( \frac{a}{b} \Biggr)^{(m-l)}\Bigl[
x^l - (-1)^{l} \Bigl] + (-1)^{m} \Biggl( \frac{a}{b}
\Biggr)^{m}\cdot J_{0}(x, 0)
\nonumber \\
\end{eqnarray}

\subsection{Calculation of the integrals $K_n(x,w)$.}
        \label{integrationK}

Now we calculate the integrals $K_n(x,w)$ defined in
(\ref{IntegralK}) by
\begin{eqnarray}
\label{IntegralKApp} K_{n}(x, w) \equiv \int_{-1}^{x}
\frac{T_{n}(\tilde{x})}{(a+b\tilde{x})^{(1+3w)}} \, d\tilde{x} \,\,,
\end{eqnarray}

To this end we use the following representation for Chebyshev
polynomials with odd and even integer subindex,
\begin{eqnarray}
\label{RepresentationPCodd} T_{2m+1}(\tilde x) = \frac{2m+1}{2}
\cdot \sum_{l=0}^{m} \frac{(-1)^{(m+l)}(m+l)!}{(m-l)! \cdot (2l+1)!}
\cdot (2\tilde x)^{2l+1}  \qquad \mbox{for} \,\, m \geq 0 ,\nonumber \\
\end{eqnarray}

\begin{eqnarray}
\label{RepresentationPCeven} T_{2m}(\tilde x) = m \cdot
\sum_{l=0}^{m} (-1)^{(m+l)} \cdot \frac{(m+l+1)!}{(m-l)! \cdot
(2l)!} \cdot (2\tilde x)^{2l}  \qquad \mbox{for} \,\, m \geq 1
\end{eqnarray}

Introducing (\ref{RepresentationPCodd}) and
(\ref{RepresentationPCeven}) in (\ref{IntegralKApp}) and it doing
the explicit integration we have the closed forms for the integrals
with odd integer subindex, valid for $n \geq 0$, $w \neq (n+1)/3$,
\begin{eqnarray}
\label{closedsolutionKodd} K_{2n+1}(x, w) & = \frac{(2n+1)}{2b}
\cdot \sum_{m=0}^{n} \sum_{l=1}^{2m+1} (-1)^{(n+3m-l+1)} \cdot
\frac{(n+m)! \cdot (2)^{2m+1}}{(n-m)!\cdot l!} \cdot \nonumber
\\
& \quad \qquad \qquad \qquad \quad \quad \cdot
\Biggl(\frac{a}{b}\Biggr)^{(2m+1-l)} \cdot \Biggl[
\prod_{k=l}^{2m+1} \frac{1}{(k-3w)} \Biggr] \cdot \nonumber
\\
& \quad \qquad \qquad \qquad \quad \quad \cdot \Biggl[
\frac{x^l}{(a+bx)^{3w}} -(-1)^{l} \Biggr] \nonumber
\\
& + \frac{(2n+1)}{2} \cdot \sum_{m=0}^{n} \frac{(-1)^{(n+3m+1)}\cdot
(n+m)! \cdot (2)^{2m+1}}{(n-m)!\cdot (2m+1)!}
\Biggl(\frac{a}{b}\Biggr)^{(2m+1)}
\nonumber \\
& \nonumber \\ & \quad \qquad \qquad \qquad \cdot
\Biggl[\prod_{k=1}^{2m+1} \frac{k}{(k-3w)} \Biggr] \cdot K_{0}(x,w)
\end{eqnarray}
meanwhile the integrals with even integer subindex, valid for $n
\geq 1$, $w \neq n/3$, are
\begin{eqnarray}
\label{closedsolutionKeven} K_{2n}(x, w) & = \frac{n}{b} \cdot
\sum_{m=1}^{n} \sum_{l=1}^{2m} (-1)^{(n+3m-l)} \cdot \frac{(n+m-1)!
\cdot (2)^{2m}}{(n-m)!\cdot l!} \Biggl(\frac{a}{b}\Biggr)^{(2m-l)}
\cdot \nonumber
\\
& \quad \qquad \qquad \quad \cdot \Biggl[ \prod_{k=l}^{2m}
\frac{1}{(k-3w)} \Biggr] \cdot \Biggl[ \frac{x^l}{(a+bx)^{3w}} -
(-1)^{l} \Biggr] \nonumber
\\
& + \Biggl[ 1 + n \sum_{m=1}^{n} \frac{(-1)^{3m} (n+m-1)!
(2)^{2m}}{(n-m)!\cdot (2m)!} \Biggl(\frac{a}{b}\Biggr)^{2m}\cdot
\prod_{k=1}^{2m} \frac{k}{(k-3w)} \Biggr] \nonumber \\
& \nonumber \\ & \cdot (-1)^{n} \cdot K_{0}(x,w)
\end{eqnarray}
where the initial function $K_{0}(x,w)$ is given by
\begin{equation}
\label{Integral-initialK} K_{0}(x, w) = \left\{
\begin{array}{rl}
\frac{1}{b} \ln (a+bx) & \mbox{if} \quad w = 0,\\
\frac{1}{3wb} \Biggl[ 1 - \frac{1}{(a+bx)^{3w}} \Biggr] & \mbox{if}
\quad w \neq 0.
\end{array} \right.
\end{equation}


        \section{Reconstruction of the interaction using Chebyshev polynomials up to order $5$.}
        \label{Reconstruction5}

To simplify our analysis and to show how the method works we do the
reconstruction taking a expansion in terms of Chebyshev polynomials
up to order $N=5$. The first step is to calculate the first fifth
integrals $J_n(x,w)$:
\begin{eqnarray}
\label{Integrals-J5} J_{1}(x, w) &=& \frac{1}{b(1-3w)} \Biggl\{
\Biggl[
\frac{x}{(a+bx)^{3w}} + 1 \Biggr] - a J_{0}(x, w) \Biggr\} \\
J_{2}(x, w) &=&  \frac{1}{b(2-3w)} \Biggl[ \frac{x^2}{(a+bx)^{3w}} -
1 \Biggr] + \nonumber \\
&& \frac{2a}{b^2 (2-3w)(1-3w)} \Biggl\{ aJ_{0}(x, w) - \Biggl[
\frac{x}{(a+bx)^{3w}} + 1 \Biggr] \Biggr\}\\
J_{3}(x, w) &=& \frac{1}{b(3-3w)} \Biggl[ \frac{x^3}{(a+bx)^{3w}} +
1 \Biggr] -
\nonumber \\
&& \frac{3a}{b^2 (3-3w)(2-3w)} \Biggl[ \frac{x^2}{(a+bx)^{3w}} - 1
\Biggr] + \nonumber \\
&& \frac{6a^2}{b^3 (3-3w)(2-3w)(1-3w)} \cdot \nonumber \\
&& \Biggl\{ \Biggl[ \frac{x}{(a+bx)^{3w}} + 1 \Biggr] - aJ_{0}(x,w)
\Biggr\} \\
J_{4}(x, w) &=&  \frac{1}{b(4-3w)} \Biggl[ \frac{x^4}{(a+bx)^{3w}} -
1 \Biggr] - \nonumber \\
&& \frac{4a}{b^2 (4-3w)(3-3w)} \Biggl[ \frac{x^3}{(a+bx)^{3w}} + 1
\Biggr] + \nonumber \\
&& \frac{12a^2}{b^3 (4-3w)(3-3w)(2-3w)} \Biggl[
\frac{x^2}{(a+bx)^{3w}} - 1
\Biggr] + \nonumber \\
&& \frac{24a^3}{b^4 (4-3w)(3-3w)(2-3w)(1-3w)} \cdot \nonumber \\
&& \Biggl\{ aJ_0(x,w) - \Biggl[ \frac{x}{(a+bx)^{3w}} + 1
\Biggr] \Biggr\}  \\
J_{5}(x, w) &=& \frac{1}{b(5-3w)} \Biggl[ \frac{x^5}{(a+bx)^{3w}} +
1 \Biggr] - \nonumber \\
&& \frac{5a}{b^2 (5-3w)(4-3w)} \Biggl[ \frac{x^4}{(a+bx)^{3w}} - 1
\Biggr] + \nonumber \\
&& \frac{20a^2}{b^3 (5-3w)(4-3w)(3-3w)} \Biggl[
\frac{x^3}{(a+bx)^{3w}} + 1
\Biggr] - \nonumber \\
&& \frac{60a^3}{b^4 (5-3w)(4-3w)(3-3w)(2-3w)} \Biggl[
\frac{x^2}{(a+bx)^{3w}} - 1
\Biggr] + \nonumber \\
&& \frac{120a^4}{b^5 (5-3w)(4-3w)(3-3w)(2-3w)(1-3w)} \cdot \nonumber \\
&& \Biggl\{ \Biggl[ \frac{x}{(a+bx)^{3w}} + 1 \Biggr] - aJ_{0}(x,w)
\Biggr\}
\end{eqnarray}
where we have the definitions,
\begin{eqnarray}
\label{def-constants}
b &=& \frac{z_{max}}{2}  \\
a &=& 1 + \frac{z_{max}}{2} \\
x &=& \frac{2\,z}{z_{max}} - 1
\end{eqnarray}
\begin{eqnarray}
a + bx &=& 1 + z
\end{eqnarray}
At the other hand, the first fifth Chebyshev polynomials are:
\begin{eqnarray}
\label{Chebyshev}
T_{0}(x) &=& 1 \\
T_{1}(x) &=& x  \\
T_{2}(x) &=& 2x^{2} - 1   \\
T_{3}(x) &=& 4x^{3} - 3x  \\
T_{4}(x) &=& 8x^{4} - 8x^{2} + 1 \\
T_{5}(x) &=& 16x^{5} - 20x^{3} + 5x
\end{eqnarray}
using these polynomials we find the relation between the integrals
(\ref{Integral-J}) and (\ref{IntegralKApp}),
\begin{eqnarray}
\label{relation-integrals}
K_{0}(x, w) &=& J_{0}(x, w) \\
K_{1}(x, w) &=& J_{1}(x, w) \\
K_{2}(x, w) &=& 2 J_{2}(x, w) - J_{0}(x, w) \\
K_{3}(x, w) &=& 4 J_{3}(x, w) - 3 J_{1}(x, w) \\
K_{4}(x, w) &=& 8 J_{4}(x, w) - 8 J_{2}(x, w)+ J_{0}(x, w) \\
K_{5}(x, w) &=& 16 J_{5}(x, w) - 20 J_{3}(x, w) + 5 J_{1}(x, w)
\end{eqnarray}
The general solutions (\ref{eq:Omega12})-(\ref{eq:Omega13}) up to
order $N$ can be written as
\begin{eqnarray}
\label{eq:Omegamatter5} \Omega^{\star}_{DM}(z) &=& (1+z)^{3}\left[
{\Omega_{DM}^0} - \frac{z_{max}}{2} \sum_{n=0}^{N}
\lambda_{n} \, \cdot K_{n}(x, 0) \right],\\
\label{eq:Omegaenergyfinal} \Omega^{\star}_{DE}(z) &=&
(1+z)^{3(1+w)}\left[ {\Omega_{DE}^0} + \frac{z_{max}}{2}
\sum_{n=0}^{N} \lambda_{n} \, \cdot K_{n}(x, w) \right]  \,,
\end{eqnarray}
Finally, the Hubble parameter is written as,
\begin{equation}
\label{hubbleAppendix} H^{2}\left(z \right) = H^2_{0}
\left[\Omega_{b}^{0}{(1+z)}^{3} + \Omega_{r}^{0}{(1+z)}^{4} +
\Omega^{\star}_{DM}(z) + \Omega^{\star}_{DE}(z) \right],
\end{equation}
With this formulation we do the reconstruction of the coupling
function ${\rm I}_{\rm Q}(z)$ for $N = 1, 2, 3, 4, 5$ respectively
using the using the type Ia Supernova SCP Union2 data set sample.





\section*{References.}

\begin{thebibliography}{120}


\bibitem{Riess1998}
SUPERNOVA SEARCH TEAM collaboration, A.G. Riess et al.,
\textit{Observational evidence from supernovae for an accelerating
universe and a cosmological constant, Astron. J.} \textbf{116}
(1998) 1009 [astro-ph/9805201] [SPIRES]; \\
SUPERNOVA COSMOLOGY PROJECT collaboration, S. Perlmutter et al.,
\textit{Measurements of $\Omega$ and $\Lambda$ from 42 high-redshift
supernovae, Astrophys. J.} \textbf{517} (1999) 565
[astro-ph/9812133] [SPIRES]; \\
SUPERNOVA SEARCH TEAM collaboration, J.L. Tonry et al.,
\textit{Cosmological results from high-z supernovae, Astrophys. J.}
\textbf{594} (2003) 1 [astro-ph/0305008] [SPIRES]; \\
SUPERNOVA SEARCH TEAM collaboration, A.G. Riess et al., \textit{Type
Ia supernova discoveries at $z > 1$ from the Hubble space telescope:
evidence for past deceleration and constraints on dark energy
evolution, Astrophys. J.} \textbf{607} (2004) 665 [astro-ph/0402512]
[SPIRES]; \\
SNLS collaboration, P. Astier et al., \textit{The supernova legacy
survey: measurement of $\Omega_M$, $\Lambda$ and w from the first
year data set, Astron. Astrophys.} \textbf{447} (2006) 31
[astro-ph/0510447] [SPIRES]; \\
A.G. Riess et al., \textit{New Hubble space telescope discoveries of
type Ia supernovae at $z > 1$: narrowing constraints on the early
behavior of dark energy, Astrophys. J.} \textbf{659} (2007) 98
[astro-ph/0611572] [SPIRES]; \\
T.M. Davis et al., \textit{Scrutinizing exotic cosmological models
using ESSENCE supernova data combined with other cosmological
probes, Astrophys. J.} \textbf{666} (2007) 716 [astro-ph/0701510]
[SPIRES]; \\
ESSENCE collaboration, W.M. Wood-Vasey et al., \textit{Observational
constraints on the nature of the dark energy: first cosmological
results from the ESSENCE supernova survey, Astrophys. J.}
\textbf{666} (2007) 694 [astro-ph/0701041] [SPIRES]; \\
M. Kowalski et al., \textit{Improved cosmological constraints from
new, old and combined supernova datasets, Astrophys. J.}
\textbf{686} (2008) 749 [arXiv:0804.4142] [SPIRES].

\bibitem{AmanullahUnion22010}
R. Amanullah et al., \textit{Spectra and light curves of six type Ia
supernovae at $0.511 < z < 1.12$ and the Union2 compilation,
Astrophys. J.} \textbf{716} (2010) 712 [arXiv:1004.1711] [SPIRES].



\bibitem{WMAP}
WMAP collaboration, C.L. Bennett et al., \textit{First year
Wilkinson Microwave Anisotropy Probe (WMAP) observations:
preliminary maps and basic results, Astrophys. J. Suppl.}
\textbf{148} (2003) 1 [astro-ph/0302207] [SPIRES]; \\
WMAP collaboration, D.N. Spergel et al., \textit{First year
Wilkinson Microwave Anisotropy Probe (WMAP) observations:
Determination of cosmological parameters, Astrophys. J. Suppl.}
\textbf{148} (2003) 175 [astro-ph/0302209] [SPIRES]; \\
WMAP collaboration, D.N. Spergel et al., \textit{Wilkinson Microwave
Anisotropy Probe (WMAP) three year results: implications for
cosmology, Astrophys. J. Suppl.} \textbf{170} (2007) 377
[astro-ph/0603449][SPIRES]. \\
WMAP collaboration, G. Hinshaw et al., \textit{Five-year Wilkinson
Microwave Anisotropy Probe (WMAP) observations: data processing, sky
maps, basic results, Astrophys. J. Suppl.} \textbf{180} (2009) 225
[arXiv:0803.0732] [SPIRES]. \\
WMAP collaboration, E. Komatsu et al., \textit{Five-Year Wilkinson
Microwave Anisotropy Probe (WMAP) Observations: Cosmological
Interpretation, Astrophys. J. Suppl.} \textbf{180} (2009) 330
[arXiv:0803.0547] [SPIRES].

\bibitem{Komatsu2011}
WMAP collaboration, E. Komatsu et al., \textit{Seven-Year Wilkinson
Microwave Anisotropy Probe (WMAP) Observations: Cosmological
Interpretation, Astrophys. J. Suppl.} \textbf{192} (2011) 18
[arXiv:1001.4538] [SPIRES].


\bibitem{SDSS}
SDSS collaboration, K. Abazajian et al., \textit{ The First data
release of the Sloan digital sky survey, Astron. J.} \textbf{126}
(2003) 2081 [astro-ph/0305492] [SPIRES]. \\
SDSS collaboration, K. Abazajian et al., \textit{ The Second data
release of the Sloan digital sky survey, Astron. J.} \textbf{128}
(2004) 502 [astro-ph/0403325] [SPIRES]. \\
SDSS collaboration, K. Abazajian et al., \textit{The Third Data
Release of the Sloan Digital Sky Survey, Astron. J.} \textbf{129}
(2005) 1755 [astro-ph/0410239] [SPIRES]. \\
SDSS collaboration, M. Tegmark et al., \textit{Cosmological
parameters from SDSS and WMAP, Phys. Rev.} \textbf{D 69} (2004)
103501 [astro-ph/0310723] [SPIRES]. \\
SDSS collaboration, D. J. Eisenstein et al., \textit{Detection of
the baryon acoustic peak in the large-scale correlation function of
SDSS luminous red galaxies, Astrophys. J.} \textbf{633} (2005) 560
[astro-ph/0501171] [SPIRES]. \\
SDSS collaboration, M. Tegmark et al., \textit{Cosmological
Constraints from the SDSS Luminous Red Galaxies, Phys. Rev.}
\textbf{D 74} (2006) 123507 [astro-ph/0608632] [SPIRES]. \\
SDSS collaboration, K. Abazajian et al., \textit{The Seventh Data
Release of the Sloan Digital Sky Survey, Astrophys. J. Suppl.}
\textbf{182} (2009) 543 [astro-ph/0812.0649] [SPIRES].

\bibitem{ReidSDSS}
SDSS collaboration, B. A. Reid et al., \textit{Baryon Acoustic
Oscillations in the Sloan Digital Sky Survey Data Release 7 Galaxy
Sample, Mon. Not. Roy. Astron. Soc.} \textbf{401} (2010) 2148
[astro-ph.CO/0907.1660] [SPIRES]. \\
SDSS collaboration, B. A. Reid et al., \textit{Cosmological
Constraints from the Clustering of the Sloan Digital Sky Survey DR7
Luminous Red Galaxies, Mon. Not. Roy. Astron. Soc.} \textbf{404}
(2010) 60 [astro-ph.CO/0907.1659] [SPIRES].


\bibitem{Peebles1988}
P. J. E. Peebles and B. Ratra, \textit{Cosmology with a Time
Variable Cosmological Constant, Astrophys. J.} \textbf{325} (1988)
L17.

\bibitem{Ratra1988}
B. Ratra and P. J. E. Peebles, \textit{Cosmological Consequences of
a Rolling Homogeneous Scalar Field, Phys. Rev.} \textbf{D 37} (1988)
3406.

\bibitem{Wetterich1988}
C. Wetterich, \textit{Cosmology and the Fate of Dilatation Symmetry,
Nucl. Phys.} \textbf{B 302} (1988) 668.

\bibitem{Ratra1995}
B. Ratra and P. J. E. Peebles, \textit{Inflation in an open
universe, Phys. Rev.} \textbf{D 52} (1995) 1837.

\bibitem{Wetterich1995}
C. Wetterich, \textit{An asymptotically vanishing time dependent
cosmological "constant", Astron. Astrophys.} \textbf{301} (1995) 321
[hep-th/9408025].

\bibitem{Steinhardt1997}
P. J. Steinhardt, \textit{Cosmological Challenges for the 21th
century}, in \textit{Critical Problems in Physics}, V. L. Fitch and
D. R. Marlow eds., Princeton University Press, Princeton U.S.A.
(1997), pg. 123.

\bibitem{Carroll1998}
S. M. Carroll, \textit{Quintessence and the rest of the world, Phys.
Rev. Lett.} \textbf{81} (1998) 3067 [astro-ph/9806099].

\bibitem{Caldwell1998}
R. R. Caldwell, R. Dave and P. J. Steinhardt, \textit{Quintessential
cosmology: Novel models of cosmological structure formation,
Astrophys. Space Sci.} \textbf{261} (1998) 303.

\bibitem{Zlatev1999}
J. Zlatev, L.-M. Wang and P. J. Steinhardt,
\textit{Quintessence, cosmic coincidence, and the cosmological
constant, Phys. Rev. Lett.} \textbf{82} (1999) 896
[astro-ph/9807002].

\bibitem{Steinhardt-Science}
N. A. Bahcall, J. P. Ostriker, S. Perlmutter, P. J. Steinhardt, The
\textit{Cosmic triangle: Assessing the state of the universe,
Science} \textbf{284} (1999) 1481 [astro-ph/9906463].

\bibitem{Steinhardt1999}
P. J. Steinhardt, L.-M. Wang and I. Zlatev, \textit{Cosmological
tracking solutions, Phys. Rev.} \textbf{D 59} (1999) 123504
[astro-ph/9812313].

\bibitem{Wang2000}
L.-M. Wang, R. R. Caldwell, J. P. Ostriker and P. J. Steinhardt,
\textit{Cosmic concordance and quintessence, Astrophys. J.}
\textbf{530} (2000) 17 [astro-ph/9901388].

\bibitem{Chimento2000}
L. P. Chimento, A. S. Jakubi and D. Pavon, \textit{Enlarged
quintessence cosmology, Phys. Rev.} \textbf{D 62} (2000) 063508
[astro-ph/0005070].

\bibitem{Sahni-2002}
V. Sahni, \textit{The Cosmological constant problem and
quintessence, Class. Quant. Grav.} \textbf{19} (2002) 3435
[astro-ph/0202076].

\bibitem{stefano-copeland-2002}
P. S. Corasaniti and E. J. Copeland, \textit{Constraining the
quintessence equation of state with SnIa data and CMB peaks, Phys.
Rev.} \textbf{D 65} (2002) 043004 [astro-ph/0107378].

\bibitem{Peebles2003}
P. J. E. Peebles and B. Ratra, \textit{The Cosmological constant and
dark energy, Rev. Mod. Phys.} \textbf{75} (2003) 559
[astro-ph/0207347].

\bibitem{Steinhardt2003}
P. J. Steinhardt, \textit{A quintessential introduction to dark
energy, Phil. Trans. Roy. Soc. Lond.} \textbf{A 361} (2003) 2497.

\bibitem{stefano-copeland-2003}
P. S. Corasaniti and E. J. Copeland, \textit{A Model independent
approach to the dark energy equation of state, Phys. Rev.} \textbf{D
67} (2003) 063521 [astro-ph/0205544].

\bibitem{Sahni2004}
V. Sahni, \textit{Dark matter and dark energy, Lect. Notes Phys.}
\textbf{653} (2004) 141 [astro-ph/0403324].

\bibitem{Alam-Sahni-Starobinsky-2004}
U. Alam, V. Sahni and A. A. Starobinsky, \textit{Is there supernova
evidence for dark energy metamorphosis?, Mon. Not. Roy. Astron.
Soc.} \textbf{354} (2004) 275 [astro-ph/0311364].

\bibitem{Alam-Sahni-Starobinsky-JCAP-2004}
U. Alam, V. Sahni and A. A. Starobinsky, \textit{The Case for
dynamical dark energy revisited, JCAP} \textbf{06} (2004) 008
[astro-ph/0403687].

\bibitem{stefano-copeland-2004}
P. S. Corasaniti, M. Kunz, D. Parkinson, E. J. Copeland and B. A.
Bassett, \textit{The Foundations of observing dark energy dynamics
with the Wilkinson Microwave Anisotropy Probe, Phys. Rev.} \textbf{D
70} (2004) 083006 [astro-ph/0406608].

\bibitem{Miao-Li-2011}
M. Li, X.-D. Li, S. Wang and Y. Wang, \textit{Dark Energy, Commun.
Theor. Phys.} \textbf{56} (2011) 525 [astro-ph.CO/1103.5870].

\bibitem{Copeland-Sami-2006}
E. J. Copeland, M. Sami and S. Tsujikawa, \textit{Dynamics of dark
energy, Int. J. Mod. Phys.} \textbf{D 15} (2006) 1753
[hep-th/0603057].


\bibitem{Weinberg1989}
S. Weinberg,\textit{The Cosmological Constant Problem, Rev. Mod.
Phys.} \textbf{61} (1989) 1.

\bibitem{Carroll1992}
S. M. Carroll, W. H. Press and E. L. Turner, \textit{The
Cosmological constant, Ann. Rev. Astron. Astrophys.} \textbf{30}
(1992) 499.

\bibitem{Weinberg1998}
S. Weinberg, \textit{Likely values of the cosmological constant,
Astrophys. J.} \textbf{492} (1998) 29 [astro-ph/9701099].

\bibitem{Sahni-Starobinsky-2000}
V. Sahni and A. A. Starobinsky, \textit{The Case for a positive
cosmological Lambda term, Int. J. Mod. Phys.} \textbf{D 9} (2000)
373 [astro-ph/9904398].

\bibitem{Weinberg2000}
S. Weinberg, \textit{A Priori probability distribution of the
cosmological constant, Phys. Rev.} \textbf{D 61} (2000) 103505
[astro-ph/0002387].

\bibitem{Carroll2001}
S. M. Carroll, \textit{The Cosmological constant, Living Rev. Rel.}
\textbf{4} (2001) 1 [astro-ph/0004075].

\bibitem{Padmanabhan2003}
T. Padmanabhan, \textit{Cosmological constant: The Weight of the
vacuum, Phys. Rept.} \textbf{380} (2003) 235 [hep-th/0212290].


\bibitem{Vikhlinin2009}
A. Vikhlinin et al., \textit{Chandra Cluster Cosmology Project III:
Cosmological Parameter Constraints, Astrophys. J.} \textbf{692}
(2009) 1060 [astro-ph/0812.2720].

\bibitem{Rozo2010}
E. Rozo et al., \textit{Cosmological Constraints from the SDSS
maxBCG Cluster Catalog, Astrophys. J.} \textbf{708} (2010) 645
[astro-ph.CO/0902.3702].



\bibitem{Amendola2000}
L. Amendola, \textit{Coupled quintessence, Phys. Rev.} \textbf{D 62}
(2000) 043511 [astro-ph/9908023] [SPIRES].

\bibitem{Amendola2000MNRAS}
L. Amendola,\textit{ Perturbations in a coupled scalar field
cosmology, Mon. Not. Roy. Astron. Soc.} \textbf{312} (2000) 521
[astro-ph/9906073] [SPIRES].

\bibitem{Billyard-Coley2000}
A. P. Billyard and A. A. Coley,\textit{Interactions in scalar field
cosmology, Phys. Rev.} \textbf{D 61} (2000) 083503
[astro-ph/9908224].

\bibitem{zimdahl-pavon-PLB-2001}
W. Zimdahl, D. Pavon and L. P. Chimento \textit{Interacting
quintessence, Phys. Lett.} \textbf{B 521} (2001) 133
[astro-ph/0105479].

\bibitem{Amendola2003}
L. Amendola and C. Quercellini, \textit{Tracking and coupled dark
energy as seen by WMAP, Phys. Rev.} \textbf{D 68} (2003) 023514
[astro-ph/0303228] [SPIRES].

\bibitem{Chimento-Jakubi2003}
 L. P. Chimento, A. S. Jakubi, D. Pavon and W. Zimdahl,
\textit{Interacting quintessence solution to the coincidence
problem, Phys. Rev.} \textbf{D 67} (2003) 083513 [astro-ph/0303145].

\bibitem{amendola2004}
L. Amendola, \textit{Linear and non-linear perturbations in dark
energy models, Phys. Rev.} \textbf{D 69} (2004) 103524
[astro-ph/0311175].

\bibitem{farrar}
G. R. Farrar and P. J. E. Peebles, \textit{Interacting dark matter
and dark energy, Astrophys. J.} \textbf{604} (2004) 1
[astro-ph/0307316].

\bibitem{zimdahl-pavon-GRG-2004}
W. Zimdahl and D. Pavon, \textit{Statefinder parameters for
interacting dark energy, Gen. Rel. Grav.} \textbf{36} (2004) 1483
[gr-qc/0311067].

\bibitem{herrera-pavon-zimdahlGRG-2004}
R. Herrera, D. Pavon and W. Zimdahl, \textit{Exact solutions for the
interacting tachyonic - dark matter system, Gen. Rel. Grav.}
\textbf{36} (2004) 2161 [astro-ph/0404086].

\bibitem{pavon}
D. Pavon and W. Zimdahl, \textit{Holographic dark energy and cosmic
coincidence, Phys. Lett.} \textbf{B 628} (2005) 206 [gr-qc/0505020].

\bibitem{pietroni2}
D. Comelli, M. Pietroni and A. Riotto, \textit{Dark energy and dark
matter, Phys. Lett.} \textbf{B 571} (2003) 115 [hep-ph/0302080].

\bibitem{campo-herrera2005}
S. Campo, R. Herrera and D. Pavon, \textit{Soft coincidence in late
acceleration, Phys. Rev.} \textbf{D 71} (2005) 123529
[astro-ph/0506482].

\bibitem{Guo-Cai2005}
Z.-K. Guo, R.-G. Cai and Y.-Z. Zhang, \textit{Cosmological evolution
of interacting phantom energy with dark matter, JCAP} 05 (2005) 002
[astro-ph/0412624].

\bibitem{zimdahl-IJMP-2005}
W. Zimdahl, \textit{Interacting dark energy and cosmological
equations of state, Int. J. Mod. Phys.} \textbf{D 14} (2005) 2319
[gr-qc/0505056].

\bibitem{sussman-quiros-osmel-2005}
R. A. Sussman, I. Quiros and O. M. Gonzalez, \textit{Inhomogeneous
models of interacting dark matter and dark energy, Gen. Rel. Grav.}
\textbf{37} (2005) 2117 [astro-ph/0503609].

\bibitem{campo-olivares2006}
S. Campo, R. Herrera, G. Olivares and D. Pavon, \textit{Interacting
models of soft coincidence, Phys. Rev.} \textbf{D 74} (2006) 023501
[astro-ph/0606520].

\bibitem{olivares-atrio2006}
G. Olivares, F. Atrio-Barandela and D. Pavon,\textit{Matter density
perturbations in interacting quintessence models, Phys. Rev.}
\textbf{D 74} (2006) 043521 [astro-ph/0607604].

\bibitem{Huey-Wandelt-2006}
G. Huey and B. D. Wandelt, \textit{Interacting quintessence. The
Coincidence problem and cosmic acceleration, Phys. Rev.} \textbf{D
74} (2006) 023519 [astro-ph/0407196].

\bibitem{chimento-pavon-PRD-2006}
L. P. Chimento and D. Pavon, \textit{Dual interacting cosmologies
and late accelerated expansion, Phys. Rev.} \textbf{D 73} (2006)
063511 [gr-qc/0505096].

\bibitem{barrow-clifton-2006}
J. D. Barrow and T. Clifton, \textit{Cosmologies with energy
exchange, Phys. Rev.} \textbf{D 73} (2006) 103520 [gr-qc/0604063].

\bibitem{corasaniti2006}
S. Das, P. S. Corasaniti and J. Khoury, \textit{Super-acceleration
as signature of dark sector interaction, Phys. Rev.} \textbf{D 73}
(2006) 083509 [astro-ph/0510628].

\bibitem{shinji-2006}
L. Amendola, S. Tsujikawa and M. Sami, \textit{Phantom damping of
matter perturbations, Phys. Lett.} \textbf{B 632} (2006) 155
[astro-ph/0506222] [SPIRES].

\bibitem{berger-shojaei-PRD73-2006}
M. S. Berger and H. Shojaei, \textit{Interacting dark energy and the
cosmic coincidence problem, Phys. Rev.} \textbf{D 73} (2006) 083528
[gr-qc/0601086].

\bibitem{berger-shojaei-PRD74-2006}
M. S. Berger and H. Shojaei, \textit{An Interacting Dark Energy
Model for the Expansion History of the Universe, Phys. Rev.}
\textbf{D 74} (2006) 043530 [astro-ph/0606408].

\bibitem{macorra2007}
A. de la Macorra, \textit{Interacting Dark Energy: Decay into
Fermions, Astropart. Phys.} \textbf{28} (2007) 196
[astro-ph/0702239].

\bibitem{campo-herrera-pavon2008}
S. Campo, R. Herrera and D. Pavon, \textit{Toward a solution of the
coincidence problem, Phys. Rev.} \textbf{D 78} (2008) 021302(R)
[astro-ph/0806.2116].

\bibitem{corasaniti2008}
P. S. Corasaniti, \textit{Slow-Roll Suppression of Adiabatic
Instabilities in Coupled Scalar Field-Dark Matter Models, Phys.
Rev.} \textbf{D 78} (2008) 083538 [astro-ph/0808.1646].

\bibitem{Bean-Flanagan-Trodden-2008}
R. Bean, E. E. Flanagan and M. Trodden, \textit{Adiabatic
instability in coupled dark energy-dark matter models, Phys. Rev.}
\textbf{D 78} (2008) 023009 [astro-ph/0709.1128].

\bibitem{berger-shojaei-PRD77-2008}
M. S. Berger and H. Shojaei, \textit{Possible Equilibria of
Interacting Dark Energy Models, Phys. Rev.} \textbf{D 77} (2008)
123504 [gr-qc/0712.2450].

\bibitem{brookfield2008}
A. W. Brookfield, C. van de Bruck and L. M. H. Hall, \textit{New
interactions in the dark sector mediated by dark energy, Phys. Rev.}
\textbf{D 77} (2008) 043006 [astro-ph/0709.2297].

\bibitem{jesus-santos-alcaniz-2008}
J. F. Jesus, R. C. Santos, J. S. Alcaniz and J. A. S. Lima,
\textit{New coupled quintessence cosmology, Phys. Rev.} \textbf{D
78} (2008) 063514 [astro-ph/0806.1366].

\bibitem{pettorino2008}
V. Pettorino and C. Baccigalupi, \textit{Coupled and Extended
Quintessence: theoretical differences and structure formation, Phys.
Rev.} \textbf{D 77} (2008) 103003 [astro-ph/0802.1086].

\bibitem{valiviita-majerotto-maartens-2008}
J. Valiviita, E. Majerotto and R. Maartens, \textit{Instability in
interacting dark energy and dark matter fluids, JCAP} \textbf{07}
(2008) 020 [astro-ph/0804.0232].

\bibitem{Calvao-Joras-2008}
M. Quartin, M. O. Calvao, S. E. Joras, R. R. R. Reis and I. Waga,
\textit{Dark Interactions and Cosmological Fine-Tuning, JCAP}
\textbf{05} (2008) 007 [astro-ph/0802.0546].

\bibitem{JianHe-Wang-Abdalla2009}
J.-H. He, B. Wang, E. Abdalla, \textit{Stability of the curvature
perturbation in dark sectors' mutual interacting models, Phys.
Lett.} \textbf{B 671} (2009) [gr-qc/0807.3471].

\bibitem{campo-herrera-pavon2009}
S. Campo, R. Herrera and D. Pavon, \textit{Interacting models may be
key to solve the cosmic coincidence, JCAP} \textbf{01} (2009) 020
[gr-qc/0812.2210].

\bibitem{chimento-forte-kremer-2009}
L. P. Chimento, M. I. Forte, G. M. Kremer, \textit{Cosmological
model with interactions in the dark sector, Gen. Rel. Grav.}
\textbf{41} (2009) 1125 [astro-ph/0711.2646].

\bibitem{chimento-2010-interactions}
L. P. Chimento, \textit{linear and nonlinear interactions in the
dark sector, Phys. Rev.}, \textbf{D81} (2010) 043525.
[astro-ph.CO/0911.5687].

\bibitem{abdalla}
S. Micheletti, E. Abdalla, B. Wang, \textit{A Field Theory Model for
Dark Matter and Dark Energy in Interaction, Phys. Rev.} \textbf{D
79} (2009) 123506 [gr-qc/0902.0318] [SPIRES].

\bibitem{cabral-maartens-schaefer2009}
G. Caldera-Cabral, R. Maartens and B. M. Schaefer, \textit{The
Growth of Structure in Interacting Dark Energy Models, JCAP}
\textbf{07} (2009) 027 [astro-ph.CO/0905.0492].

\bibitem{Gavela-Hernandez-Lopez2009}
M. B. Gavela, D. Hernandez, L. Lopez Honorez, O. Mena and S.
Rigolin, \textit{Dark coupling, JCAP} \textbf{07} (2009) 034
[astro-ph/0901.1611].

\bibitem{Lopez-Mena-Grigoris2010}
L. Lopez Honorez, O. Mena and G. Panotopoulos, \textit{Higher-order
coupled quintessence, Phys. Rev.} \textbf{D 82} (2010) 123525
[astro-ph.CO/1009.5263].

\bibitem{Gavela-Lopez-Mena-2010}
M. B. Gavela, L. Lopez Honorez, O. Mena and S. Rigolin, \textit{Dark
Coupling and Gauge Invariance, JCAP} \textbf{11} (2010) 044
[astro-ph.CO/1005.0295].

\bibitem{macorra2010}
U. Filobello and A. de la Macorra, \textit{Accelerating Cosmological
Models for an Interacting Tachyon, Nuovo Cim.} \textbf{B 125} (2010)
315 [hep-ph/0909.4241].

\bibitem{valiviita-majerotto-maartens-2010-1}
E. Majerotto, J. Valiviita and R. Maartens, \textit{Adiabatic
initial conditions for perturbations in interacting dark energy
models, Mon. Not. Roy. Astron. Soc.} \textbf{402} (2010) 2344
[astro-ph.CO/0907.4981].

\bibitem{fabris-fraga2010}
J. C. Fabris, B. Fraga, N. Pinto-Neto and W. Zimdahl,
\textit{Transient cosmic acceleration from interacting fluids, JCAP}
04 (2010) 008 [astro-ph.CO/0910.3246].

\bibitem{Lip2011}
S. Z. W. Lip, \textit{Interacting Cosmological Fluids and the
Coincidence Problem, Phys. Rev.} \textbf{D 83} (2011) 023528
[gr-qc/1009.4942].


\bibitem{Curbelo-Tame-Quiros2006}
R. Curbelo, T. Gonzalez and I. Quiros, \textit{Interacting phantom
energy and avoidance of the big rip singularity, Class. Quant.
Grav.} \textbf{23} (2006) 1585 [astro-ph/0502141].

\bibitem{macorra2008}
A. de la Macorra, \textit{Interacting dark energy: Generic
cosmological evolution for two scalar fields, JCAP} \textbf{01}
(2008) 030 [astro-ph/0703702].

\bibitem{macorra2008}
A. de la Macorra and U. Filobello, \textit{Interacting Tachyon:
Generic cosmological evolution for a tachyon and a scalar field,
Phys. Rev.} \textbf{D 77} (2008) 023531 [hep-th/0705.2059].

\bibitem{tame-quiros2008}
T. Gonzalez and I. Quiros, \textit{Exact models with non-minimal
interaction between dark matter and (either phantom or quintessence)
dark energy, Class. Quant. Grav.} \textbf{25} (2008) 175019
[gr-qc/0707.2089].

\bibitem{cabral-maartens2008}
C. G. Boehmer, G. Caldera-Cabral, R. Lazkoz and R. Maartens,
\textit{Dynamics of dark energy with a coupling to dark matter,
Phys. Rev.} \textbf{D 78} (2008) 023505 [gr-qc/0801.1565].

\bibitem{cabral-maartens-urena2009}
G. Caldera-Cabral, R. Maartens and L. A. Ure\~{n}a-L\'{o}pez,
\textit{Dynamics of interacting dark energy, Phys. Rev.} \textbf{D
79} (2009) 063518 [gr-qc/0812.1827].

\bibitem{cabral-maartens2010}
C. G. Boehmer, G. Caldera-Cabral, N. Chan, R. Lazkoz and R.
Maartens, \textit{Quintessence with quadratic coupling to dark
matter, Phys. Rev.} \textbf{D 81} (2010) 083003 [gr-qc/0911.3089].

\bibitem{saridakis2009}
M. Saridakis, \textit{Phase space analysis of interacting phantom
cosmology, JCAP} \textbf{04} (2009) 001 [gr-qc/0812.1117].


\bibitem{pasqui}
L. Amendola, C. Quercellini, D. T. Valentini and A. Pasqui,
\textit{Constraints on the interaction and selfinteraction of dark
energy from cosmic microwave background, Astrophys. J.} \textbf{583}
(2003) L53 [astro-ph/0205097].

\bibitem{amendola-piazza-2004}
L. Amendola, M. Gasperini and F. Piazza, \textit{Fitting type Ia
supernovae with coupled dark energy, JCAP} \textbf{09} (2004) 014
[astro-ph/0407573].

\bibitem{pavon-sen-zimdahl2004}
D. Pavon, S. Sen and W. Zimdahl, \textit{CMB constraints on
interacting cosmological models, JCAP} \textbf{05} (2004) 009
[astro-ph/0402067].

\bibitem{olivares-atrio2005}
G. Olivares, F. Atrio-Barandela and D. Pavon,\textit{Observational
constraints on interacting quintessence models, Phys. Rev.}
\textbf{D 71} (2005) 063523 [astro-ph/0503242].

\bibitem{Cai2005} R. G. Cai and A. Wang, \textit{Cosmology with
interaction between phantom dark energy and dark matter and the
coincidence problem, JCAP} \textbf{03} (2005) 002 [hep-th/0411025].

\bibitem{wang}
B. Wang, Y. G. Gong and E. Abdalla, \textit{Transition of the dark
energy equation of state in an interacting holographic dark energy
model, Phys. Lett.} \textbf{B 624} (2005) 141 [hep-th/0506069]
[SPIRES].

\bibitem{Lin}
B. Wang, C. Y. Lin and E. Abdalla, \textit{Constraints on the
interacting holographic dark energy model, Phys. Lett.} \textbf{B
637} (2006) 357 [hep-th/0509107] [SPIRES].

\bibitem{Lee-Liu-2006}
S. Lee, G.-C. Liu and K.-W. Ng, \textit{Constraints on the coupled
quintessence from cosmic microwave background anisotropy and matter
power spectrum, Phys. Rev.} \textbf{D 73} (2006) 083516
[astro-ph/0601333].

\bibitem{mainini-bonometto-2007}
R. Mainini and S. Bonometto, \textit{Limits on coupling between dark
components, JCAP} \textbf{06} (2007) 020 [astro-ph/0703303].

\bibitem{Guo-Ohta-2007}
Z.-K. Guo, N. Ohta and S. Tsujikawa, \textit{Probing the Coupling
between Dark Components of the Universe, Phys. Rev.} \textbf{D 76}
(2007) 023508 [astro-ph/0702015].

\bibitem{Micheletti}
B. Wang, J. Zang, C. Y. Lin, E. Abdalla and S. Micheletti,
\textit{Interacting Dark Energy and Dark Matter: Observational
Constraints from Cosmological Parameters, Nucl. Phys.} \textbf{B
778} (2007) 69 [astro-ph/0607126] [SPIRES].

\bibitem{rosenfeld}
L. Amendola, G. C. Campos and R. Rosenfeld, \textit{Consequences of
dark matter-dark energy interaction on cosmological parameters
derived from SNIa data, Phys.Rev.}, \textbf{D 75} (2007) 083506
[astro-ph/0610806] [SPIRES].

\bibitem{bertolami-pedro-delliou-2007}
O. Bertolami, F. G. Pedro and M. Le Delliou, \textit{Dark
Energy-Dark Matter Interaction and the Violation of the Equivalence
Principle from the Abell Cluster A586, Phys. Lett.} \textbf{B 654}
(2007) 165 [astro-ph/0703462].

\bibitem{bertolami-pedro-delliou-2008}
O. Bertolami, F. G. Pedro and M. Le Delliou, \textit{Dark
Energy-Dark Matter Interaction from the Abell Cluster A586},
[astro-ph/0801.0201].

\bibitem{olivares-atrio2008}
G. Olivares, F. Atrio-Barandela and D. Pavon,\textit{Dynamics of
Interacting Quintessence Models: Observational Constraints, Phys.
Rev.} \textbf{D 77} (2008) 063513 [astro-ph/0706.3860].

\bibitem{wu-gong-wang-alcaniz-2008}
Q. Wu, Y. Gong, A. Wang and J. S. Alcaniz, \textit{Current
constraints on interacting holographic dark energy, Phys. Lett.}
\textbf{B 659}(2008) 34 [astro-ph/0705.1006].

\bibitem{feng}
C. Feng, B. Wang, E. Abdalla and R.-K. Su, \textit{Observational
constraints on the dark energy and dark matter mutual coupling,
Phys. Lett.} \textbf{B 665} (2008) 111 [astro-ph/0804.0110]
[SPIRES].

\bibitem{He-Wang-2008}
J.-H. He and B. Wang, \textit{Effects of the interaction between
dark energy and dark matter on cosmological parameters, JCAP}
\textbf{06} (2008) 010 [astro-ph/0801.4233].

\bibitem{Bean-Flanagan-Laszlo-Trodden-2008}
R. Bean, E. E. Flanagan, I. Laszlo and M. Trodden,
\textit{Constraining Interactions in Cosmology's Dark Sector, Phys.
Rev.} \textbf{D 78} (2008) 123514 [astro-ph/0808.1105].

\bibitem{Schafer2008}
B. M. Sch\"{a}fer, \textit{The integrated Sachs-Wolfe effect in
cosmologies with coupled dark matter and dark energy, Mon. Not. Roy.
Astron. Soc.} 388 (2008) 1403 [astro-ph/0803.2239].

\bibitem{bertolami-pedro-delliou-2009}
O. Bertolami, F. G. Pedro and M. Le Delliou, \textit{The Abell
Cluster A586 and the Equivalence Principle, Gen. Rel. Grav.}
\textbf{41} (2009) 2839 [astro-ph/0705.3118].

\bibitem{Jun-Qing2009}
J.-Q. Xia, \textit{Constraint on coupled dark energy models from
observations, Phys. Rev.} \textbf{D 80} (2009) 103514
[astro-ph.CO/0911.4820].

\bibitem{He-Wang-Zhang-2009}
J.-H. He, B. Wang and P. Zhang, \textit{Imprint of the interaction
between dark sectors in large scale cosmic microwave background
anisotropies, Phys. Rev.} \textbf{D 80} (2009) 063530
[gr-qc/0906.0677].

\bibitem{He-Wang-Jing-2009}
J.-H. He, B. Wang and Y. P. Jing, \textit{Effects of dark sectors'
mutual interaction on the growth of structures, JCAP} \textbf{07}
(2009) 030 [gr-qc/0902.0660].

\bibitem{koyama-maartens2009}
K. Koyama, R. Maartens and Y. S. Song, \textit{Velocities as a probe
of dark sector interactions, JCAP} \textbf{10} (2009) 017
[astro-ph.CO/0907.2126].

\bibitem{valiviita-majerotto-maartens-2010-2}
J. Valiviita, R. Maartens and E. Majerotto, \textit{Observational
constraints on an interacting dark energy model, Mon. Not. Roy.
Astron. Soc.} \textbf{402} (2010) 2355 [astro-ph.CO/0907.4987].

\bibitem{izquierdo-pavon-2010}
G. Izquierdo and D. Pavon, \textit{Limits on the parameters of the
equation of state for interacting dark energy, Phys. Lett.}
\textbf{B 688} (2010) 115 [astro-ph.CO/1004.2360].

\bibitem{abramo}
E. Abdalla, L. R. Abramo and J. C. C. de Souza, \textit{Signature of
the interaction between dark energy and dark matter in observations,
Phys. Rev.} \textbf{D 82} (2010) 023508 [gr-qc/0910.5236] [SPIRES].

\bibitem{He-Wang-Abdalla-2010}
J.-H. He, B. Wang, E. Abdalla and D. Pavon, \textit{The Imprint of
the interaction between dark sectors in galaxy clusters, JCAP}
\textbf{12} (2010) 022 [gr-qc/1001.0079].

\bibitem{cao-liang2010}
S. Cao and N. Liang, \textit{Testing the phenomenological
interacting dark energy with observational H(z) data},
[astro-ph.CO/1012.4879].

\bibitem{Lopez-Beth-Mena-Verde2010}
L. Lopez Honorez, B. A. Reid, O. Mena, L. Verde and R. Jimenez,
\textit{Coupled dark matter-dark energy in light of near Universe
observations, JCAP} \textbf{09} (2010) 029 [astro-ph.CO/1006.0877]

\bibitem{Martinelli-Lopez-Mena-2010}
M. Martinelli, L. Lopez Honorez, A. Melchiorri and O. Mena, Future
\textit{CMB cosmological constraints in a dark coupled universe,
Phys.Rev.} \textbf{D 81} (2010) 103534 [astro-ph.CO/1004.2410].

\bibitem{bernardis-melchiorri2011}
F. De Bernardis, M. Martinelli, A. Melchiorri, O. Mena and A.
Cooray, \textit{Future weak lensing constraints in a dark coupled
universe, Phys. Rev.} \textbf{D 84} (2011) 023504
[astro-ph.CO/1104.0652].

\bibitem{bertolami-pedro-delliou-2011}
O. Bertolami, F. G. Pedro and M. Le Delliou, \textit{Testing the
interaction of dark energy to dark matter through the analysis of
virial relaxation of clusters Abell Clusters A586 and A1689 using
realistic density profiles}, [astro-ph.CO/1105.3033].

\bibitem{abramo2}
J. He, B. Wang and E. Abdalla, \textit{Testing the interaction
between dark energy and dark matter via latest observations, Phys.
Rev.} \textbf{D 83} (2011) 063515 [astro-ph.CO/1012.3904] [SPIRES].

\bibitem{Xu-He-Wang-2011}
X.-D. Xu, J.-H. He and B. Wang, \textit{Breaking parameter
degeneracy in interacting dark energy models from observations,
Phys. Lett.} \textbf{B 701} (2011) 513 [astro-ph.CO/1103.2632].

\bibitem{cao-liang-zhu-2011}
S. Cao, N. Liang and Z.-H. Zhu, \textit{Interaction between dark
energy and dark matter: observational constraints from H(z), BAO,
CMB and SNe Ia}, [astro-ph.CO/1105.6274].

\bibitem{Cai-Su}
R. G. Cai and Q. Su, \textit{On the Dark Sector Interactions, Phys.
Rev.} \textbf{D 81} (2010) 103514 [astro-ph.CO/0912.1943] [SPIRES].

\bibitem{He-Zhang}
Y.-H. Li and X. Zhang, \textit{Running coupling: Does the coupling
between dark energy and dark matter change sign during the
cosmological evolution?, Eur. Phys. J.} \textbf{C 71} (2011) 1700
[astro-ph.CO/1103.3185] [SPIRES].

\bibitem{pavon1999GRGThermo}
D. Pavon and B. Wang, \textit{Le Chatelier-Braun principle in
cosmological physics, Gen. Rel. Grav.}, \textbf{41} (2009) 1
[gr-qc/0712.0565].

\bibitem{Mbaldi}
M. Baldi, \textit{Early massive clusters and the bouncing coupled
dark energy, MNRAS}, \textbf{420} (2011) 430
[astro-ph.CO/1107.5049].


\bibitem{Simon2005}
J. Simon, L. Verde and R. Jimenez, \textit{Constraints on the
redshift dependence of the dark energy potential, Phys. Rev.}
\textbf{D 71} (2005) 123001 [astro-ph/0412269].

\bibitem{Martinez2008}
E. F. Martinez and L. Verde, \textit{Prospects in Constraining the
Dark Energy Potential, JCAP} \textbf{08} (2008) 023
[astro-ph/0806.1871].


\bibitem{Rosenfeld2007}
R. Rosenfeld, \textit{Reconstruction of interacting dark energy
models from parameterizations, Phys. Rev.} \textbf{D 75} (2007)
083509 [astro-ph/0701213].



\bibitem{Freedman2001}
HST Collaboration, W. L. Freedman et al, \textit{Final results from
the Hubble Space Telescope key project to measure the Hubble
constant, Astrophys. J.} \textbf{553} (2001) 47 [astro-ph/0012376].


\bibitem{nesseris1}
S. Nesseris and L. Perivolaropoulos, \textit{Tension and systematics
in the Gold06 SnIa dataset, JCAP} \textbf{02} (2007) 025
[astro-ph/0612653] [SPIRES].

\bibitem{nesseris2}
S. Nesseris and L. Perivolaropoulos, \textit{Crossing the phantom
divide: theoretical implications and observational status, JCAP}
\textbf{01} (2007) 018 [astro-ph/0610092] [SPIRES].

\bibitem{nesseris3}
J. C. Bueno Sanchez, S. Nesseris and L. Perivolaropoulos,
\textit{Comparison of recent SnIa datasets, JCAP} \textbf{11} (2009)
029 [astro-ph.CO/0908.2636] [SPIRES].

\bibitem{wei}
H. Wei, \textit{Tension in the recent type Ia supernovae datasets,
Phys. Lett.} \textbf{B 687} (2010) 286 [astro-ph.CO/0906.0828]
[SPIRES].


\bibitem{kunz-corasaniti}
B. A. Bassett, P. S. Corasaniti, M. Kunz, \textit{The Essence of
quintessence and the cost of compression. Astrophys. J.}
\textbf{617} (2004) L1-L4 [astro-ph/0407364] [SPIRES].



\bibitem{Stern2010-2}
D. Stern, R. Jimenez, L. Verde, S. A. Stanford ans M. Kamionkowski,
\textit{Cosmic Chronometers: Constraining the Equation of State of
Dark Energy. II. A Spectroscopic Catalog of Red Galaxies in Galaxy
Clusters, Astrophys. J. Suppl.} \textbf{188} (2010) 280
[astro-ph.CO/0907.3152].

\bibitem{Riess-Hubble}
A. G. Riess et al., \textit{A Redetermination of the Hubble Constant
with the Hubble Space Telescope from a Differential Distance Ladder,
Astrophys. J.} \textbf{699} (2009) 539 [astro-ph.CO/0905.0695].

\bibitem{Gaztanaga}
E. Gaztanaga, A. Cabre and L. Hui, \textit{Clustering of Luminous
Red Galaxies IV: Baryon Acoustic Peak in the Line-of-Sight Direction
and a Direct Measurement of H(z), Mon. Not. Roy. Astron. Soc.}
\textbf{399} (2009) 1663 [astro-ph/0807.3551].

\bibitem{Stern2010-1}
D. Stern, R. Jimenez, L. Verde, M. Kamionkowski and S. A. Stanford,
\textit{Cosmic Chronometers: Constraining the Equation of State of
Dark Energy. I: H(z) Measurements, JCAP} \textbf{02} (2010) 008
[astro-ph.CO/0907.3149].




\bibitem{Sasaki1996}
S. Sasaki, \textit{A New Method to Estimate Cosmological Parameters
Using the Baryon Fraction of Clusters of Galaxies, Publ. Astron.
Soc. Jap.} \textbf{48} (1996) L119 [astro-ph/9611033].

\bibitem{Allen2002} S. W. Allen, R. W. Schmidt and A. C. Fabian,
\textit{Cosmological constraints from the x-ray gas mass fraction in
relaxed lensing clusters observed with Chandra, Mon. Not. Roy.
Astron. Soc.} \textbf{334} (2002) L11 [astro-ph/0205007].

\bibitem{Allen2004}
S. W. Allen, R. W. Schmidt, H. Ebeling, A. C. Fabian and L. Van
Speybroeck, \textit{Constraints on dark energy from Chandra
observations of the largest relaxed galaxy clusters, Mon. Not. Roy.
Astron. Soc.} \textbf{353} (2004) 457 [astro-ph/0405340].

\bibitem{Rapetti2005}
D. Rapetti, S. W. Allen and J. Weller, \textit{Constraining dark
energy with X-ray galaxy clusters, supernovae and the cosmic
microwave background, Mon. Not. Roy. Astron. Soc.} \textbf{360}
(2005) 555 [astro-ph/0409574].

\bibitem{Allen2008}
S. W. Allen, D. A. Rapetti, R. W. Schmidt, H. Ebeling, G. Morris and
A. C. Fabian, \textit{Improved constraints on dark energy from
Chandra X-ray observations of the largest relaxed galaxy clusters,
Mon. Not. Roy. Astron. Soc.} \textbf{383} (2008) 879
[astro-ph/0706.0033].

\bibitem{Rapetti2008}
D. Rapetti and S. W. Allen, \textit{The prospects for constraining
dark energy with future X-ray cluster gas mass fraction
measurements, Mon. Not. Roy. Astron. Soc.} \textbf{388} (2008) 1265
[astro-ph/0710.0440].


\bibitem{Cueva-Nucamendi} F. Cueva Solano and U. Nucamendi, In
Preparation (2010).






\end{thebibliography}
\end{document}